\documentclass[structabstract]{aa}
\usepackage{graphicx}
\usepackage{txfonts}
\usepackage{upgreek}
\usepackage{tipa}
\usepackage{verbatim}
\usepackage{natbib}
\usepackage{color}
\begin{document}

\newcommand{\hoA}{o-H$_2^{18}$O 1$_{10}$-1$_{01}$}
\newcommand{\hoB}{o-H$_2^{17}$O 1$_{10}$-1$_{01}$}
\newcommand{\hoC}{o-H$_2$O 1$_{10}$-1$_{01}$}
\newcommand{\hoD}{p-H$_2$O 2$_{11}$-2$_{02}$}
\newcommand{\hoE}{p-H$_2$O 2$_{02}$-1$_{11}$}
\newcommand{\hoF}{p-H$_2^{18}$O 2$_{02}$-1$_{11}$}
\newcommand{\hoG}{o-H$_2^{18}$O 3$_{12}$-3$_{03}$}
\newcommand{\hoH}{o-H$_2$O 3$_{12}$-3$_{03}$}
\newcommand{\hoI}{p-H$_2^{18}$O 1$_{11}$-0$_{00}$}
\newcommand{\hoJ}{p-H$_2^{17}$O 1$_{11}$-0$_{00}$}
\newcommand{\hoK}{p-H$_2$O 1$_{11}$-0$_{00}$}
\newcommand{\hoL}{o-H$_2$O 2$_{21}$-2$_{12}$}
\newcommand{\hoM}{o-H$_2^{17}$O 2$_{12}$-1$_{01}$}
\newcommand{\hoN}{o-H$_2$O 2$_{12}$-1$_{01}$}
\newcommand{\hoO}{p-H$_2^{18}$O 3$_{13}$-3$_{20}$}
\newcommand{\hoP}{p-H$_2$O 5$_{24}$-4$_{31}$}
\newcommand{\kms}{km~s$^{-1}$}
\newcommand{\ms}{m~s$^{-1}$}
\newcommand{\cms}{cm~s$^{-1}$}
\newcommand{\halp}{H$\alpha$}
\newcommand{\msun}{$\rm M_{\odot}$}
\newcommand{\etal}{et~al.~}
\newcommand{\vsini}{$v~sin~i$}
\newcommand{\ctemp}{$^{\circ}$C}
\newcommand{\ktemp}{$^{\circ}$K}
\newcommand{\be}{\begin{equation}}
\newcommand{\ee}{\end{equation}}
\newcommand{\bd}{\begin{displaymath}}
\newcommand{\ed}{\end{displaymath}}
\newcommand{\bi}{\begin{itemize}}
\newcommand{\ei}{\end{itemize}}
\newcommand{\bfig}{\begin{figure}}
\newcommand{\efig}{\end{figure}}
\newcommand{\bc}{\begin{center}}
\newcommand{\ec}{\end{center}}
\newcommand{\hii}{{H\scriptsize{II}}}
\newcommand{\vlsr}{V$_{\mathrm{LSR}}$}
\newcommand{\vtur}{V$_{\textrm{\tiny{tur}}}$}
\newcommand{\vexp}{V$_{\textrm{\tiny{exp}}}$}
\newcommand{\vinfall}{V$_{\textrm{\tiny{infall}}}$}
\newcommand{\coa}{$^{12}\mathrm{CO}$}
\newcommand{\cob}{$^{13}\mathrm{CO}$}
\newcommand{\coc}{$\mathrm{C}^{18}\mathrm{O}$}
\newcommand{\lsun}{L$_{\odot}$~}
\newcommand{\lfir}{L$_{\textrm{\tiny{FIR}}}$}
\newcommand{\agua}{$X_{\textrm{\tiny{H$_2$O}}}$}
\newcommand{\ratioop}{o$/$p}
\newcommand{\ratiosept}{$X_{\textrm{\tiny{$^{18}$O$/$$^{17}$O}}}$}
\newcommand{\ratiohuit}{$X_{\textrm{\tiny{$^{16}$O$/$$^{18}$O}}}$}
\newcommand{\watersept}{H$_2^{17}$O}
\newcommand{\waterhuit}{H$_2^{18}$O}
\newcommand{\water}{H$_2^{16}$O}
\newcommand{\lsol}{L$_\odot$\,}
\newcommand{\Msol}{M$_\odot$\,}

\def\etal{et al.$\;$}


\def\kms{km\thinspace s$^{-1}$}
\def\Lsun{L$_\odot$}
\def\Msun{M$_\odot$}
\def\ms{m\thinspace s$^{-1}$}
\def\percc{cm$^{-3}$}

\title{The Herschel-HIFI view of mid-IR quiet massive protostellar objects\thanks{Herschel is an ESA space observatory with science instruments provided by European-led Principal Investigator consortia and with important participation from NASA.}}

\titlerunning{The HIFI-HSO view of mid-IR quiet massive protostellar objects}

\subtitle{}

\author{F. Herpin,
	\inst{1,2}
	\and
	L. Chavarr\'{\i}a
	\inst{1,2,3}
	\and
	T. Jacq
	\inst{1,2}
	\and
	J. Braine
	\inst{1,2}
	\and
	F. van der Tak
	\inst{4}
	\and
	F. Wyrowski
	\inst{5}
	\and
	E. F. van Dishoeck
	\inst{6,7}
	\and
	A. Baudry
	\inst{1,2}
	\and
	S. Bontemps
	\inst{1,2}
	\and
	L. Kristensen
	\inst{8}
	\and
	M. Schmalzl
	\inst{6}
	\and
	J. Mata
	\inst{1}
}

\institute{
Univ. Bordeaux, LAB, UMR 5804, F-33270, Floirac, France.
\and 
CNRS, LAB, UMR 5804, F-33270, Floirac, France, 
	\email{herpin$@$obs.u-bordeaux1.fr}
\and
CONICYT-Universidad de Chile, Camino del Observatorio 1515, Santiago, Chile.
\and 
SRON Netherlands Institute for Space Research, PO Box 800, 9700AV, Groningen, The Netherlands
\and 
Max-Planck-Institut f\"ur Radioastronomie, Auf dem H\"ugel 69, 53121 Bonn, Germany
\and  
Leiden Observatory, Leiden University, PO Box 9513, 2300 RA, Leiden, The Netherlands
\and  
Max Planck Institut f\"ur Extraterrestrische Physik, Giessenbachstrasse 1, 85748 Garching, Germany
\and  
Harvard-Smithsonian Center for Astrophysics, 60 Garden Street, Cambridge, MA 02138, USA}

\date{Received; accepted}
\abstract
   {}  
   {We present Herschel/HIFI observations of 14 water lines in a small sample of galactic massive protostellar objects: NGC6334I(N), DR21(OH), IRAS16272-4837, and IRAS05358+3543. Using water as a tracer of the structure and kinematics, we aim to individually study each of these objects, to estimate the amount of water around them, but to also shed light on the high-mass star formation process.}
   {We analyze the gas dynamics from the line profiles using Herschel-HIFI observations acquired as part of the WISH key-project of 14 far-IR water lines (\water, \watersept, H$_2^{18}$O), and several other species. Then through modeling of the observations using the RATRAN radiative transfer code, we estimate outflow, infall, turbulent velocities, molecular abundances, and investigate any correlation with the evolutionary status of each source.}
   {The four sources (plus previously studied W43-MM1) have been ordered in terms of evolution based on their SED: NGC64334I(N)~$\rightarrow$ W43-MM1~$\rightarrow$ DR21(OH)~$\rightarrow$ IRAS16272-4837~$\rightarrow$ IRAS05358+3543. The molecular line profiles exhibit a broad component coming from the shocks along the cavity walls associated with the protostars, and an infalling (or expansion for IRAS05358+3543) and passively heated envelope component, with highly supersonic turbulence likely increasing with the distance from the center. Accretion rates between 6.3 $\times$ $10^{-5}$  and 5.6 $\times$ $10^{-4}$ \msun yr$^{-1}$ are derived from the infall observed in three of our sources. The outer water abundance is estimated to be at the typical value of a few $10^{-8}$ while the inner abundance varies from $1.7\times10^{-6}$ to $1.4\times10^{-4}$ with respect to H$_2$ depending on the source. 
}
   {We confirm that regions of massive star formation are highly turbulent and that the turbulence likely increases in the envelope with the distance to the star. The inner abundances are lower than the expected $10^{-4}$ perhaps because our observed lines do not probe deep enough into the inner envelope, or because photodissociation through protostellar UV photons is more efficient than expected. We show that the higher the infall/expansion velocity in the protostellar envelope, the higher is the inner abundance, maybe indicating that larger infall/expansion velocities generate shocks that will sputter water from the ice mantles of dust grains in the inner region. High-velocity water must be formed in the gas-phase from shocked material.
}


\maketitle

%

\section{Introduction}
\label{intro}
The importance of high-mass stars (M $> 8$ M$_{\odot}$) in the matter cycle in the Universe and the evolution of galaxies is well known  \citep[e.g.][]{zinnecker2007,tan2014}. As powerful sources of UV and driving strong outflows they deeply influence their environment. Despite this fundamental role, their formation is still not well understood due to their rareness, often large distances and embeddedness. 

\begin{table*}
\begin{small}
\caption{Source list and parameters }             
\label{source_list}      
\centering                          
\begin{tabular}{lccccccccccc}        
\hline\hline                 
Object & RA & Dec$^{a}$  & d$^{a}$ &  ${V_{LSR}}^{a}$ & L$_{\textrm{bol}}$$^{a}$   &  $M_{\textrm{env}}$$^{a}$ &  ${F_{12}}^{b}$ & ${F_{21}}^{b}$ & $L^{0.6}{M_{\textrm{env}}}^{-1}$ & $\lambda_{F_\textrm{\tiny max}}$ & $F_{35}/F_\textrm{\tiny total}^{c}$ \\   
                      & [h m s] & [$^{\circ}$ ' ''] &   [kpc] & [km s$^{-1}$]    &  [$10^3L_{\odot}$] &  [$M_{\odot}$]  & [Jy]  & [Jy] & & [$\mu$m] & \\
                      \hline
 NGC6334I(N)  &17 20 55.2 & --35 45 04.1  & 1.7 & -3.3  & 1.9 & 3826 &  0.0 & 0.6 & 0.02& 220.6& 0.7 \\
  W43-MM1  &   18 47 47.0 &   --01 54 28.0 &    5.5 &     +98.8 &    23 &    7550 &      0.0 &    0.7 &    0.05&    131.7&    0.5\\
  DR21(OH)  &   20 39 00.8 &   +42 22 48.0  &    1.5 &    -3.1 &    13 &    472 &     0.3 &    1.3 &    0.6&    100.0 &    1.7\\ 
  IRAS16272-4837  &   16 30 58.7 &    --48 43 55.0  &    3.4 &    -46.2 &    24 &    2170 &     0.5 &    11.6 &    0.2 &    108.0&    3.5\\
  IRAS05358+3543  &   05 39 13.1 &   +35 45 50.0  &    1.8 &    -17.6 &    6.3 &    142 &     0.2 &    17.2 &    1.4 &    81.1 &    7.7\\
\hline                                  
\end{tabular}
\tablefoot{$^{(a)}$ Taken from van der Tak et al. (2013) based on JCMT$/$SCUBA and APEX$/$LABOCA data. $^{(b)}$ The $F_{12}$ and $F_{21}$ are the MSX values for the 12 and 21 $\mu$m flux density (or values estimated from the SED fitting for W43-MM1 and NGC6334I(N)), corrected for the distance to each source ($F_{\lambda}\times (d/1.5$kpc$)^2$. $^{(c)}$ $F_{35}/F_\textrm{\tiny total}$ is the ratio of the integrated flux over the range 0-35 microns and the total integrated flux.}
\end{small}
\end{table*}

Empirically, high-mass star formation may be divided into several stages  \citep[e.g.][]{vandertak2000a,beuther2007,mottram2011} leading to the (not unique) following evolutionary sequence: from the initial stage massive pre-stellar cores (PSC's), to high-mass protostellar objects (HMPO's), hot molecular cores (HMC's), and  finally to the more evolved ultra-compact HII regions (UCHII). While pre-stellar cores represent a quiet phase with no significant luminosity \citep[][]{ragan2012} or activity \citep[infall, outflow, or maser activity,][]{shipman2014}, the HMPO stage is characterized by the presence of an active protostar exhibiting infall of a massive envelope onto the central star and strong outflows. Then, the temperature of the inner regions of the protostellar envelope will increase, going beyond the evaporation limits of molecules on the grains, hence enriching the envelope with complex molecules leading to the formation of a HMC. Later when the star gets hot enough to emit significant Lyman continuum, it will ionize the surrounding gas, leading to the formation of an UCHII. Based on the definition of \citet{motte2007}, HMPO sources can be divided in two different types depending on their 21 $\mu$m flux: the mid-IR-quiet dense cores ($F_{21}<10$ Jy) and the more evolved mid-IR bright ones ($F_{21}>10$ Jy). The distinction of mid-IR quiet and bright being different evolutionary stages is not obvious since geometry plays a significant role, and this still uncertain scenario may evolve and be constrained, e.g. by further observations. In this paper, the mid-IR quiet HMPOs will be studied. 

The "classical" issue in massive star formation about how the accretion of matter can overcome radiative pressure is now well addressed by models considering a protostar-disk system \citep[e.g.,][]{yorke2002, krumholz2005, banerjee2007}. More recently, the last generation simulations of \citet{kuiper2010,kuiper2015} demonstrate that disk accretion and protostellar outflows enable the accretion process to go on longer and then to reach final star masses well above the upper mass limit of spherically symmetric accretion. The two main theoretical scenarios both require the presence of a disk and high accretion rates: (a) a monolithic collapse scenario \citep[][]{mckee2002, mckee2003}, also called turbulent core model; and (b) a highly dynamical competitive accretion model involving the formation of a cluster \citep[][]{bonnel2006}. The turbulent core model implies supersonic turbulence in the protostellar envelope, while the competitive accretion model predicts cores that are subsonic, but still embedded in a supersonic envelope \citep[see][]{krumholz2009}. Moreover, massive star formation triggered by converging turbulent flows is predicted by numerical simulations \citep[e.g.][]{vazquez2007,heitsch2008} and proposed for several objects \citep[e.g. DR21(OH),][]{csengeri2011}. 

Several studies have been previously done on water chemistry in massive protostars, well before the golden age of IR satellites. \citet{jacq1990} observed the $3_{13}-2_{20}$ transition of \waterhuit~at 203 GHz, one of the very few thermal water lines observable from ground, toward several sources, including one presented in this paper, DR21(OH). They concluded that the water abundance is typically of order $10^{-5}$ (relative to H$_2$) or less in hot dense regions (where $T>100$ K). While in cold regions water is mostly found as ice on dust grains, at temperatures $T>$100 K the gas-phase water abundance increases by several orders of magnitudes as the ice evaporates \citep[][]{fraser2001, aikawa2008}. A second increase occurs at $T\ge$ 250 K, when gas-phase reactions drive all available oxygen into water \citep[e.g.,][]{charnley1997,vandishoeck2013}. The launch of ISO and SWAS satellites offered the possibility to observe for the first time ro-vibrational fundamental bands of water in absorption against bright infrared continuum sources. Hence, the inner water abundance has been probed toward AFGL2591 by \citet{helmich1996} and estimated to 2-6 $10^{-5}$, comparable to that of solid H$_2$O \citep[][]{vandishoeck1996}. Some uncertainties remain on the water content due to the low spectral resolution of the ISO data. The outer abundance (in region where T$<$ 100 K) has been estimated to 0.8-13 $10^{-9}$ by \citet{snell2000} using SWAS satellite data (from observation of the 557 GHz water line) in several sources (e.g. AFGL2591). Later \citet{boonman2003} combined ISO-SWS$/$LWS$/$SWAS data to derive an abundance profile in a set of massive protostars, including W3IRS5 and AFGL2591. Their preferred scenario is a model with ice evaporation occurring at T$\sim90-110$ K in the inner part ($\chi_{in}\sim10^{-4}$) with a low outer water abundance ($\chi_{out}\sim10^{-8}$). This jump-like radial profile for water has been confirmed by \citet{vandertak2006} using new interferometric observations from the ground of the 203 GHz \waterhuit~line, and by the results of time-dependent gas-grain chemical modeling by \citet{kazmierczak2015}. Observing the same \waterhuit~line in younger sources including two of our sample, \citet{marseille2010} have not been able to constrain the inner abundance, but have derived a global water abundance of 5 $10^{-8}$ and 5 $10^{-7}$, respectively for the HMPOs W43-MM1 and DR21(OH). 

As shown by \citet{chavarria2010}, \citet{herpin2012}, or \citet{wiel2013}, from single-dish observations it is only possible to trace the dynamics of gas  in the deeply embedded phase of star formation using spectrally resolved emission-line profiles. This has been done before with species like CS by \citet{vandertak2000a} and \citet{shirley2003}. Infall motion and supersonic turbulence in massive dense clouds have been further studied via HCN and CS surveys by \citet{wu2010}, or using HCO$^+$ and N$_2$H$^+$ by \citet{schlingman2011}. In addition, spectroscopic observations of the molecular content of the gas surrounding the protostellar object shed light on the complex chemistry occurring in these environments \citep[e.g.,][]{herpin2009,benz2010,bruderer2010,wyrowski2010}. Within the CHESS Herschel program \citep[][]{ceccarelli2010}, HIFI spectral survey of AFGL2591 has been performed, and the chemical structure of its protostellar envelope has been modeled by \citet{kazmierczak2015}. More generally, understanding the chemical evolution of HM protostars is now a very active field: for instance \citet{gerner2014} have shown that the chemical composition evolves along with the evolutionary stages.

\begin{table*}
\caption{Herschel/HIFI observed water line transitions in the source sample.}             
\label{table_transitions}      
\centering                          
\begin{tabular}{lccccccccc}        
\hline\hline                 
Species$/$transitions & Frequency$^b$ &  Wavelength & $E_u$    &  $n_{crit}$$^c$ & HIFI & Beam  &  $\eta_{\textrm{mb}}$  &  $T_{\textrm{sys}}$   &  rms$^d$ \\   
                      &    [GHz]     &       [$\mu$m] &      [K]  &  [cm$^{-3}$] &band &  [\arcsec]  &  & [K]  & [mK] \\
\hline                        
   \hoA$^a$    & 547.6764     &  547.4 &  60.5   & $3.5\times10^7$ & 1a & 37.8  & 0.62 & 80 & 27\\
   \hoB    & 552.0209    &  543.1  & 61.0    & $3.5\times10^7$ &  1a & 37.8 &  0.62 & 70  & 28 \\     
   \hoF     & 994.6751  &  301.4  &  100.6 &  $5.8\times10^7$ & 4a & 21.1   &  0.63 & 290  & 44  \\
   \hoG    & 1095.6274  &  273.8  &  248.7  & $1.6\times10^8$ &  4b & 19.9  &  0.63 & 380   & 32  \\
   \hoI    & 1101.6982  &  272.1  &   52.9  & $1.9\times10^8$ &  4b & 19.9 &  0.63 &  390   & 20   \\
   \hoJ    & 1107.1669  &  272.1  &   52.9  & $1.9\times10^8$  &   4b & 19.9 &  0.63 &  380   & 32 \\
   \hoM    & 1662.4644  &  180.3  &  113.6  & $5.6\times10^8$& 6b  & 12.7 &  0.58 & 1410   & 215   \\
\hline
   \hoC$^a$    & 556.9361    &  538.3  &   61.0  &  $3.5\times10^7$ & 1a & 37.1  & 0.62 & 80   & 27   \\
   \hoD    & 752.0332    &  398.6  &  136.9  & $7.1\times10^7$ & 2b & 28.0   &  0.64 &  90   & 36 \\
     \hoP    & 970.3150    &   309.0 &  598.8  & $9.0\times10^6$ & 4a & 21.8   &  0.63 &  620   & 25 \\
   \hoE     & 987.9268 &  303.5  &   100.8 &  $5.8\times10^7$ & 4a & 21.3  & 0.63 &  340   & 79 \\
   \hoH    & 1097.3651  &  273.2  &  249.4  & $1.6\times10^8$ & 4b  & 19.9  &  0.63 & 380   & 32   \\
   \hoK    & 1113.3430  &  269.0  &  53.4  & $1.9\times10^8$ &4b & 19.7   &  0.63 & 395   & 20 \\
   \hoL    & 1661.0076  &  180.5  &  194.1  &  $3.1\times10^8$ &6b & 12.7  &  0.58 & 1410   & 215 \\
   \hoN    & 1669.9048  &  179.5  &   114.4  &  $5.6\times10^8$ & 6b  & 12.6  &  0.58 & 1410   & 215  \\
\hline                                  
\end{tabular}
\tablefoot{$^a$ This line was mapped in OTF mode. $^b$ Frequencies are from \citet{pearson1991}. $^c$ At 100 K. $^d$ The rms is the noise in $\delta \nu=1.1$MHz.}
\end{table*}

Probing star formation using spectroscopic observations of  water was the major goal of the guaranteed-time key program {\it Water In Star-forming regions with Herschel}  \citep[WISH,][]{vandishoeck2011}. Water was indeed one of the main drivers of the Herschel Space Observatory mission \citep[hereafter Herschel,][]{Pilbratt2010} and particularly of the HIFI spectroscopy instrument \citep[][]{deGraauw2010}. The WISH program aims at characterizing the dynamics of the different components of the gas surrounding the central massive core and intends to  measure the amount of cooling that water lines provide. 

\begin{figure}
\centering
\includegraphics[width=7.5cm]{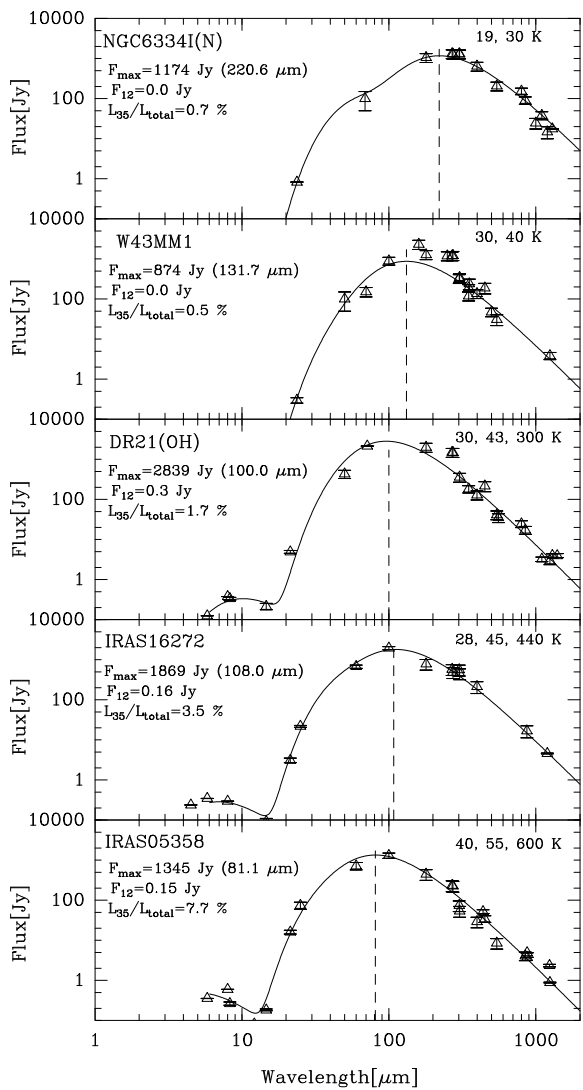}
\caption{Spectral energy distributions obtained from the 1D model overlaid on observed flux densities of each source. Peak flux densities in the millimeter range have been adjusted to fit the radial extent of the model. Each caption gives the temperatures of the black body components (top right), the maximum flux $F_\mathrm{max}$,  and the corresponding wavelength of the SED, the flux $F_{12}$ at 12$\mu$m, and the contribution of the hot part ($F_{35}$, integrated flux for $\lambda <$35 $\mu$m) to the total integrated flux $F_\mathrm{total}$ (top left).}
\label{Figsed}%
\end{figure}

A sample of 19 massive protostars, covering all phases of high-mass star formation has been observed within WISH.  In addition, 4 pre-stellar cores have been studied by \citet{shipman2014}. The velocity profiles of the low-excitation H$_2$O lines toward this sample have been presented by \citet{vandertak2013}, without detailed modeling. They decomposed HIFI water line spectra into three distinct physical components: (i) dense cores (protostellar envelopes) usually seen as medium or narrow absorption$/$emission; (ii)  outflows seen as broader features; and (iii) absorptions by foreground clouds along the line of sight. More generally, the line profiles obtained in low- \citep[][]{kristensen2010}, intermediate- \citep[][]{johnstone2010}, and high-mass \citep[see W3IRS5,][]{chavarria2010} young stellar objects exhibit similar velocity components \citep[][]{sanjose2013,mottram2014}: a broad (full width at half maximum FWHM $\sim$ 25 \kms), a medium ($\sim$5-10 \kms) one, and a narrower component ($<$5 \kms).

In this paper, we focus on the analysis of the water observations toward the mid-IR quiet massive protostars of the WISH sample. A similar study has been done by Choi {\em et al.} (in preparation) for mid-IR bright sources. Using the high velocity resolution of the HIFI instrument we study the dynamics of the gas, estimate the infall and turbulent velocities present in the protostellar envelopes, and derive the H$_2$O abundances in these sources. Sections \ref{sec:observations} and \ref{sec:sample} present our observations and source sample respectively. Results coming from Gaussian (water and other species) line fittings are given in Sect. \ref{sec:results}, while detailed line analysis is given in Sect. \ref{analysis}. We then model the observations using a radiative transfer code in Sect. \ref{sec:model}. We estimate the outflow and infall velocities, turbulent velocity, molecular abundances, and the physical structure of the sources. We finally discuss (Sect. \ref{sec:discussion}) the results in terms of massive-star formation scenarios and compare them to previous studies, in particular \citet{herpin2012}.

\section{Observations}
\label{sec:observations}

Fourteen water lines (see~Table \ref{table_transitions}) have been observed with HIFI at frequencies between 547 and 1670 GHz toward the whole source sample in 2010 and 2011 (list of observation identification numbers, {\em obsids}, are given in Appendix \ref{AOR_list}). An additional high-energy water line at 970.3150 GHz has been observed toward DR21(OH). The observations are part of the WISH GT-KP. 

Data were taken simultaneously in H and V polarizations using both the acousto-optical Wide-Band Spectrometer (WBS) with 1.1 MHz resolution and the digital auto-correlator or High-Resolution Spectrometer (HRS) providing higher spectral resolution (125 kHz). We used the Double Beam Switch observing mode with a throw of 3'. The off positions have been inspected and do not show any H$_2$O or significant continuum emission. The frequencies, energy of the upper levels, system temperatures, integration times and {\it rms} noise level at a given spectral resolution for each of the lines are provided in Table \ref{table_transitions}. Calibration of the raw data into the $T_A$ scale was performed by the in-orbit system \citep[][]{roelfsema2012}; conversion to $T_{mb}$ was done using the latest beam efficiency estimate from October 2014\footnote{http://herschel.esac.esa.int/twiki/pub/Sandbox/TestHifiInfoPage/} given in Table~\ref{table_transitions} and a forward efficiency of 0.96. HIFI receivers are double sideband with a sideband ratio close to unity \citep[][]{roelfsema2012}. The flux scale accuracy is estimated to be around 10\% for bands 1 and 2, 15\% for bands 3 and 4, and 20 \% in bands 6 and 7$^1$. The frequency calibration accuracy is 20 kHz and 100 kHz (i.e. better than 0.06 \kms), respectively for HRS and WBS observations. Data calibration was performed in the Herschel Interactive Processing Environment \citep[HIPE,][]{ott2010} version 12.1. Further analysis like Gaussian fit was done within the CLASS\footnote{http://www.iram.fr/IRAMFR/GILDAS/} package. These lines are not expected to be polarized, thus, after inspection, data from the two polarizations were averaged together. For all observations, possible contamination from lines in the image sideband of the receiver has been checked and none was found. Because HIFI is operating in double-sideband, the measured continuum level has been divided by a factor of 2 (in the Figures and the Tables) to be directly compared to the single sideband line profiles (this is justified because the sideband gain ratio is close to 1). 

\section{The source sample}
\label{sec:sample}

Five sources are studied here: NGC6334I(N), DR21(OH), IRAS16272-4837, and IRAS05358+3543 whose entire water observations set is analyzed for the first time in this paper while W43-MM1 results have been presented in \citet{herpin2012}.  The source coordinates, luminosity, distance, and velocities are given in Table \ref{source_list}. The position observed corresponds to the peak of the mm continuum emission from the literature \citep[see][]{vandertak2013}. The selected sources are mid IR-quiet dense cores \citep[][]{motte2007}, with bolometric luminosities 0.19--2.4~$10^4$~L$_{\odot}$ at distances 1.7--5.5~kpc and sizes (radius) in mm continuum of $\sim$0.26--0.82~pc \citep[see Appendix C in][]{vandertak2013}, hence larger than a 20$\arcsec$ beam (0.15-0.55 pc at these distances). Even if these massive dense cores are expected to be fragmented on small scales, only a small fraction of the total mass is in individual fragments, except for the dominant central source (see references in the following for each source). As a consequence the large-scale, average-density profile of the cores applies well down to the smallest structures observable with HIFI. 

\begin{figure*}
   \begin{minipage}[c]{0.46\linewidth}
     \includegraphics[width=7.9cm]{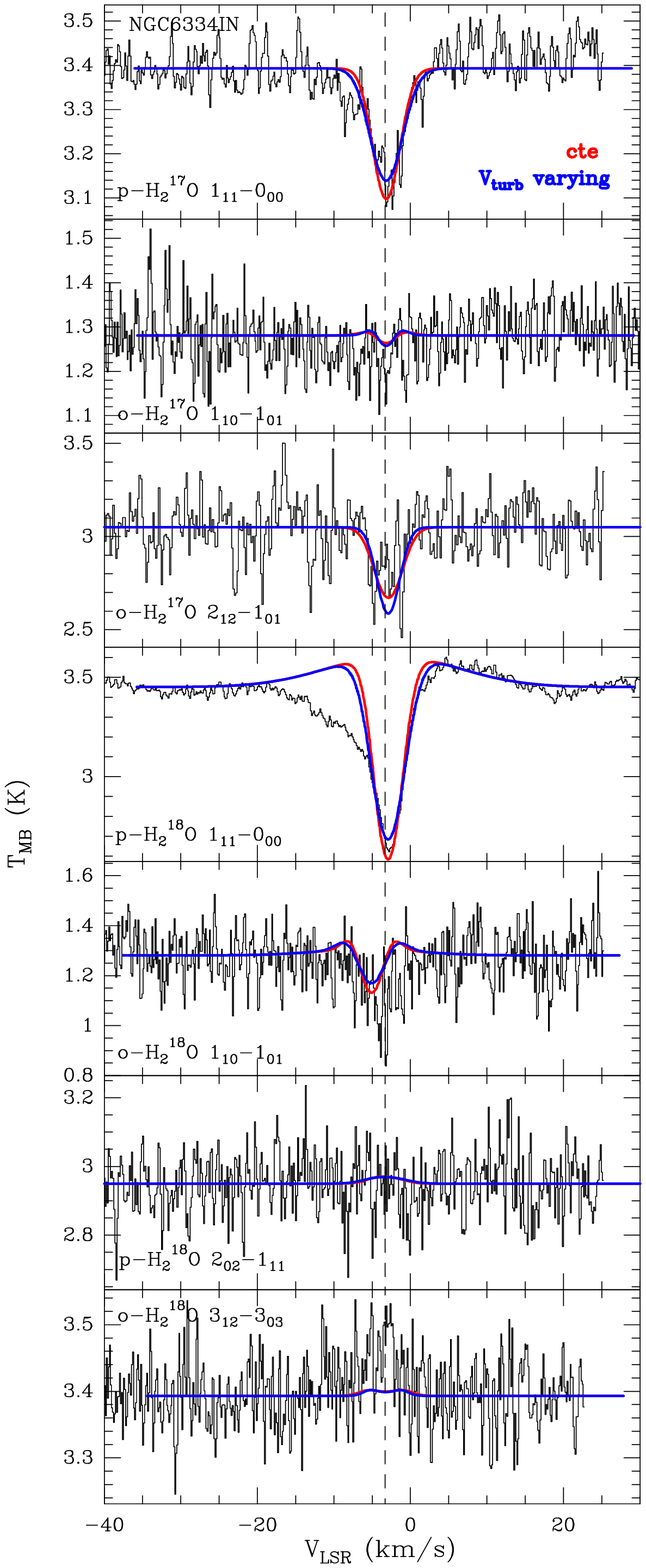}
   \end{minipage} \hfill
   \begin{minipage}[c]{1.96\linewidth}
      \includegraphics[width=7.9cm]{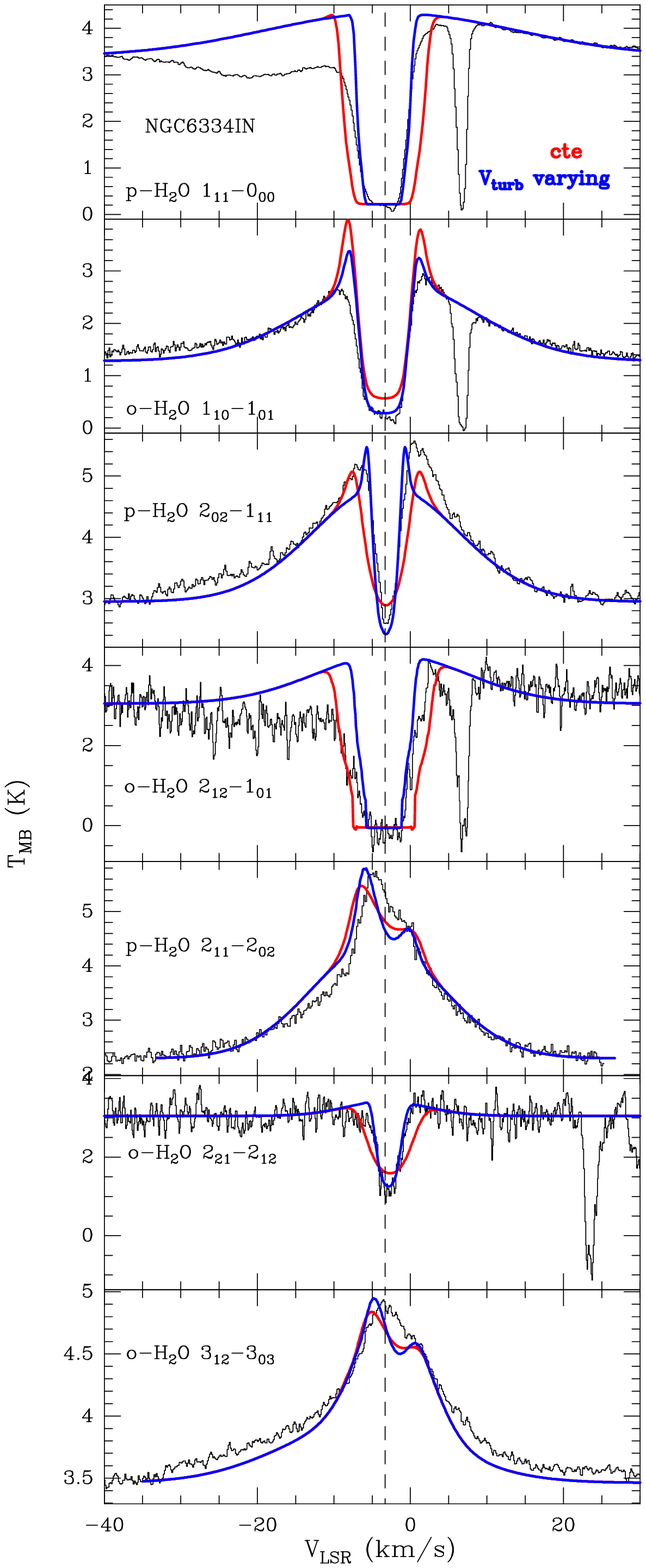}
   \end{minipage}
\caption{HIFI spectra of \watersept~and \waterhuit~({\em Left}) and \water~({\em Right}) lines (in black) with continuum for NGC6334I(N). The best-fit radiative transfer models are shown as red and blue lines over the spectra, respectively for constant parameters ($V_{inf}=-0.7$ and $V_{turb}=2.5$ \kms) and varying turbulent velocity. Vertical dotted lines indicate the \vlsr. The spectra have been smoothed to 0.2 \kms, and the continuum divided by a factor of two.}
\label{FigNGC6334IN}%
\end{figure*}

\begin{figure*}
   \begin{minipage}[c]{0.5\linewidth}
     \includegraphics[width=7.8cm]{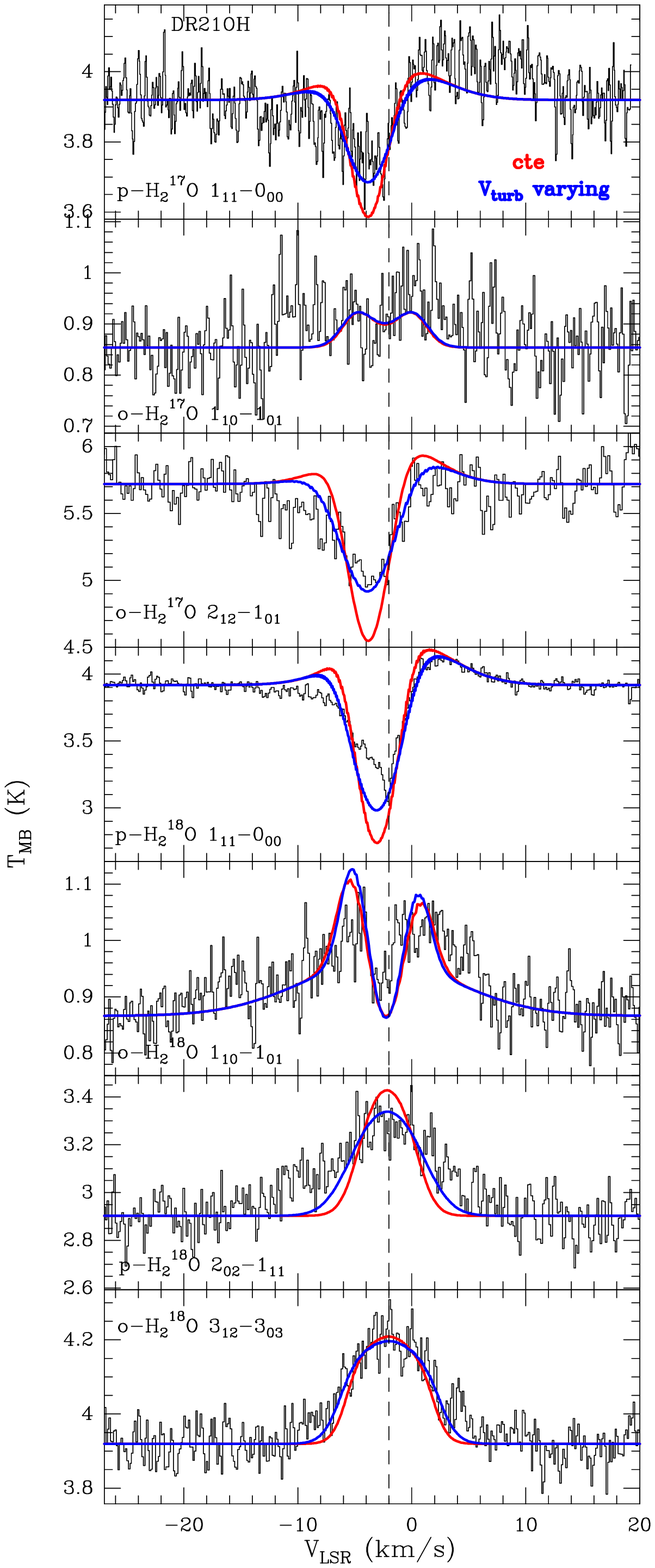}
   \end{minipage} \hfill
   \begin{minipage}[c]{1.96\linewidth}
      \includegraphics[width=7.7cm]{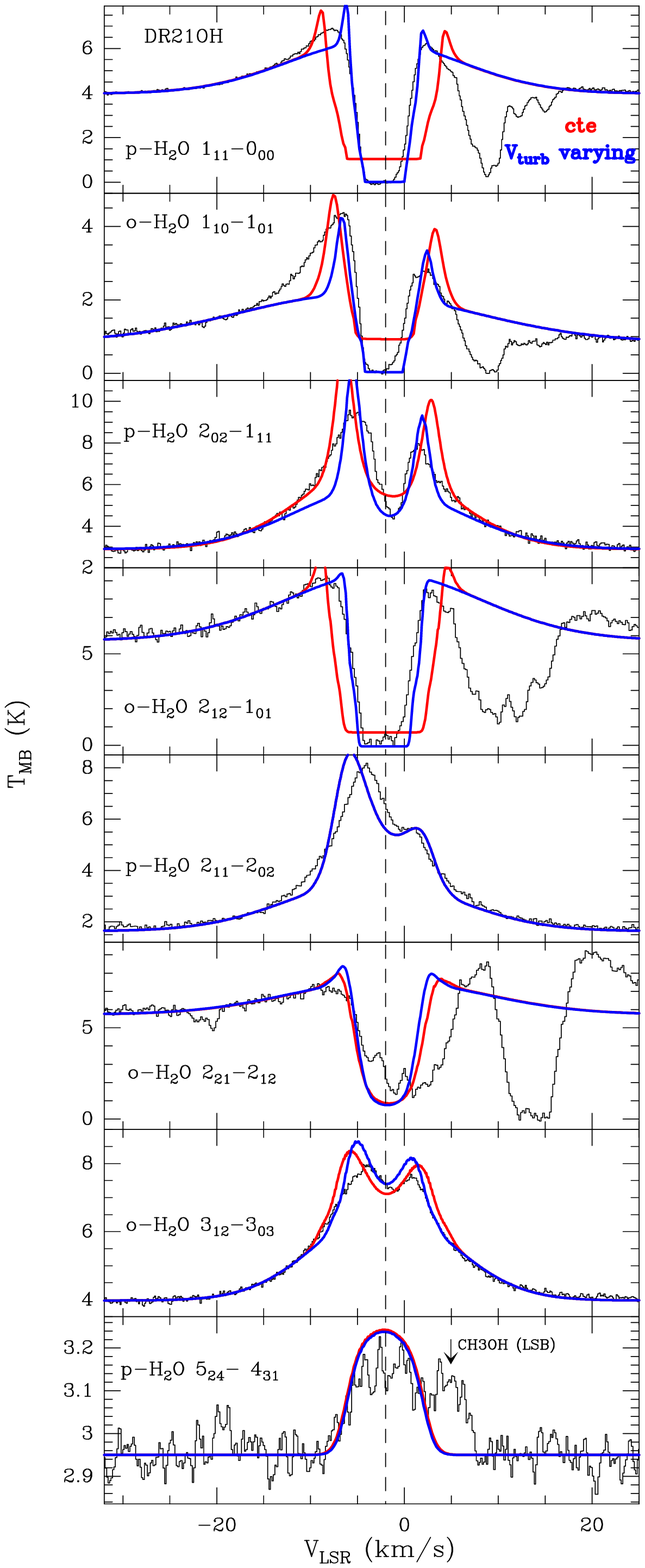}
   \end{minipage}
\caption{As in Figure \ref{FigNGC6334IN} but for DR21(OH) ($V_{inf}=-1.5$ and $V_{turb}=2.5$ \kms~for the constant model). The \hoP line is blended with the CH$_3$OH line at 957.995737 GHz from the LSB.}
\label{FigDR21OH}
\end{figure*}

\begin{figure*}
   \begin{minipage}[c]{0.5\linewidth}
     \includegraphics[width=7.5cm]{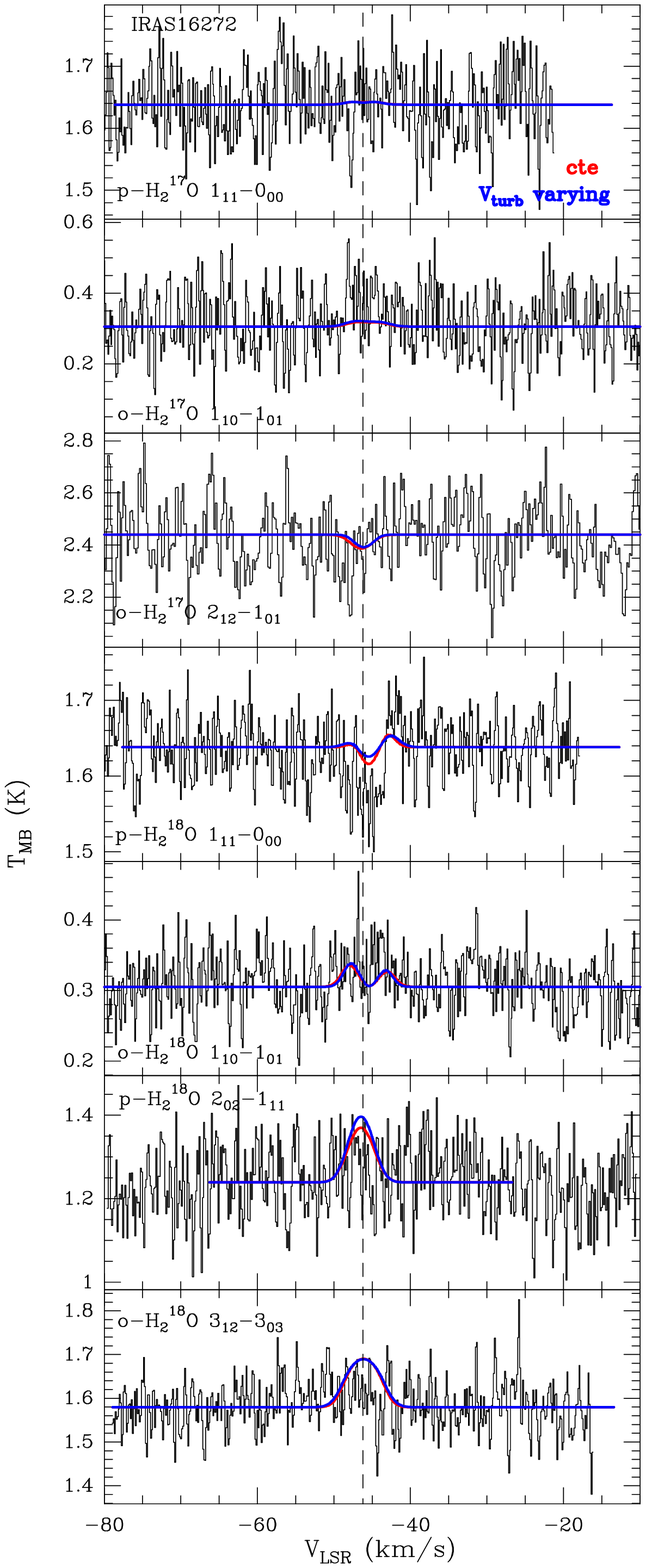}
   \end{minipage} \hfill
   \begin{minipage}[c]{0.96\linewidth}
      \includegraphics[width=7.5cm]{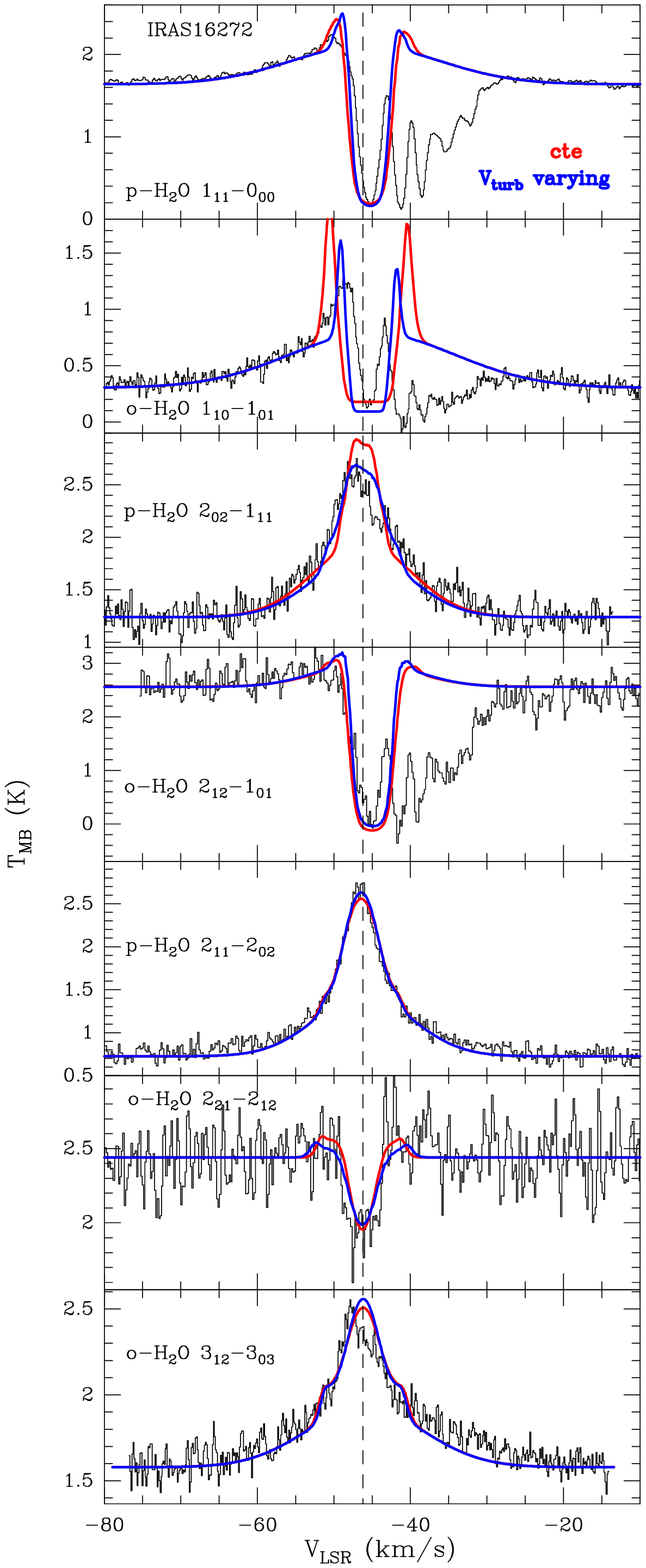}
   \end{minipage}
\caption{As in Figure \ref{FigNGC6334IN} but for IRAS16272 ($V_{inf}=-0.2$ and $V_{turb}=2.2$ \kms~for the constant model).}
\label{Fig16272}
\end{figure*}

\begin{figure*}
   \begin{minipage}[c]{0.36\linewidth}
     \includegraphics[width=8.cm]{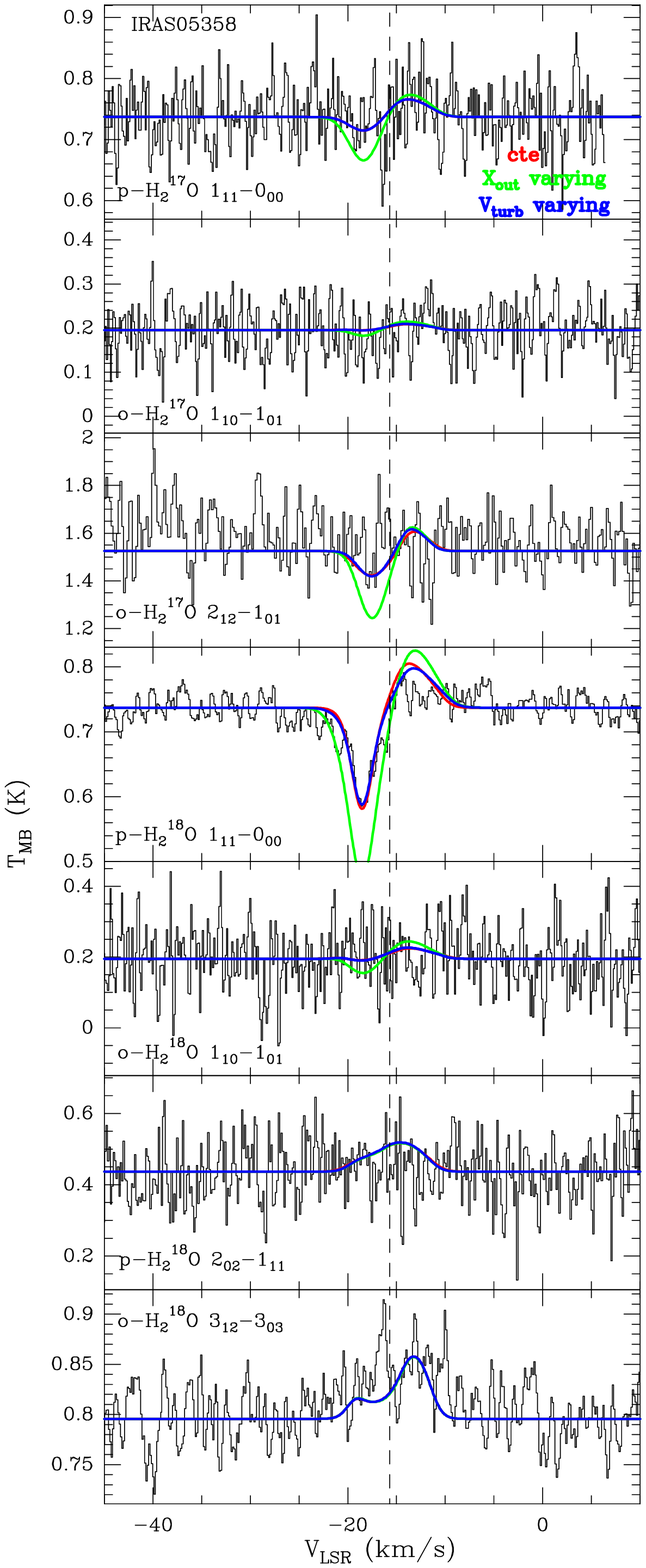}
   \end{minipage} \hfill
   \begin{minipage}[c]{2.46\linewidth}
      \includegraphics[width=14.3cm]{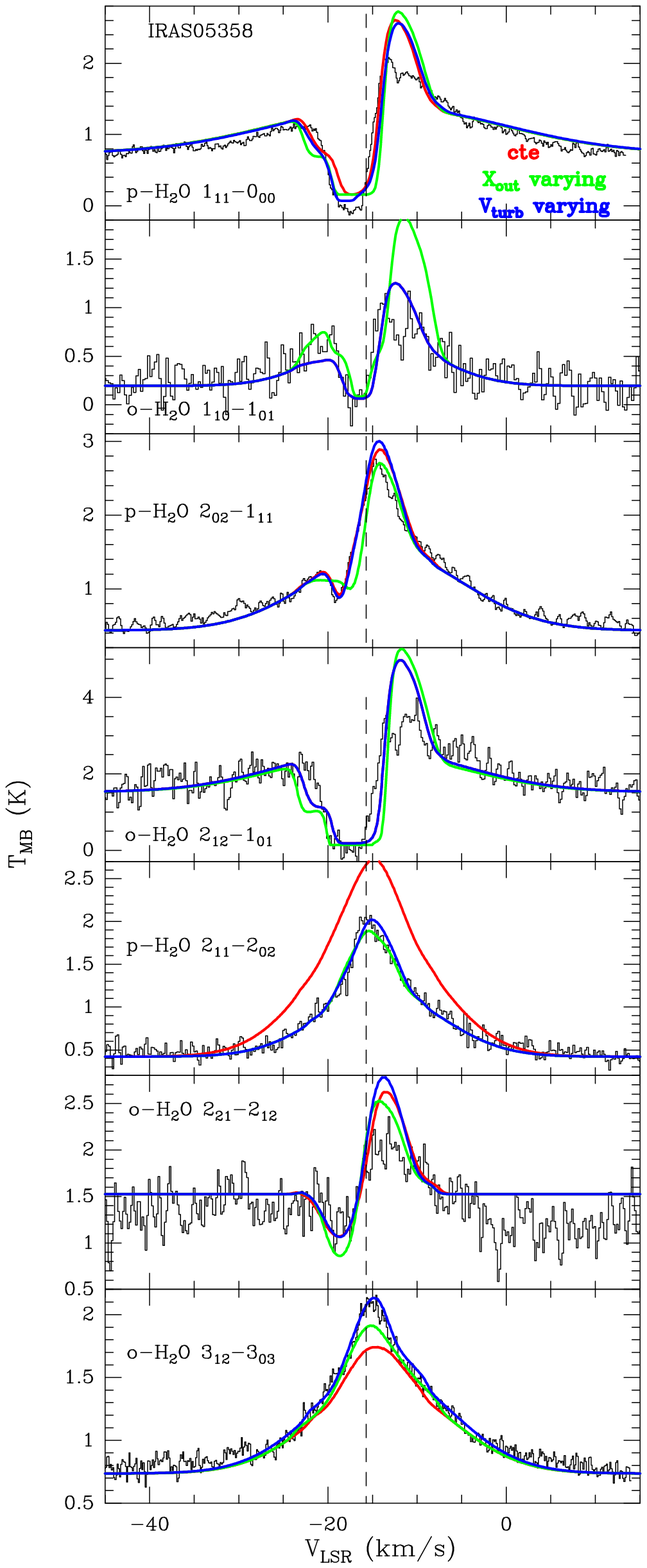}
   \end{minipage}
\caption{As in Figure \ref{FigNGC6334IN} but for IRAS05358 ($V_{inf}=+3.0$ and $V_{turb}=2.0$ \kms~for the constant model), more (in green) a varying outer abundance model (see Sect. \ref{abundance_section})}
\label{Fig05358}
\end{figure*}

The NGC6334 I(N) source, located in the northern part of the filament in the central region of NGC6334, has been extensively studied at mm$/$submm wavelengths \citep[e.g.][]{hunter2006,hunter2014}. \citet{brogan2009} imaged NGC6334 I(N) at $\sim 2\arcsec$ angular resolution using the SMA. They detected a cluster of compact sources, SMA1-SMA7 (most of them within the HIFI beams), in the 1.3 mm dust continuum emission, with gas masses of 6-74 \Msol. Spectral line data show evidence of infall accelerating with depth into SMA1 and the presence of multiple outflows. Note that parameters for this source and NGC6334I were exchanged in \citet{vandishoeck2011}. Observation with HIFI of NGC6334I have been presented by \citet{emprechtinger2013}.

DR21(OH) is located in the Cygnus X region, in the dense DR 21 filamentary ridge, where active star formation and global infall motions are observed \citep[][]{csengeri2011,hennemann2012}. Infall signatures in DR21(OH) are observed in low-$J$ CS lines on the same spatial scale covered by our observations \citep[][]{chandler1993}. \citet{girart2013} defined this rich molecular source as a highly fragmented, magnetized, turbulent dense core. The DR21(OH) core is formed by two main dusty condensations, MM 1 and MM 2, split into a cluster of dusty sources at scales of 1000 AU \citep[][]{zapata2012}. MM1 contains a hot core and shows centimeter continuum emission \citep[][]{araya2009}. Very active and powerful outflows are detected.

In contrast, less information is available about the more luminous ($L \sim2.4~10^4$~\lsol) source IRAS16272-4837. From MSX observations and studies by  \citet{garay2007}, this source can definitely be classified as a massive star-forming region in a very early evolutionary stage, a mid-IR quiet HMPO ($F_{21}=2.2$ Jy). No radio continuum emission was detected. The 1.2 mm emission arises from a central component surrounded by more extended gas \citep[region of $41\arcsec \times 25 \arcsec$,][]{faundez2004} and the line profiles observed suggest that the molecular gas is undergoing infalling motions.

IRAS05358+3543  is a relatively low-luminosity ($L\sim6.3~10^3$~\lsol) and nearby (1.8~kpc) massive dense core, composed of four main sources, all within a box of 4\arcsec $\times$6\arcsec\ \citep[][]{leurini2007,palau2014}, hence within the Herschel telescope beam at all frequencies. Two of these sources are part of a protobinary system  with a dynamical age of $\sim3.6~10^4$~yr \citep[][]{beuther2007}. Several molecular outflows are observed \citep[][]{beuther2002c}, but no infall is detected  \citep[][]{herpin2009}. The integrated gas mass is estimated to 142 \Msol by \citet{vandertak2013}. \citet{leurini2007} suggest that the main source mm1a harbours a hot core with $T \sim 220$ $(75 < T < 330)$ K and may contain a massive circumstellar disk. Because of its 21 microns flux density (11.9 Jy), this source is rather between the mid-IR quiet and mid-IR bright HMPO stages.

Continuum emission from our sources is determined with well-sampled observations from various telescopes including IRAS (archive\footnote{http://irsa.ipac.caltech.edu/applications/IRAS/ISSA/}), Spitzer (archive\footnote{http://sha.ipac.caltech.edu/applications/Spitzer/SHA/}), MSX (archive\footnote{http://irsa.ipac.caltech.edu/Missions/msx.html}), JCMT \citep[archive\footnote{http://www.cadc-ccda.hia-iha.nrc-cnrc.gc.ca/en/jcmt/}, and][]{vallee2006, sandell2000, mccutcheon2000}, KAO \citep[][]{harvey1986, lester1985}, CSO \citep[][]{motte2003}, SMA \citep[][]{beuther2007, hunter2006}, APEX, SEST \citep[][]{garay2007, munoz2007}, IRAM-30m \citep[][]{beuther2002a, motte2007}, IRAM-PdB \citep[][]{beuther2007}, VLA \citep[][]{beuther2002c, rodriguez2007}, ATCA \citep[][]{walsh1998, beuther2008}, OVRO \citep[][]{woody1989}, and Herschel-HIFI/PACS \citep[][]{vandertak2013}. In particular, flux densities below 35 $\mu$m come from Spitzer and MSX observations. Hence, the spectral energy distribution (SED) for these sources is particularly well constrained. Following the method of \citet{herpin2009}, we propose a rough evolutionary classification of our 5 objects from the fitted SEDs shown in Fig.~\ref{Figsed} (the reference spatial resolution is the beam of the observation at 1.1 or 1.2 mm, i.e. 11\arcsec with IRAM-30m for IRAS05358, W43MM1, and DR21(OH), and 24 \arcsec with JCMT or SEST for the two other sources), using the following parameters:
\begin{itemize}
  \item flux density at 12 $\mu$m,
   \item flux density at 21 $\mu$m,
  \item wavelength and flux density of the maximum continuum emission,
  \item contribution of the hot part ($\lambda <$35 $\mu$m) to the total integrated flux.
\end{itemize}
In addition, we also use the evolutionary tracer $L^{0.6}{M_{\textrm{env}}}^{-1}$ first introduced by \citet{bontemps1996} for low-mass objects: this quantity increases with the evolutionary status of the source. It is assumed that less evolved sources are colder, hence the SED peaks at longer wavelength with weaker flux. As the massive core evolves, it becomes warmer, thereby heating the dust. As a consequence, the contribution of the flux at shorter wavelength ($F_{35}$, the integrated flux density for $\lambda <$35 $\mu$m) increases. 

Comparison of these quantities (Table \ref{source_list}) leads to the following evolutionary sequence going from youngest to older: NGC64334I(N)~$\rightarrow$ W43-MM1~$\rightarrow$ DR21(OH)~$\rightarrow$ IRAS16272-4837~$\rightarrow$ IRAS05358+3543. The fact that three different estimates give the same order lends credibility to this sequence. We adopt this sequence for the following discussion. Nevertheless, we note that the order of DR21(OH) and IRAS16272-4837 can be inverted if one uses the $L^{0.6}{M_{\textrm{env}}}^{-1}$ criterion only (see values in Table \ref{source_list}) or if we consider that DR21(OH) harbours a hot core as IRAS05358.

\section{Results}
\label{sec:results}

\begin{table*}
  \caption{Observed line emission parameters for the detected lines toward NGC6334I(N). $V$ is the Gaussian component peak velocity. $\Delta V$ is the velocity full width at half-maximum (FWHM) of the narrow, medium and broad components. FWZI is the full-width at zero intensity. The opacity $\tau$ is from absorption lines.}
\begin{center}
\label{table_param_6334}      
\begin{tabular}{lcccccccccc} \hline \hline
{\bf Line}  & $T_{mb}$ & $T_{cont}$ & FWZI & $V_{nar}$ & $\Delta V_{nar}$ & $V_{med}$ & $\Delta V_{med}$ & $V_{br}$ & $\Delta V_{br}$  & $\tau$ \\ 
 & [K] & [K] & [\kms] & [\kms] & [\kms] & [\kms] & [\kms] &  [\kms] &[\kms]  & \\ 
\hline                        
   \hoA     &  1.10 & 1.28  & 14.6 & & & -3.8$\pm$0.4$^a$ & 6$\pm$1& & & 0.15$\pm$0.03\\ 
   \hoG   & 0.09 & 3.42 & 13.7 & & & -3.9$\pm$0.4& 6$\pm$1& && \\
   \hoI$^b$     & 2.63 & 3.45 & 45.6 & -2.9 $\pm$0.2$^a$& 4.5$\pm$0.2 &  & & &&  0.27$\pm$0.05\\
   \hoJ$^b$    & 3.10 & 3.42 & 24.2 & & & -3.0$\pm$0.2$^a$ &5.7$\pm$0.5 & && 0.10$\pm$0.02\\
   \hoM   & 2.56 & 3.05 & 5.0 & -3.3$\pm$0.3$^a$&4.2$\pm$0.5 & & & & &0.17$\pm$0.06 \\
\hline
   \hoC   & 0.0 & 1.28 & 92.2 & &  & -3.5$\pm$0.1$^a$& 5.8$\pm$0.2& -2.0$\pm$0.2 & 26.1$\pm$0.6 & $>5$\\
   \hoD      & 5.85 & 2.30 & 90.4 & & &-3.2$\pm$0.1 & 7.3$\pm$0.3 & -4.4$\pm$0.2& 29.4$\pm$0.7&\\
   \hoE      & 2.31 & 2.95 & 58.1 & -3.0$\pm$0.1$^a$&3.1$\pm$0.1& -3.0$\pm$0.2 & 10.0$\pm$0.3 & -5.7$\pm$0.2 & 29.6$\pm$0.5  &0.24$\pm$0.05\\
   \hoH   & 4.91 & 3.48 & 82.6 & &  & -1.9$\pm$0.1& 9.6$\pm$0.3& -6.1$\pm$0.3 & 27.6$\pm$0.6 &\\
   \hoK$^b$    & 0.06 & 3.48 & 98.0  & & & -3.9$\pm$0.2$^a$&6.1$\pm$0.3 & -3$\pm$2& 30$\pm$1 & 4.1$\pm$0.6\\
   \hoL    & 1.20 & 3.05 & 6.0 & -2.8$\pm$0.2$^a$& 2.9$\pm$0.4& &  & &&0.9$\pm$0.3 \\
   \hoN$^b$    & 0.0 & 3.05 & 27.4 & &  & -3.9$\pm$0.2$^a$& 6.8$\pm$0.4& & &$>5$\\
\hline
\end{tabular}
\end{center}
\tablefoot{$^a$ in absorption, $^b$ WBS data}
\end{table*}

\begin{table*}
  \caption{Observed line emission parameters for the detected lines toward DR21(OH). $V$ is the Gaussian component peak velocity. $\Delta V$ is the velocity full width at half-maximum (FWHM) of the narrow, medium and broad components. FWZI is the full-width at zero intensity. The opacity $\tau$ is from absorption lines.}
\begin{center}
\label{table_param_DR21OH}      
\begin{tabular}{lcccccccccc} \hline \hline
{\bf Line}  & $T_{mb}$ & $T_{cont}$ & FWZI & $V_{nar}$ & $\Delta V_{nar}$ & $V_{med}$ & $\Delta V_{med}$ & $V_{br}$ & $\Delta V_{br}$  & $\tau$ \\ 
 & [K] & [K] & [\kms] & [\kms] & [\kms] & [\kms] & [\kms] &  [\kms] &[\kms]  & \\ 
\hline                        
   \hoA     & 1.04 & 0.87 & 40.0 & -2.7 $\pm$0.2$^a$ & 1.4 $\pm$0.3&  & &-1.5$\pm$0.3 & 16.3$\pm$0.8 & \\ 
    \hoB   & 0.96 & 0.85 & 19.8 & & & & & -1.6$\pm$0.8& 15$\pm$2&\\
    \hoF   & 3.36 & 2.90 & 24.4 & & & & & -2.5$\pm$0.2& 12.0$\pm$0.4&\\
  \hoG   & 4.25 & 3.92 & 21.2 & & & -1.9$\pm$0.1& 8.6$\pm$0.3& && \\
   \hoI    & 3.04 & 3.92 & 35.4 & -2.1$\pm$0.1$^a$& 0.8$\pm$0.1& -2.5$\pm$0.2$^a$& 5.7$\pm$0.2& 0.$\pm$0.6& 10.$\pm$0.6 & 0.25$\pm$0.05 \\
   \hoJ    & 3.77 & 3.92 & 26.9 & & & -3.6$\pm$0.2$^a$&5.3$\pm$0.7 & 0$\pm$1& 12$\pm$2&0.04$\pm$0.01\\
   \hoM   & 5.12 & 5.72 & 25.7 & -4.2$\pm$0.2$^a$& 4.4$\pm$0.4& & & && 0.11$\pm$0.04\\
\hline
   \hoC   & 0.0 & 0.85 & 85.7 & &  & -2.7$\pm$0.2$^a$& 5.6$\pm$0.1& -1.9$\pm$0.2 & 22.4$\pm$0.3 &$>5$\\
   \hoD      & 8.19 & 1.65 & 62.7 & & &-3.0$\pm$0.1 & 8.3$\pm$0.1 & -3.1$\pm$0.1& 22.3$\pm$0.3&\\
  \hoP$^b$      & 3.25 & 2.95 & 26.0 & & & -1.4$\pm$0.3 &  9.9$\pm$0.6 && &\\
   \hoE      & 9.40 & 2.90 & 51.3 & -1.7$\pm$0.1$^a$ & 3.7$\pm$0.1& -2.6$\pm$0.1& 8.2$\pm$0.1  & -2.9$\pm$0.1& 20.4$\pm$0.2 & \\
   \hoH   & 7.93 & 3.98 & 52.5 & -1.7$\pm$0.1& 2.9$\pm$0.2 & -2.5$\pm$0.1& 9.9 $\pm$0.1& -1.9$\pm$0.1& 20.4$\pm$0.3&\\
   \hoK    & 0.0 & 3.98 & 51.7 & -3.5$\pm$0.2$^a$ & 4.6$\pm$0.2& & & -1.7$\pm$0.1& 21.6$\pm$0.2  & $>5$\\
   \hoL$^c$    & 1.06 & 5.73 & 10.8 & & & 0.3$\pm$0.2$^a$& 6.3$\pm$0.4 &  &   &1.7$\pm$0.5\\
   \hoN$^b,c$    & 0.0 & 5.73 & 50.2 & -2.4$\pm$0.2$^a$ &  4.1$\pm$0.2 & & & -1.9$\pm$0.1&24.1$\pm$0.5 & $>5$\\
\hline
\end{tabular}
\end{center}
\tablefoot{$^a$ in absorption, $^b$ WBS data, $^c$ blended with \hoN}
\end{table*}

\begin{table*}
  \caption{Observed line emission parameters for the detected lines toward IRAS16272. $V$ is the Gaussian component peak velocity. $\Delta V$ is the velocity full width at half-maximum (FWHM) of the narrow, medium and broad components. FWZI is the full-width at zero intensity. The opacity $\tau$ is from absorption lines. Values within brackets stand for tentative detection.}
\begin{center}
\label{table_param_16272}      
\begin{tabular}{lcccccccccc} \hline \hline
{\bf Line}  & $T_{mb}$ & $T_{cont}$ & FWZI & $V_{nar}$ & $\Delta V_{nar}$ & $V_{med}$ & $\Delta V_{med}$ & $V_{br}$ & $\Delta V_{br}$  & $\tau$ \\ 
 & [K] & [K] & [\kms] & [\kms] & [\kms] & [\kms] & [\kms] &  [\kms] &[\kms]  & \\ 
\hline                        
   \hoA     & (0.34) & (0.30)  & (8.1) & & &-45.1$\pm$0.7 &6.$\pm$1 & & & \\ 
   \hoI     & 1.54 & 1.64 & 5.7 & -46.0 $\pm$0.3$^a$& 3.8$\pm$0.4 & & & & & 0.06$\pm$0.01 \\
\hline
   \hoC   & 0.12 & 0.30 & 80.2 & -45.8$\pm$0.1$^a$& 2.1$\pm$0.1 & -47$\pm$1& 7$\pm$1& -45.0$\pm$0.6 & 28$\pm$2 &0.9$\pm$0.1\\
   \hoD      & 2.80  & 0.73 & 59.9 & & &-46.6$\pm$0.1 & 5.0$\pm$0.1 & -46.3$\pm$0.1& 20.1$\pm$0.1&\\
   \hoE     & 2.67 & 1.24 & 36.0 & -47.7$\pm$0.2&2.8$\pm$0.4 & &  & -47.3$\pm$0.2&20.9$\pm$0.1 &\\
   \hoH   & 2.50& 1.58 & 46.0 & & & -46.7$\pm$0.1& 5.0$\pm$0.3& -45.0$\pm$0.3&23.7$\pm$0.7 \\
   \hoK$^b$    & 0.20 & 1.64 & 63.7 & -45.3$\pm$0.2$^a$& 3.0$\pm$0.2& & & -45.3$\pm$0.2& 24.0$\pm$0.3  & 2.1$\pm$0.3\\
   \hoL    & 1.90 & 2.44 & 15.9 & -46.5$\pm$0.2$^a$& 3.5$\pm$0.2& &  & && 0.25$\pm$0.05\\
   \hoN$^b$    & 0.0 & 2.44 & 50.0 & -45.7$\pm$0.1$^a$& 3.1$\pm$0.1 & & & (-47 $\pm$2) & (30$\pm$5) & $>5$\\
  \hline 
\end{tabular}
\end{center}
\tablefoot{$^a$ in absorption, $^b$ WBS data}
\end{table*}

\begin{table*}
  \caption{Observed line emission parameters for the detected water lines toward IRAS05358. $V$ is the Gaussian component peak velocity. $\Delta V$ is the velocity full width at half-maximum (FWHM) of the narrow, medium and broad components. FWZI is the full-width at zero intensity. The opacity $\tau$ is from absorption lines.}
\begin{center}
\label{table_param_05358}      
\begin{tabular}{lcccccccccc} \hline \hline
{\bf Line}  & $T_{mb}$ & $T_{cont}$ & FWZI & $V_{nar}$ & $\Delta V_{nar}$ & $V_{med}$ & $\Delta V_{med}$ & $V_{br}$ & $\Delta V_{br}$  & $\tau$ \\ 
 & [K] & [K] & [\kms] & [\kms] & [\kms] & [\kms] & [\kms] &  [\kms] &[\kms]  & \\ 
\hline                        
   \hoI     & 0.58 & 0.74 & 21.8 & -18.1 $\pm$0.1$^a$& 2.5$\pm$0.3 & & & &  &0.24$\pm$0.06\\
   \hoG   & 0.87 & 0.80 & 14.1 & & & -14.1$\pm$0.4 & 9.0$\pm$0.8& & & \\
\hline
   \hoC   & 0.0 & 0.20 & 26.4 & -16.9$\pm$0.3$^a$& 3.0$\pm$0.2 & & & -14.9$\pm$0.3 & 13.9$\pm$0.7 & $>5$ \\
   \hoD      & 2.17 & 0.42 & 52.1 & -15.3$\pm$0.1& 4.5$\pm$0.2& &  & -13.8$\pm$0.2& 17.0$\pm$0.4 & \\
     \hoE      & 2.72 & 0.44 & 55.6 & -18.7$\pm$0.1$^a$ &1.7$\pm$0.2 &-13.9$\pm$0.1 & 5.0$\pm$0.2 & -12.9$\pm$0.4 & 23.6$\pm$0.5 & \\
   \hoH   & 2.18 & 0.74 & 56.2 & & & -15.1$\pm$0.1& 6.2$\pm$0.3& -13.7$\pm$0.2&21.0$\pm$0.2 \\
     \hoK    & 0.0 & 0.74 & 50.7 & -16.5$\pm$0.3$^a$& 4.9$\pm$0.3& -14.6$\pm$0.3& 5.9$\pm$0.3& -13.3$\pm$0.3& 20.6$\pm$0.3 & $>5$ \\
   \hoL    & 1.0 & 1.52 & 25.5 & -18.1$\pm$0.3$^a$& 3.1$\pm$0.5 & &  & -12.6$\pm$0.4 & 10.9$\pm$0.8 & 0.4$\pm$0.1 \\
   \hoN    & 0.0& 1.52 & 26.3 & &  & -17.7$\pm$0.2$^a$& 5.6$\pm$0.3 & -13.8$\pm$0.4 &20.3$\pm$0.8  &$>5$ \\
\hline
\end{tabular}
\end{center}
\tablefoot{$^a$ in absorption.}
\end{table*}

The spectra including continuum emission are presented in Fig. \ref{FigNGC6334IN}-\ref{Fig05358} for the rare isotopologues (\watersept, H$_2^{18}$O) and \water~for all sources, except W43-MM1 which was presented in \citet{herpin2012}. Spectra of the H$_2$O 1$_{11}$-0$_{00}$ (and H$_2^{18}$O), 2$_{02}$-1$_{11}$, and 2$_{12}$-1$_{01}$ lines have also been included in \citet{vandertak2013}. In addition, Fig. \ref{Figbyproducts13CO}-\ref{FigbyproductsH2S} in Appendix B display lines from other species than water that are serendipitously covered in our data. In most cases, we show the HRS spectra, except for the ground-state \hoI, \hoK, and \hoN~lines, where WBS spectra were used since the velocity range covered by the HRS was insufficient to show the broad component. In a few cases (e.g. \hoN~spectra in NGC6334I(N)), the detected absorption is slightly below the continuum but within the flux uncertainties (see Sect. \ref{sec:observations}).

Several foreground clouds \citep[][]{vandertak2013} contribute to the spectra in NGC6334I(N), DR21(OH) and IRAS16272 in terms of water absorption at V$_{lsr}$ shifted with respect to source velocity in the \hoC, \hoK~ and \hoN~lines spectra (absorption is also visible in the \hoL~spectra in DR21(OH)). These foreground clouds are not analyzed here.

\subsection{Velocity components}
\label{sec:vel}

For each transition, we derive the peak, or minimum (in case of absorption), main-beam and continuum temperatures, the full-width at zero intensity \citep[FWZI, see][for details]{mottram2014}, half power line-widths for the different line components from multi-component Gaussian fits, and opacities for lines in absorption. Line parameters are given in Tables~\ref{table_param_6334}-\ref{table_param_05358}.  An example of the Gaussian fits is given for IRAS05358 in Appendix C. Our results are consistent with \citet{vandertak2013} and \citet{sanjose2015a} and do not depend on the adopted method.

We will follow the terminology adopted in previous WISH papers \citep[e.g.][]{johnstone2010,kristensen2010,herpin2012}: narrow ($<$5 \kms), medium (FWHM$\simeq$5-10 \kms), and broad (FWHM$\simeq$20-35 \kms). The narrow component centered at the source velocity is characterized by small FWHM and offset velocity, and is called {\em envelope component}, i.e. emission from the quiescent envelope, the signature of the passively heated envelope. Both broad and medium components arise in {\em cavity shocks}, i.e. shocks along the cavity walls according to the model of \citet{mottram2014}. The medium component is a narrower version of the broad component \citep[also called {\em narrow outflow} by][]{vandertak2013}, coming from a thin layer (1-30 AU) along the outflow cavity where non-dissociative shocks occur.  A different physical component, called the medium offset component, with a velocity offset of at least a few km/s, is observed in low-mass objects and is associated with spot shocks, i.e., dissociative shocks in the jet itself or at the base of the outflow \citep[][]{mottram2014}. This component is seen in the HM data by \citet{vandertak2013} as a narrow outflow in absorption in the 1113 and 1669 GHz water lines and slightly offset from the envelope velocity.

\subsection{Water Lines}

\subsubsection{Rare isotopologues}

The more evolved objects, IRAS05358 and IRAS16272, exhibit fewer and weaker rare isotopologue lines, with no \watersept~detection. This is not due to a lower source luminosity or a larger distance as can be inferred from Table \ref{source_list}: e.g. IRAS16272-4837 exhibits a luminosity similar to that of W43-MM1 and is closer to us. The para ground state line \hoI~is detected, as well as the \hoG~in IRAS05358, but the \hoA~line is not detected in IRAS16272 and IRAS05358. The fact that no \watersept~and no \hoA~line are detected in IRAS16272 and IRAS05358 is likely due to lower water column densities, hence insufficient S$/$N ratio. 

On the other hand, \watersept~lines are detected in the two other objects. The strongest \watersept~and \waterhuit~lines are found in DR21(OH). In this source, all observed rare isotopologue lines are detected.

The \hoI, \hoM, and \hoJ~lines appear in absorption, and broad signatures from the cavity shocks (red component in emission while the blue one is absorbed) are observed for DR21(OH) and NGC6334I(N). The \hoA~line is in absorption for NGC6334I(N) while it is a blend of absorption (at line center) and emission in DR21(OH). The two other detected \waterhuit~lines, $2_{02}-1_{11}$ and $3_{12}-3_{03}$, arising from more excited energy levels, are in emission. 

Most of the \watersept$/$\waterhuit~line profiles for all sources but W43-MM1 exhibit an envelope component in absorption. We interpret this absorption as resulting from cold material in front of the passively heated envelope. A medium component is detected in several lines too.

\subsubsection{\water}

All \water~lines listed in Table \ref{table_transitions} are detected toward all 4 sources and the global shape of the line profile for each line is similar from source to source (including W43-MM1): ground state lines are deeply absorbed and all lines (except \hoP) exhibit a broad component (blue component in absorption for NGC6334IN). In addition, line profiles consist of a narrow or medium component depending on the source and the line. The \hoE~and \hoD~lines are asymmetric (see section  \ref{sec:asym}) due to the infall or expansion of the gas. The \hoL~line is in absorption except for IRAS05358 for which some emission is present too. The \hoH~line is dominated by the broad and medium components in emission. 

For the only source, DR21(OH), where the \hoP~line has been observed, a medium component, slightly redshifted (by 1.7 \kms), is observed in emission, but blended with a methanol line from the lower sideband at 957.995737 GHz. 

\begin{figure}
\centering
\includegraphics[width=7.5cm]{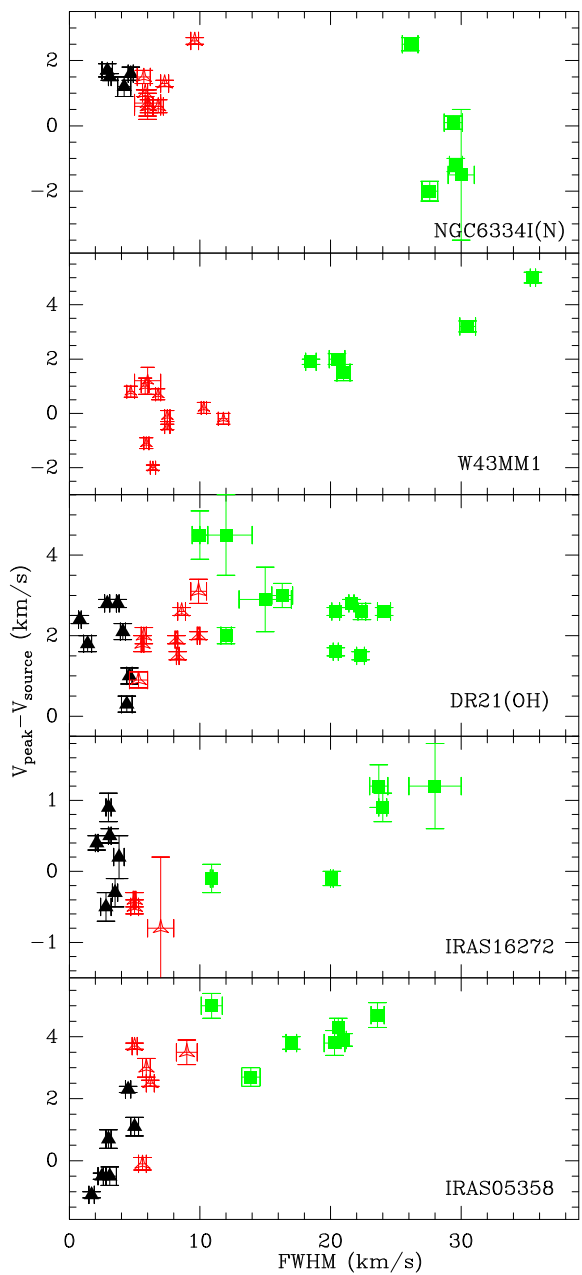}
\caption{FWHM versus offset of the peak from the source velocity for the gaussian velocity components found in the water (\water~and \waterhuit) line profiles. Filled stars (black), empty stars (red), and filled squares (green) represent respectively the narrow, medium, and broad components.}
\label{vel_FWHM}%
\end{figure}

Broad, medium, and narrow velocity components are detected in all sources. We observe for the envelope component a similar mean FWHM value of 3 \kms~for IRAS16272 ($\pm0.5$) and NGC6334I(N) ($\pm0.1$), 3.5($\pm1.3$) \kms~for IRAS05358, and 3.8($\pm0.7$) \kms~for DR21(OH). The width of the medium component is clearly larger for the less evolved sources ($7.1\pm1.5$, $8\pm2$, and $8.0\pm1.8$ \kms~respectively for NGC6334IN, W43-MM1, and DR21(OH)) than for IRAS16272 ($5.7\pm1.3$ \kms) and IRAS05358 ($5.7\pm0.5$ \kms). The FWHM of the broad component decreases from more than 25 \kms~for NGC6334I(N) and W43-MM1 to roughly 20 \kms~for the three other objects.  Figure \ref{vel_FWHM} shows that no or small velocity offset is observed for different components: only IRAS05358 and W43-MM1 exhibit offset increasing with FWHM (see section \ref{kine} for a trend analysis).  

\subsection{Other species}

Several other species have been detected toward all sources within the 4 GHz wide WBS spectra (see Tables \ref{byproduct_05358}-\ref{byproduct_DR21OH}): CH$_3$OH, $^{13}$CO (J=5-4 and 10-9), C$^{18}$O (J=9-8), CS (11-10), and H$_2$S ($3_{0,3}-2_{1,2}$). These lines are detected in all sources in emission. 

In addition to the bulk of methanol lines with $E_u\leq200$ K observed in the entire sample, lines involving upper energy level up to 291, 434, and 514 K are detected respectively in NGC6334I(N), W43-MM1/IRAS16272, and DR21(OH). These lines exhibit an envelope and$/$or a medium component (and for one or two lines a broad component) similar to what is observed for water lines.

Both detected $^{13}$CO J=5-4 and 10-9  lines exhibit similar profiles, even if the 5-4 line in DR21(OH) is much more self-absorbed at the center because of higher  opacity. In addition to the C$^{18}$O J=9-8 line, the C$^{18}$O J=10-9 transition has been detected toward all sources but IRAS16272. For each species we note that the line widths are similar for both observed transition, and the velocity components derived from the Gaussian fitting are consistent with \citet{sanjose2013} (even if our narrow and medium components are only one single component for them in a few cases as they only distinguish between FWHM smaller or larger than 7.5 \kms). 

The water cation H$_2$O$^+$ ($1_{11}-0_{00}$, $J=3/2-1/2)$ is detected, in absorption, in all sources, except NGC6334I(N): its velocity components indicate that it likely originates from the envelope or outflow for IRAS05358 and W43-MM1 \citep[see also][for other high-mass sources]{benz2010,wyrowski2010}, while for DR21(OH) and IRAS16272 the absorption is redshifted by 5-10 \kms~and then very likely associated to the foreground clouds described before. In addition, the H$_3$O$^+$ cation is detected in absorption in W43-MM1, but is weaker than the H$_2$O$^+$ line, which, as stressed by \citet{wyrowski2010}, is unexpected. 

More species are detected toward the 3 less evolved sources: deuterated water (HDO) and methyl formate (CH$_3$OCHO) for instance (see Appendix \ref{sec:byproducts}). But more generally, DR21(OH) is the richest source with twice as many lines detected: the rare isotopologue $^{13}$CS, many more methanol lines, dimethyl ether (CH$_3$OCH$_3$), CH$^+$ and OH$^+$ in absorption, H$_2$CO, $^{34}$SO, OS$^{18}$O, and many lines of SO$_2$.

All these lines peak at a mean velocity of ${V_{LSR}}=-15.9\pm0.3$, $46.5\pm1.9$, $-3.3\pm0.8$, $-4.0\pm0.6$, and $98.4\pm1.2$ \kms~respectively for IRAS05358, IRAS16272, DR21(OH), NGC6334I(N), and W43-MM1, hence similar to the source velocity, except for IRAS05358 whose lines are slightly red-shifted.

\begin{table*}
\caption{\water~integrated line intensity ratios (not corrected for different beam sizes) for all sources for the broad velocity component. If available the ratio for the medium velocity component is given and noted (m). Optically thin and thick ratios are calculated for $T_{ex}=100$ and 300 K.}             
\label{opacity_ratio}      
\centering                          
\begin{tabular}{ccccccccc}        
\hline\hline    
Transitions & \multicolumn{5}{c}{Observed ratio} &  \multicolumn{2}{c}{LTE} & $\theta_1/\theta_2$$^a$ \\   
                     &   &  &  &  &  & \multicolumn{2}{c}{(100/300)} &   \\
                      &   NGC6334I(N) & W43MM1 & DR21(OH) & IRAS16272 & IRAS05358 & Thin & Thick &   \\
                      \hline
$1_{10}-1_{01}/2_{12}-1_{01}$ &  & & $0.56\pm0.06$ &   &  $0.20\pm0.08$& 0.6$/$0.4 & 1.3$/$1.1 & 3.0 \\
$1_{11}-0_{00}/2_{02}-1_{11}$ &  & & $1.03\pm0.05$ &    $0.9\pm0.1$  &  $0.56\pm0.06$ & 2.4$/$1.7 & 1.3$/$1.1 & 0.9 \\
(m) &  & &     & &  $1.2\pm0.3$ & &  &  \\
$2_{11}-2_{02}/2_{02}-1_{11}$ &  $1.14\pm0.05$ &  $1.04\pm0.06$ &  $0.88\pm0.05$ &  $0.94\pm0.05$  & $0.60\pm0.05$ & 1.4$/$1.8 & 1.2$/$1.1 & 1.3 \\
 (m) &  & &  $0.38\pm0.05$&   $3.0\pm0.1$   &    $0.70\pm0.06$& &  &  \\
\hline                                 
\end{tabular}
\tablefoot{$^a$ Beam size ratio }
\end{table*}

\section{Analysis}
\label{analysis}
\subsection{Kinematics}
\label{kine}
For each source, we have searched for a correlation between FWHM and the velocity of the peak of various water components (see Fig.\ref{vel_FWHM}). Such a correlation (both increasing together) has been found by \citet{mottram2014} for Class 0 and I low-mass protostars. A general correlation is observed for IRAS05358 and W43-MM1: the velocity offset of each component increases with FWHM, hence the broad component is more and more red-shifted as the FWHM increases. This could mean that for outflows we see stronger red lobes ($T_{ex}$ higher in the inner part) than blue lobes (where we see the outer part with lower $T_{ex}$). This trend is seen within the set of envelope and broad components for IRAS05358 but only clearly for the broad component in W43-MM1. For IRAS16272, a correlation is observed for the broad component while nothing is seen for DR21(OH). NGC6334I(N) tends to show a decrease of $V_{peak}-V_{source}$ with FWHM. Surprisingly, the broad component tends to be blue-shifted with respect to the two other velocity categories. Whatever the source, no clear trend is seen for the medium component. 

Hence, for W43-MM1 and IRAS05358 the different components globally lie in different regions of the FWHM vs. offset parameter space. We can then conclude that these components are formed under different conditions. Obviously, the same conclusion applies to the broad and narrow$/$medium components for NGC6334I(N) and perhaps for IRAS16272, but on the contrary no significant trend is observed for DR21(OH).

\subsection{Outflow}

We have used RADEX \citep[][]{vandertak2007} to roughly estimate the column density of the H$_2$O in the outflow component for each source, assuming an isothermal homogenous material, shielding envelope emission. Following \citet{vandertak2010} and \citet{herpin2012}, we adopt $n$(H$_2)=3\times10^4$cm$^{-3}$ and $T_{kin}=200$ K, but also test neighboring values. The water column density necessary to retrieve the observed intensities and reproduce the observed line ratios for the broad component as derived from the Gaussian fitting (see Sect. \ref{sec:vel} and Tables \ref{table_param_6334}-\ref{table_param_05358}) is $10^{17}, 1.2\times 10^{17}, 4\times10^{16}$, and $8\times10^{16}$ cm$^{-2}$, respectively for NGC6334I(N), DR21(OH) (with $T_{kin}=300$ K for this source), IRAS16272, and IRAS05358. These values are not sensitive to variations by less than 25\% of $T_{kin}$. In Sect. \ref{meth} the formation of water in the outflow is studied.

\subsection{Line asymmetries}
\label{sec:asym}

Compared to optically thin lines (e.g. C$^{18}$O 9-8 and $^{13}$CO 10-9 in most cases), most of the water-line profiles observed in our sources show clear asymmetries, revealing gas motions. Indeed, outflows, infall, and rotation can produce very specific line profiles with characteristic signatures \citep[see][and references therein]{fuller2005}. 

For all sources but IRAS05358, blue asymmetric (optically thick) lines, i.e. inverse P-Cygni profiles, are observed, hence likely indicating infalling material. In some circumstances, outflow or rotation could also produce a blue asymmetric line profile along a particular line of sight to the source. In IRAS05358 all lines but \hoD~and \hoH~have stronger red-shifted emission than blue-shifted emission (see Sec. \ref{kine}) with a strong self-absorption dip at the source velocity, i.e. a P-Cygni profile, typical of expansion. 

For IRAS16272, only the \hoE~line exhibits this asymmetry, the other lines being either not optically thick enough, or contaminated by foreground clouds (for ground-state lines). Toward DR21(OH), a marked inverse P-Cygni profile is observed for the  \hoD~and \hoE~lines, and weakly for \hoH, whose almost symmetric double-horn profile might be produced by the outflow \citep[][]{fuller2005}. The \hoD~ and \hoH~line profiles, which have a slightly stronger blue peak than the red one, exhibit a strong self-absorption dip at the source velocity. The \watersept~ and \waterhuit~absorption lines do not show clear asymmetry, but are blue-shifted relative to the source velocity, as expected in the case of infall. 

The case of NGC6334I(N) is less clear cut as the absorption of the blue component of the outflow in several lines complicates the interpretation: the infall is only clearly seen in the \hoD~line (blue asymmetric profile).

\subsection{Opacities and integrated line intensity ratios}
\label{opa_lines}

Depending on whether the line is in absorption or in emission, two different methods are applied to derive the opacities. For the absorption lines, we estimate the opacities at the maximum of absorption from the line-to-continuum ratio in Tables \ref{table_param_6334}-\ref{table_param_05358} using 
\begin{equation}
\tau = -{\textrm {ln}}(\frac{T_{mb}}{T_{cont}})
\end{equation} 
and assuming that the continuum is completely covered by the absorbing layer.

In all sources, even in central regions, all rare isotopologue lines are optically thin ($\tau<1$). For the three sources (W43-MM1, DR21(OH), and NGC6334I(N)) exhibiting \watersept~lines, the opacities are comparable, i.e. 0.11-0.17 and 0.04-0.1 for the $2_{12}-1_{01}$ and $1_{11}-0_{00}$ lines respectively. These opacities decrease from the less evolved (NGC6334I(N)) to the more evolved object (DR21(OH)). Interestingly, the \waterhuit$/$\watersept~opacity ratio for the $1_{11}-0_{00}$ line is close to the rare isotopologue ratio (4, see Sect. \ref{sec:model}) for W43-MM1, while it is slightly different for the two other objects (2.7$\pm$1.1 and 6.2$\pm$2.8). Except for IRAS16272 ($\tau=0.06$), the opacity of the \hoI~line is around 0.3 for all sources, larger than what is estimated for the \hoA~line (0.15-0.20) in W43-MM1 and NGC6334I(N).

In contrast, all the \water~ground state lines in absorption are optically thick, even totally absorbed for the 1113 and 557 GHz lines. More generally, IRAS16272 is the source with the lower opacities while DR21(OH) has the highest opacity. The \hoL~line is optically thin in the less evolved sources IRAS05358 and IRAS16272. 

In order to study the excitation and physical conditions of the water-emitting gas, we now focus on the opacities of the lines in emission. We first look for \water$/$\waterhuit~line pairs in our sample. Only the $3_{12}-3_{03}$ line exhibits emission for both \water~and \waterhuit~(a blend of emission and absorption prevents us from any accurate comparison of other lines). This is only observed in IRAS05358, DR21(OH), and NGC6334I(N). We adopt the following standard abundance ratios (same ratios for all the lines):  4.5 for \waterhuit$/$\watersept~\citep[][]{thomas2008}, and 3 for ortho$/$para-H$_2$O.  Based on \citet{wilson1994}, the $^{16}$O$/^{18}$O~abundance ratio depends on the distance from the Galactic center (while the \waterhuit$/$\watersept~ratio is constant). From the Wilson and Rood results and the distance adopted for IRAS05358, IRAS16272, NGC6334I(N) and DR21(OH) (see Table \ref{source_list}) we derive an  \water$/$\waterhuit~abundance ratio of 642, 363, 437 and 531, respectively. From the integrated intensity ratios for the $3_{12}-3_{03}$ line, and assuming that both isotopes have the same excitation temperature, we estimate an optical depth of order 15-30 for the \water~line, the \waterhuit~line being optically thin (according to our models, see Sect. \ref{method}).

For sources or components for which \waterhuit~data are not detected or usable, we follow the method described by \citet{mottram2014}. We use the ratio of the integrated intensity of the different components in pairs of \water~lines which share a common level. In Table \ref{opacity_ratio} we compare the ratios obtained for either the medium or broad component of the gas (again we restrict ourselves to the profiles with obvious components in emission) to the LTE ratios in the optically thin and thick regimes following \citet{goldsmith1999}, for $T_{ex}=300$ K like \citet{mottram2014} (note that some of the line ratios values in Mottram's paper are incorrect and we have then corrected the ratios from those given in Mottram et al. to the values listed in Table \ref{opacity_ratio}), and also for $T_{ex}=100$ and 500 K. The thin and thick line ratios for $T_{ex}=300$ and 500 K differ by only 10\% which is not significant compared to the observed ratios. Decreasing $T_{ex}$ has a larger impact, but does not change the interpretation. 
The line ratios must be corrected by a factor reflecting the different beam sizes of the observations ($\theta_1/\theta_2$, see Table \ref{opacity_ratio}). This correction factor is different depending on either the emission comes from a point source, $(\theta_1/\theta_2)^2$, fills the beam in one axis and is point-like in the other, $\theta_1/\theta_2$, or if the emitting region covers both axes (no correction). For our high-mass objects (mainly because of the large distance), one can only assume that the broad component for the ground-state lines covers the beam (based on HIFI maps, see Jacq et al., in preparation) while the other line emission should be smaller than the beam. Hence, we will consider the cases $\theta_1/\theta_2$ and $(\theta_1/\theta_2)^2$. Intermediate cases apply if one line is optically thick but the other is optically thin, or if the transitions are sub-thermally excited at temperatures less than 300 K.

For the broad component, nearly all line ratios are close to the optically thick limit (after beam correction). IRAS05358 is a difficult case because the $1_{10}-1_{01}/2_{12}-1_{01}$ ratio is close to the optically thin limit, even considering the large difference in beam size and likely emitting regions. The thick case is probably valid for the $1_{11}-0_{00}/2_{02}-1_{11}$ ratio even if the beam correction factor is close to 1. Nevertheless, considering the high critical densities of these water transitions (see Table \ref{table_transitions}), these lines can be sub-thermally excited. The medium velocity component in IRAS05358 is obviously in the optically thick limit while the conclusion is uncertain for DR21(OH) and IRAS16272. Likely the LTE conditions do not apply to the medium component in these sources.

\begin{figure}
\centering
\includegraphics[width=7.5cm]{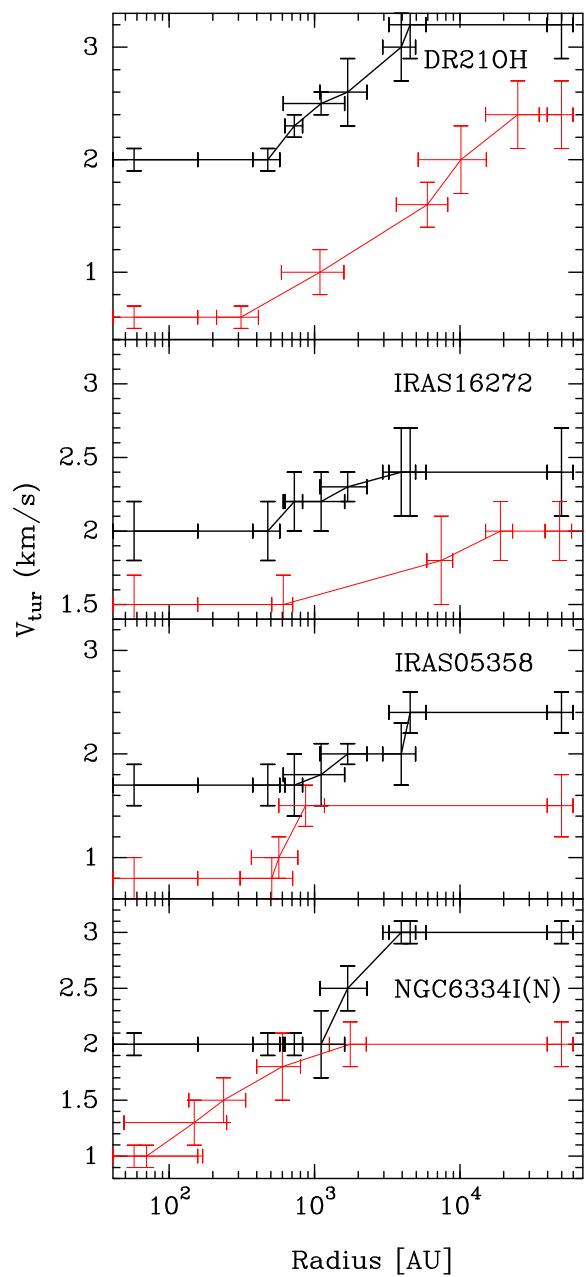}
\caption{Variation of the turbulent velocity $V_{turb}$ with the distance $R$ (AU) to the central object as derived from the model for water (in black) and CS (in red, see Appendix D). Error bars illustrate the range of values without any impact on the model.}
\label{Vturb_plot}%
\end{figure}

\section{Modeling}
\label{sec:model}
While the previous sections have presented the different velocity components in the observed lines profiles and indications of infall$/$expansion and outflow in the observed massive protostellar objects, this section intends to model the full line profiles in a single spherically symmetric model. The dynamics is driven by turbulence, infalling motions and outflow components.

\subsection{Method}
\label{method}

\begin{table*}
\caption{Derived parameters from RATRAN model and accretion luminosities.}             
\label{output_mod}      
\centering                          
\begin{tabular}{lcccccccccc}        
\hline\hline                 
Object &  $\chi_{out}$ &  $\chi_{in}$&M$_{H_2O}$ & M$_{H_2O}$ & M$_{inner}$  & $V_{inf}$ & $L_{tot}/(c.v_{inf})$ &$\dot{M}_{acc}$ &$L_{acc}$\\   
                      & $10^{-8}$ & $10^{-5 }$ &   [$10^{-4}$ M$_{\odot}$] &  [$10^{-7}$ M$_{env}$] & [\% M$_{total}$] &  [\kms] &  [M$_{\odot}$.yr$^{-1}$]&  [M$_{\odot}$.yr$^{-1}$] &  [$10^3$ L$_{\odot}$]\\
                      \hline
NGC6334I(N)  & 2.3 & 0.4& 8.8 & 2.3  & 3.6 & -0.7 & $5.4\times10^{-5}$& $5.2$-$5.6\times10^{-4}$ & 3.1-3.3 \\
W43-MM1  & 6.7 & 14 & 1100 &  146 & 97.2  & -2.9 &  $1.6\times10^{-4}$& $3$-$4\times10^{-2}$ & 30 \\
DR21(OH)  & 14 & 0.5 & 7.7 & 16.3  & 11.8 & -1.5 & $1.7\times10^{-4}$& $9.6$-$11\times10^{-5}$ & 0.58-0.66 \\ 
IRAS16272  & 4.7 & 0.17 & 1.3 & 0.6  & 43.8 & -0.2 &  $2.\times10^{-3}$& $6.3$-$6.8\times10^{-5}$ & 0.38-0.41 \\
IRAS05358  & 8.8 & 1.3 &  1.1 &  7.7 & 26.7 & +3.0& N$/$A & N$/$A &N$/$A \\
\hline                                  
\end{tabular}
\end{table*}

For all sources, the envelope temperature and density structure from \citet{vandertak2013} are used as input to the 1D-radiative transfer code RATRAN \citep[][]{hogerheijde2000} in order to reproduce simultaneously all the water line profiles, following the method of \citet{herpin2012}.  The H$_2$O collisional rate coefficients are from \citet{daniel2011}.

As in our previous publications \citep[i.e.,][]{chavarria2010,marseille2010,herpin2012}, the source model has two gas components: an outflow and the proto-stellar envelope. The outflow parameters, intensity and width, come from the Gaussian fitting presented in Section \ref{sec:results}. The envelope contribution is parametrized with three input variables: water abundance ($\chi_{H_2O}$), turbulent velocity ($V_{tur}$), and infall velocity ($V_{inf}$). The width of the line is adjusted by varying $V_{tur}$. The line asymmetry is reproduced by adjusting the infall velocity parameter. The line intensity is best fitted by adjusting a combination of the abundance, turbulence, and outflow parameters. We adopt the abundance ratios presented in Sect. \ref{opa_lines}. The models assume a jump in the abundance in the inner envelope at 100 K (see Section \ref{abundance_section}) due to the evaporation of ice mantles. Table \ref{output_mod} gives the parameters used in the models.

Our modeling strategy consists in fitting first the rare isotopologue lines (\watersept~and \waterhuit) since they are optically thin (see Sect. \ref{opa_lines}). Then we model the \water~lines starting from the highest energy level (the \water~abundances are derived from the \watersept~and \waterhuit~values times the isotopic abundance ratios). Once we are able to reproduce the main features of the profiles by minimizing the residuals \citep[see][]{herpin2012} in a grid of values, we model the remaining lines using the same parameters, including the outflow component when this is justified.

\subsection{Velocity structure}
\label{sec_turb}

Thanks to the high spectral resolution HIFI observations, we have access to crucial velocity details as already explained in Sect. \ref{sec:results} which help to constrain the source dynamics. In W43-MM1 \citet{herpin2012} have shown that a turbulence increasing with circumstellar radius provided the best line model. In order to test this conclusion in the other sample sources, we first try a model where turbulent velocity (and infall or expansion) is constant with radius for all lines, and then a model in which $V_{turb}$ varies with radius. For the constant model, the best fitted $V_{turb}$ values (same for all lines) are 2.5, 2.5, 2.2, and 2.0 \kms~(and $V_{inf}$ as in Table \ref{output_mod}) respectively for DR21(OH), NGC6334I(N), IRAS16272, and IRAS05358 (line models are in red in Fig. \ref{FigNGC6334IN}-\ref{Fig05358}). Except for DR21(OH), the model with constant velocity parameters for all lines fits the \waterhuit~and \watersept~data quite well as it does for C$^{18}$O \citep[][]{sanjose2013}. In contrast, it is not possible to converge to a good model for all \water~lines, even though the line profiles of IRAS05358 and IRAS16272 are well reproduced.

Inspection of the line profiles (see Tables~\ref{table_param_6334}-\ref{table_param_05358}) shows that the width of the velocity components is not the same for all lines. As for W43-MM1 \citep[][]{herpin2012} we do not expect a model with equal velocity parameters for all lines to fit the data well. Hence, as a second step we try a model in which the turbulent velocity varies with radius. We have tested several possibilities: a power-law variation, then various step profiles based on the turbulent velocity estimated for each line from Tables~\ref{table_param_6334}-\ref{table_param_05358}). The best models are obtained using the turbulence profiles shown in black in Fig.\ref{Vturb_plot} and overplotted in blue on Fig.\ref{FigNGC6334IN}-\ref{Fig05358}. For IRAS16272, the adopted turbulent profile is quite flat and improvement of the model fit is small and significant only for the \hoC~and \hoE~lines. An increasing turbulent velocity in IRAS05358 gives a better result for at least the \hoD~and \hoH~lines, the other line profiles being less sensitive to this change. The two other sources, DR21(OH) and NGC6334I(N) are the most sensitive to the turbulent velocity profile. The relatively steep profiles leading to the best lines fitting vary from 2 up to at least 3 \kms and strongly impact the depth and width of the absorption components (all isotopes). Clearly, a model in which the turbulent velocity increases with radius works better for NGC6334I(N), DR21(OH), and IRAS05358, whereas this is less clear for IRAS16272. Moreover, as explained in Appendix \ref{sec:CS_dis}, this also applies to the CS line modeling (see red plot in Fig. \ref{Vturb_plot}).

\subsection{Abundance structure}
\label{abundance_section}

The abundance is constrained by the modeling of the entire set of observed lines. Even if, as underlined by \citet{herpin2012}, only a couple of these lines (\hoG~and \hoH) are optically thin enough to probe the inner part of the envelope, part of all water line profiles is produced by water excited in the inner part and is revealed by the high spectral resolution of these observations. Moreover, for DR21(OH) we have access to the high-excitation \hoP~line. In addition, we have applied our model to the \waterhuit~ $3_{13}-2_{20}$ line at 203.3916 GHz ($E_{up}$=204 K) observed toward IRAS05358 (not detected) and DR21(OH) by \citet{marseille2010a}. As illustrated by the analysis in \citet{visser2013} for low-mass sources, the 203 GHz line is actually very useful because it is less optically thick due to lower Einstein A-coefficient and the dust continuum is more optically thin in the inner envelope, an issue that may certainly affect the high-mass sources. For these reasons, this  sample of lines should probe the entire water region. 

All line profiles are well reproduced. No deviation from the standard $o/p$ ratio of 3 is found. The \water~abundances relative to H$_2$ (see Table \ref{output_mod}) range from $1.7\times10^{-6}$ (IRAS16272) to $1.3\times10^{-5}$ (IRAS05358) in the inner part where $T>100$ K while the outer abundances (where $T<100$K) are a few $10^{-8}$ (except DR21(OH), $1.4\times10^{-7}$). These abundances are typical of those found in HMPO's  \citep[e.g.,][]{marseille2010,herpin2012}. While the water outer abundance is close to the common values of a few $10^{-8}$ \citep[][]{marseille2010,emprechtinger2013} for all sources, the relatively broad range of values ($1.7\times10^{-6}-1.4\times10^{-4}$) derived for the inner abundance requires further discussion (see Sect.\ref{sec:dry}). Except for W43-MM1, all our estimated inner water abundances are below the predicted high water inner abundance value from \citet{fraser2001} (see Table \ref{output_mod}). 

From our RATRAN models and the physical structure adopted for our sources \citep[][]{vandertak2013}, we find that all inner abundances correspond to a region where $n_{H_2}\sim1-4\times10^{7}$ cm$^{-3}$ and $T_{dust}\sim115$ K (except IRAS05358, 180 K) while the outer abundances are found for $n_{H_2}\sim10^{6}$ cm$^{-3}$ and $T_{dust}\sim25-50$ K. The region of the envelope probed by our observations is then roughly between 500 and 10000 AU. This also applies to W43-MM1 ($n_{H_2}\sim4\times10^{7}/10^6$ cm$^{-3}$ and $T_{dust}\sim210/30$ K) where the largest inner abundance is observed. This range of distances is exactly where the turbulent velocity is increasing in our model (see Sect.\ref{sec_turb}). We stress that our Ratran modeling is 1D only so the inferred distance is just indicative. Also we assume perfect symmetry which of corse is questionable in massive objects. 

\citet{schmalzl2014} have used a small (N reactions) chemical network ({\em SWaN}) to predict realistic water abundance profiles for low-mass protostellar envelopes, which have significantly improved the HIFI water line modeling for low-mass protostellar cores. Applying this chemistry network to our high-mass protostellar objects is very difficult because several crucial input parameters are not well known or are not optimized for high-mass protostars (e.g. the large amounts of UV photons produced by the massive object). Moreover, a large range of parameters has to be explored in order to get a realistic model. Even if the abundance profile is then only illustrative, we have used SWaN with one single set of input parameters only for IRAS05358 in order to illustrate that the outer abundance varying with radius likely affects the line profiles. The output of the line modeling is shown in green on Fig.\ref{Fig05358}. The result is worse than with a step-profile but has a clear impact on some line profiles.

\section{Discussion}
\label{sec:discussion}

\subsection{Why do the inner envelopes appear so dry?}
\label{sec:dry} 

Compared to previous studies of the water content in HM protostars \citep[e.g.][]{boonman2003}, HIFI gives us access to multiple water lines, and,  hence, enables us to model the complete water spectra in a robust and quite self-consistent way. Moreover, the use of resolved spectral lines allows us to separate the various kinematic components, and thus to partly compensate for the lack of spatial resolution. Our line sample probes deep enough the inner envelope at least for DR21(OH), IRAS05358, and W43-MM1 where we include in our model the $3_{13}-2_{20}$ transition of \waterhuit~(which strongly constrains the inner water abundance as explained in Sect. \ref{abundance_section}). On the other hand, we adopt here a simplified physical model, which is only constrained by single dish dust continuum data, hence not sensitive to physical structure of the hot inner core, whereas a highly complex structure is known for most of the studied sources \citep[e.g.][]{zapata2012}. Hence the high outer ($1.4\times10^{-7}$) and low inner ($5\times10^{-6}$) abundances derived for instance for DR21(OH) do not exclude the presence of a warm inner region which is not well reproduced in our model. At the studied scale, according to the literature the same argument can be hardly applied to the other sources. Recent work by \citet{visser2013} shows that in low-mass objects the spherical geometry as adopted here is not valid on the spatial scale of the hot core: a spherical envelope model with a single power-law density profile might lead to underestimate the inner water abundance. Obviously further investigation with more realistic physical model is necessary. From the observational side, the need for high spatial resolution observation has been well shown by \citet{vandertak2006} in order to get a precise estimate of the water inner abundance. Therefore further confirmation of our results by interferometric ALMA or NOEMA observations have to be done. 

\begin{figure}
\centering
\includegraphics[width=8.5cm]{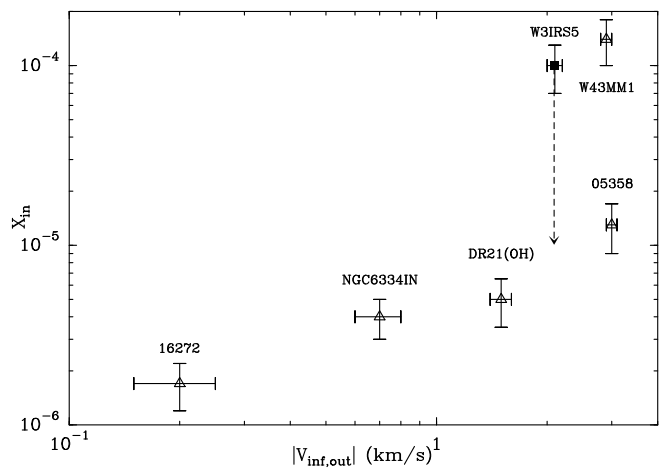}
\caption{Inner water abundance vs. the absolute value of the infall or expansion velocity as derived from our model (see Table \ref{output_mod}). The filled square symbol is for W3IRS5 from \citet{chavarria2010}, and the arrow shows the abundance for W3IRS5 from \citet{vandertak2006}.}
\label{X_Vin}%
\end{figure}

We have searched for a possible relation between kinematics and water abundances. First we investigate if a high level of turbulence could enhance (e.g. through shocks) the inner water abundance. No correlation is found between $\chi_{in}$ and the turbulent velocity ($V_{turb-min}$, $V_{turb-max}$, or $\overline{V_{turb}}$), or the outflow velocity (see Tables \ref{table_param_6334}-\ref{table_param_05358}, \ref{output_mod}, and Fig. \ref{Vturb_plot}). On the other hand, the higher the infall$/$expansion velocity ($|V_{inf,out}|$) is, the higher the inner abundance (see Fig. \ref{X_Vin}). Adding the W3IRS5 result ($10^{-4}$) from \citet{chavarria2010} confirms this trend. The trend is even strengthened if we rather use the estimate ($\sim 10^{-5}$) from \citet{vandertak2006} and Choi et al. (in preparation) for this source. Combined with the large turbulence observed in the high-mass protostellar objects, we propose that larger infall$/$expansion velocities generate shocks that will sputter water out of the dust grain mantles. Nevertheless according to \citet{neufeld2014} shock velocities of $\sim20-25$ \kms~are necessary to release water from ice mantles. 

\begin{figure}
\centering
\includegraphics[width=7.cm]{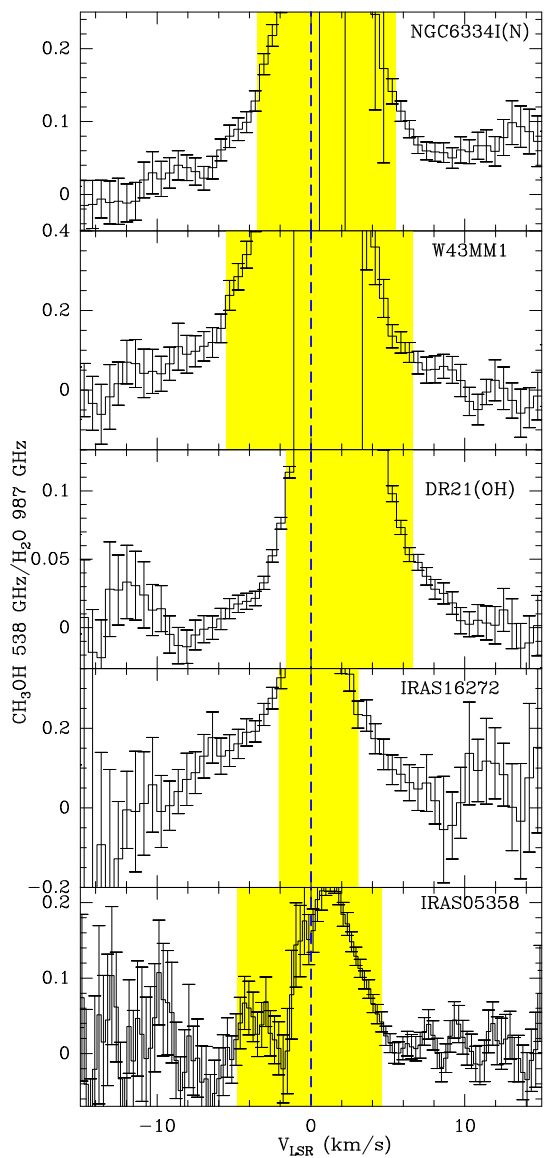}
\caption{Line ratio of the CH$_3$OH $5_{1,5}-4_{0,4}~A_+-A_+$ over the \hoE~transition, both rebinned to 0.5 \kms. The $V_{LSR}$ is shown with a dashed line. The range of velocities where the high optical depth of the water is assumed to affect the derived ratio is shown in yellow. }
\label{meth_water_ratio}%
\end{figure}

Another explanation to the low inner abundance could be that photodissociation through protostellar UV photons is more efficient than expected and thus not completely outrun by the O$\rightarrow$H$_2$O conversion. In the presence of strong UV radiation fields like the internal extreme UV radiation from the surface of a massive star \citep[e.g. $3\times10^{38}$ erg.s$^{-1}$,][]{benz2013}, water vapor is photodissociated. The physical known characteristics of our sources (see Table \ref{source_list}) do not indicate any difference in terms of FUV internal field among our sample, but we can imagine that for some reasons (e.g. self-shielding due to the thickness of the inner region) water photodissociation is more efficient in IRAS16272 than in W43-MM1 for instance. Observing good FUV irradiation tracers such as OH$^+$ or CH$^+$, or the product of the water photodissociation, i.e. OH, could help to constrain this scenario.

\subsection{Methanol and the formation route of water at higher velocities}
\label{meth}

Among the various species detected in our spectra (see Tables \ref{byproduct_05358}-\ref{byproduct_DR21OH}), methanol lines, because of their large number, are important indicators of the physical conditions in the protostellar envelope. The observed line ratio of CH$_3$OH over water as a function of the velocity in the line wings is able to differentiate between two potential formation routes of H$_2$O, gas-phase synthesis versus a sputtered origin in the outflow \citep[][]{suutarinen2014}. \citet{kempen2014} have shown that in intermediate-mass protostars organic molecules (e.g. the pure grain mantle product CH$_3$OH) likely originate from sputtering of ices due to outflow shocks and cannot form through gas-phase synthesis, as opposed to H$_2$O. Contrary to the water lines, most of the CH$_3$OH lines detected in our sample do not exhibit a broad component, but rather a medium velocity component. Hence, comparison of water and CH$_3$OH velocity components (e.g., see Fig. \ref{FigbyproductsCH3OH}) shows that methanol does not trace exactly the same gas as the water broad component, or that the abundance ratio is much smaller in the shocks. Among the detected methanol lines with a good S$/$N, we have compared the CH$_3$OH lines whose upper energy level is comparable ($E_u\leq105$ K, $5_{1,5}-4_{0,4}~A_+-A_+$, $6_{3,4}-5_{2,3}~A_+-A_+$, $6_{3,3}-5_{2,4}~A_--A_-$, and  $8_{1,7}-7_{0,7}~E$) with the \hoE~line. All transitions are assumed to fill the beam. Following \citet{kempen2014}, we do not consider a range of velocities around the line center where significant optical depth in the water lines increases the line ratio much more than physical processes would do. As explained in Sect. \ref{opa_lines}, considering the high critical densities of water transitions, we will assume effectively thin emission away from this central line region \citep[according to][the effect of water opacity on the ratio is still small]{suutarinen2014}. 

Figure \ref{meth_water_ratio} shows the ratio for the CH$_3$OH $5_{1,5}-4_{0,4}~A_+-A_+$ over the \hoE~transitions. Except for IRAS05358 where the methanol line does not show any wing, the ratio clearly drops as a function of velocity on both wings (although the number of channels in the red wing not affected by optical depth is smaller), suggesting, since methanol is produced on grain surfaces, a dominant gas-phase synthesis of H$_2$O from shocked material. Line ratios of other CH$_3$OH lines over the same H$_2$O line (or \hoD) show the same behavior. We can hence conclude that high-velocity water must be formed in the gas-phase from shocked material, i.e. not created solely through grain mantle evaporation.

\subsection{Star formation processes and evolution}

\begin{figure}
\centering
\includegraphics[width=8.5cm]{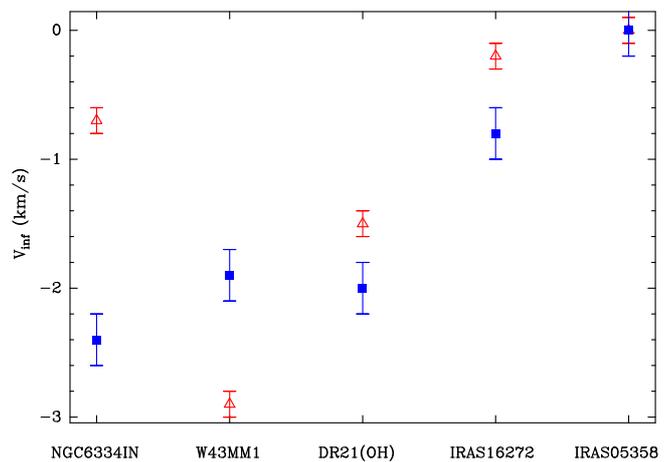}
\caption{Infall velocity vs. evolutionary sequence. Red triangles and filled blue squares represent respectively H$_2$O and CS (see Appendix D) velocities.}
\label{Vin_vary}%
\end{figure}

\begin{table*}
  \caption{Observed fluxes and luminosities for all lines and sources.}
\begin{center}
\begin{small}
\label{table_flux}      
\begin{tabular}{lcccccccc} \hline \hline
  & \multicolumn{2}{c}{NGC6334I(N)}  & \multicolumn{2}{c}{DR21(OH)} & \multicolumn{2}{c}{IRAS16272} & \multicolumn{2}{c}{IRAS05358}    \\ 
 {\bf Line}  & F & L$_{\odot}$  & F & L$_{\odot}$ & F & L$_{\odot}$ & F & L$_{\odot}$ \\ 
 &  [$10^{-21}$ W.cm$^{-2}$] & [$10^{-2}$]& [$10^{-21}$ W.cm$^{-2}$] &[$10^{-2}$]&  [$10^{-21}$ W.cm$^{-2}$] & [$10^{-2}$]&  [$10^{-21}$ W.cm$^{-2}$] & [$10^{-2}$]  \\ 
\hline                        
{\bf H$_2^{18}$O} &  & & & & & & & \\
 1$_{10}$-1$_{01}$ &  & & 1.7(0.2)& 0.12(0.02) &0.2(0.1) & 0.07(0.04)& & \\
2$_{02}$-1$_{11}$     & & & 5.7(0.6) & 0.40(0.04)& &  \\ 
3$_{12}$-3$_{03}$   &  0.4(0.2) & 0.04(0.02) &4.1(0.4) & 0.29(0.03)&  & &   0.8(0.2)   & 0.08(0.02)\\
 1$_{11}$-0$_{00}$     & & & 1.4(0.2)& 0.10(0.02)& &  \\ 
{\bf H$_2^{17}$O} &  & & & & & & & \\
1$_{10}$-1$_{01}$     & & &0.8(0.6) & 0.05(0.04) & &  \\ 
1$_{11}$-0$_{00}$    & & &2.1(0.3) & 0.15(0.02) & & \\
\hline
{\bf H$_2$O} &  & & & & & & & \\
1$_{10}$-1$_{01}$ &  20(2)& 1.8(0.2) & 28(1)& 1.97(0.07) & 5.8(0.3) & 2.1(0.1)&   5.6(0.7) &  0.57(0.07)\\
2$_{11}$-2$_{02}$     & 49(2) & 4.4(0.2)& 74(5) & 5.2(0.4) &17.4(0.8) & 6.3(0.3)&   15.5(0.8) & 1.57(0.08)  \\
5$_{24}$-4$_{31}$  &  & & 2.7(0.7)& 0.19(0.05) & & &    & \\
2$_{02}$-1$_{11}$      &  60(4)& 5.4(0.4)& 108(6) & 7.6(0.4) &19(1) & 6.9(0.4) &   37(1) & 3.7(0.1)  \\
3$_{12}$-3$_{03}$   &  34(2)& 3.1(0.2)& 87(6)& 6.1(0.4) &15(1) & 5.4(0.4) &   28(1) & 2.8(0.1)   \\
1$_{11}$-0$_{00}$ &  23(2)& 2.1(0.2)& 42(5)& 3.0(0.4) &8.9(0.6) & 3.2(0.2)&   23(1) & 2.3(0.1) \\
2$_{21}$-2$_{12}$ &  & & & & & &   14(4) & 1.4(0.4)\\
2$_{12}$-1$_{01}$ &  & & 120(20) & 8(1) &  & &   20(4) & 2.0(0.4) \\
\hline
\end{tabular}
\end{small}
\end{center}
\end{table*}

Turbulence, infall, and outflow are important ingredients of the star formation process. We confirm here that the observed molecular emission in massive protostellar objects is dominated by supersonic turbulent velocities (see Fig. \ref{Vturb_plot}) at all radii. Hence regions of massive-star formation  are highly turbulent. This supersonic turbulence is in agreement with the turbulent core model of \citet{mckee2003}, but is also consistent with the competitive accretion scenario in which local velocity dispersions are small but line-of-sight ones can be significantly larger due to the fragmentation of the cloud \citep[][]{bonnel2006}. Moreover, our results indicate that the turbulent motions tend to increase with radius, consistent again with all models \citep[see][]{bonnel2006,mckee2003}. Nevertheless, rotation and non-spherical density structure cannot be excluded as possible explanations \citep[see][]{herpin2012} and accurate estimates of the turbulence also require careful subtraction of cold foreground clouds in some cases (Jacq et al in prep). 

Assuming isotropic radiation, from the integrated fluxes of components in emission in observed lines (Table \ref{table_flux}), we estimate the lower limit to the total HIFI water luminosity by adding all individual observed luminosities to be 16.8, 33.1, 23.9, and 14.4 \lsol in our sources (evolutionary order). This confirms the low  contribution of water cooling to the total far-IR gas cooling compared to the cooling from other species \citep[][]{karska2014}. The true water emission from the inner part could be much larger but the cool envelope absorbs much of the emission. Moreover, from the modeling we estimate the total water mass in the envelope (see Table \ref{output_mod}) to roughly $8-9\times10^{-4}$ \Msol for NGC6334I(N) and DR21(OH), and $10^{-4}$ \Msol for the two other objects. The fraction of the water mass to the envelope mass differs by more than an order of magnitude from source to source ($0.6\times10^{-7}$ to $16.3\times10^{-7}$). The proportion of the mass in the inner part is only 3.6\% for NGC6334I(N) and 11.8\% for DR21(OH), but increases to 26.7\% and 43.8\% respectively for the more evolved object IRAS05358 and IRAS16272. 

If the whole envelope mass (see Table \ref{source_list}) would be collapsing, assuming free-fall accretion one would expect infall velocities from 2.1 \kms~for IRAS05358 up to 9.7 \kms~for NGC6334I(N), larger by up to a factor of 10 than what we estimate here \citep[we even detect no infall in IRAS05358 consistent with HCO$^+$ observations of][]{klassen2012}. The free-fall accretion rate \citep[monolithic collapse,][]{shu1977} is $6.3\times10^{-6}$ M$_{\odot}/$yr for sound speed of 0.3 \kms (a few $10^{-5}$ M$_{\odot}/$yr for temperatures above 100 K), hence several orders of magnitude smaller than what we derived from our observations (see Table \ref{output_mod}). We note that the CS infall velocity provides us with another estimate of the mass accretion rate (see Table \ref{output_modCS} in Appendix D), approximately 50\% lower than the values inferred from the water lines. Models of star formation based on gravoturbulent fragmentation \citep[][]{schmeja2004} predict mass accretion rates of $3.4\times10^{-5}$ M$_{\odot}/$yr (same sound speed of gas) but varying with time, while this rate is constant in the standard theory of isolated star formation from \citet{shu1977}. Of course, these accretion rates from our calculations assume spherical accretion and one central object, which is likely not true. Nevertheless, these rates are comparable to those derived from the turbulent core model for massive molecular cloud cores dominated by supersonic turbulence \citep[a few $10^{-5}-10^{-4}$ M$_{\odot}/$yr,][]{mckee2003} and from the competitive accretion model which can produce accretion rates from $10^{-9}$ to $10^{-4}$ M$_{\odot}/$yr (depending on the gas density and initial stellar, fragment, mass). Models of gravitational collapse of massive magnetized molecular cloud cores \citep[e.g., ][]{banerjee2007} generate massive star formation via high accretion rates (that can exceed $10^{-3}$ M$_{\odot}/$yr) and disk-driven outflows.

Considering a typical mass of 20 $M_{\odot}$ within a radius of 100 R$_{\star}$, we estimate the corresponding accretion luminosity of a protostar from \citet{hosokawa2010} using 
\begin{equation}
L_{acc}=G \frac{M_{\star} {\dot{M}}_{acc}}{R_{\star}}
\end{equation} 
The derived luminosities (see Table \ref{output_mod}), compared to the observed total luminosity (stellar+accretion, Table \ref{source_list}) seem unrealistically high for W43-MM1 and NGC6334I(N), but we stress that the scales of infall probed by water and the accretion onto the protostars are likely quite different: what we probe here is infall in the envelope rather than accretion onto the protostar. However, using the lower values derived from CS observations, the accretion luminosities are compatible with the observed total luminosity, so we suggest that water may not be a good accretion tracer. Moreover, the derived accretion rate, although uncertain, is high enough for W43-MM1 and NGC6334I(N) to overcome the radiation pressure due to the stellar luminosity (see Table \ref{table_flux}). This is possibly true for DR21(OH) too, but not for IRAS16272 (no accretion is detected in IRAS05358). Nevertheless, the simple comparison of radiation pressure and mass accretion rate in the simple one-dimensional collapse model view is considered as insufficient to explain the observations \citep[][]{peters2010}.

Beyond the classical monolithic collapse and competitive accretion model, \citet{peters2010} proposed the collapse of a rotating massive cloud core involving a process called fragmentation-induced starvation to reproduce the strong clumping and filamentary structures observed in collapsing cores. According to this model, the accretion is decreasing with time (between $10^{-3}$ and $10^{-5}$ M$_{\odot}/$yr), consistent with results from \citet{peters2010}. But considering the evolutionary sequence described in Sect. \ref{sec:sample}, the accretion rates derived for our sample (Table \ref{output_mod}) do not show any trend. 

Because of the different level of fragmentation$/$substructure in our source sample, comparing estimates of the mass accretion rates along the evolutionary sequence of this sample is uncertain. Nevertheless, the infall velocity derived from H$_2$O (except for NGC6334IN) and CS observations tends to decrease with the evolutionary stage of the massive object (Fig.\ref{Vin_vary}).

\section{Conclusions}
\label{sec:Conclusions}

We have presented Herschel-HIFI observations of 14 far-IR water lines (\water, \watersept, \waterhuit) toward four mid-IR quiet massive protostellar objects, assumed to be at the beginning of the high-mass star formation process, and ordered in terms of an evolutionary sequence based on their SED. We have studied the envelope kinematics (outflow, infall, turbulent velocity) from the different components identified in the line profiles of our source sample, and we have derived the water abundances using the RATRAN radiative transfer code. In addition, our analysis has been supplemented in terms of 'evolution' and water formation by the serendipitous detection of several other molecular lines, especially CS and methanol. 

The water lines have broad, medium, and narrow velocity components while no envelope component was found in W43-MM1 by \citet{herpin2012}. The more evolved sources, IRAS05358 and IRAS16272, exhibit fewer and weaker rare isotopologue lines and appear to be less rich chemically as indicated by the number of serendipitous species detected in these observations. We confirm that regions of massive star formation are highly turbulent and that turbulence tends to increase in the envelope with the distance to the star, as seen in W43-MM1 by \citet{herpin2012}. This trend is consistent with the supersonic turbulent core model leading to high mass star formation in the presence of a disk, although these constraints are also consistent with competitive accretion. 

The whole set of lines allowed us to constrain by modeling the outer water abundance to the typical value of a few $10^{-8}$ while we infer $1.7\times10^{-6}-1.4\times10^{-4}$ for the inner abundance, lower (except for W43-MM1) than expected from ice evaporation. Possible explanations could be : (i) Photodissociation of water from the massive protostar UV internal photons is more efficient than expected, or (ii) Our simple spherical envelope model underestimates the inner water abundance.  Moreover, we show that the higher the infall/expansion velocity in the protostellar envelope, the higher is the inner abundance. This fact, in addition to the observed lower infall velocity along the source evolutionary sequence, suggests that the younger sources with larger infall$/$expansion velocities may generate shocks that will sputter water from the ice mantles of dust grains in the inner region. At high velocities, water must be formed in the gas-phase from shocked material, i.e. not created solely through grain mantle evaporation.

\begin{acknowledgements}
This program was made possible thanks to the HIFI guaranteed time. HIFI has been designed and built by a consortium of
institutes and university departments from across Europe, Canada and the
United States under the leadership of SRON Netherlands Institute for Space
Research, Groningen, The Netherlands and with major contributions from
Germany, France and the US. Consortium members are: Canada: CSA,
U.Waterloo; France: CESR, LAB, LERMA, IRAM; Germany: KOSMA,
MPIfR, MPS; Ireland, NUI Maynooth; Italy: ASI, IFSI-INAF, Osservatorio
Astrofisico di Arcetri- INAF; Netherlands: SRON, TUD; Poland: CAMK, CBK;
Spain: Observatorio Astronomico Nacional (IGN), Centro de Astrobiologi
(CSIC-INTA). Sweden: Chalmers University of Technology - MC2, RSS \&
GARD; Onsala Space Observatory; Swedish National Space Board, Stockholm
University - Stockholm Observatory; Switzerland: ETH Zurich, FHNW; USA:
Caltech, JPL, NHSC.
HIPE is a joint development by the Herschel Science Ground
Segment Consortium, consisting of ESA, the NASA Herschel Science Center, and the HIFI, PACS and
SPIRE consortia. Astrochemistry in Leiden is supported by the Netherlands Research School for Astronomy (NOVA), by a Spinoza grant and grant 614.001.008 from the Netherlands Organisation for Scientific Research (NWO), and by the European Community's Seventh Framework Program FP7$/$2007-2013 under grant agreement 238258 (LASSIE). We also thank the French Space Agency CNES for financial support. We thank J. Mottram for useful comments and discussions.

\end{acknowledgements}

\bibliography{biblio}

\begin{thebibliography}{108}
\expandafter\ifx\csname natexlab\endcsname\relax\def\natexlab#1{#1}\fi

\bibitem[{{Aikawa} {et~al.}(2008){Aikawa}, {Wakelam}, {Garrod}, \&
  {Herbst}}]{aikawa2008}
{Aikawa}, Y., {Wakelam}, V., {Garrod}, R.~T., \& {Herbst}, E. 2008, \apj, 674,
  984

\bibitem[{{Araya} {et~al.}(2009){Araya}, {Kurtz}, {Hofner}, \&
  {Linz}}]{araya2009}
{Araya}, E.~D., {Kurtz}, S., {Hofner}, P., \& {Linz}, H. 2009, \apj, 698, 1321

\bibitem[{{Banerjee} \& {Pudritz}(2007)}]{banerjee2007}
{Banerjee}, R. \& {Pudritz}, R.~E. 2007, \apj, 660, 479

\bibitem[{{Benz} {et~al.}(2013){Benz}, {Bruderer}, {van Dishoeck},
  {St{\"a}uber}, \& {Wampfler}}]{benz2013}
{Benz}, A.~O., {Bruderer}, S., {van Dishoeck}, E.~F., {St{\"a}uber}, P., \&
  {Wampfler}, S.~F. 2013, Journal of Physical Chemistry A, 117, 9840

\bibitem[{{Benz} {et~al.}(2010){Benz}, {Bruderer}, {van Dishoeck}, {WISH Team},
  \& {HIFI Team}}]{benz2010}
{Benz}, A.~O., {Bruderer}, S., {van Dishoeck}, E.~F., {WISH Team}, \& {HIFI
  Team}. 2010, ArXiv e-prints

\bibitem[{{Beuther} {et~al.}(2007){Beuther}, {Leurini}, {Schilke}, {Wyrowski},
  {Menten}, \& {Zhang}}]{beuther2007}
{Beuther}, H., {Leurini}, S., {Schilke}, P., {et~al.} 2007, \aap, 466, 1065

\bibitem[{{Beuther} {et~al.}(2002{\natexlab{a}}){Beuther}, {Schilke}, {Gueth},
  {McCaughrean}, {Andersen}, {Sridharan}, \& {Menten}}]{beuther2002c}
{Beuther}, H., {Schilke}, P., {Gueth}, F., {et~al.} 2002{\natexlab{a}}, \aap,
  387, 931

\bibitem[{{Beuther} {et~al.}(2002{\natexlab{b}}){Beuther}, {Schilke}, {Menten},
  {Motte}, {Sridharan}, \& {Wyrowski}}]{beuther2002a}
{Beuther}, H., {Schilke}, P., {Menten}, K.~M., {et~al.} 2002{\natexlab{b}},
  \apj, 566, 945

\bibitem[{{Beuther} {et~al.}(2008){Beuther}, {Walsh}, {Thorwirth}, {Zhang},
  {Hunter}, {Megeath}, \& {Menten}}]{beuther2008}
{Beuther}, H., {Walsh}, A.~J., {Thorwirth}, S., {et~al.} 2008, \aap, 481, 169

\bibitem[{{Bonnell} \& {Bate}(2006)}]{bonnel2006}
{Bonnell}, I.~A. \& {Bate}, M.~R. 2006, \mnras, 370, 488

\bibitem[{{Bontemps} {et~al.}(1996){Bontemps}, {Andre}, {Terebey}, \&
  {Cabrit}}]{bontemps1996}
{Bontemps}, S., {Andre}, P., {Terebey}, S., \& {Cabrit}, S. 1996, \aap, 311,
  858

\bibitem[{{Boonman} {et~al.}(2003){Boonman}, {Doty}, {van Dishoeck}, {Bergin},
  {Melnick}, {Wright}, \& {Stark}}]{boonman2003}
{Boonman}, A.~M.~S., {Doty}, S.~D., {van Dishoeck}, E.~F., {et~al.} 2003, \aap,
  406, 937

\bibitem[{{Brogan} {et~al.}(2009){Brogan}, {Hunter}, {Cyganowski},
  {Indebetouw}, {Beuther}, {Menten}, \& {Thorwirth}}]{brogan2009}
{Brogan}, C.~L., {Hunter}, T.~R., {Cyganowski}, C.~J., {et~al.} 2009, \apj,
  707, 1

\bibitem[{{Bruderer} {et~al.}(2010){Bruderer}, {Benz}, {van Dishoeck},
  {Melchior}, {Doty}, {van der Tak}, {St{\"a}uber}, {Wampfler}, {Dedes},
  {Y{\i}ld{\i}z}, {Pagani}, {Giannini}, {de Graauw}, {Whyborn}, {Teyssier},
  {Jellema}, {Shipman}, {Schieder}, {Honingh}, {Caux}, {B{\"a}chtold},
  {Csillaghy}, {Monstein}, {Bachiller}, {Baudry}, {Benedettini}, {Bergin},
  {Bjerkeli}, {Blake}, {Bontemps}, {Braine}, {Caselli}, {Cernicharo},
  {Codella}, {Daniel}, {di Giorgio}, {Dominik}, {Encrenaz}, {Fich}, {Fuente},
  {Goicoechea}, {Helmich}, {Herczeg}, {Herpin}, {Hogerheijde}, {Jacq},
  {Johnstone}, {J{\o}rgensen}, {Kristensen}, {Larsson}, {Lis}, {Liseau},
  {Marseille}, {McCoey}, {Melnick}, {Neufeld}, {Nisini}, {Olberg}, {Parise},
  {Pearson}, {Plume}, {Risacher}, {Santiago-Garcia}, {Saraceno}, {Shipman},
  {Tafalla}, {van Kempen}, {Visser}, \& {Wyrowski}}]{bruderer2010}
{Bruderer}, S., {Benz}, A.~O., {van Dishoeck}, E.~F., {et~al.} 2010, ArXiv
  e-prints

\bibitem[{{Ceccarelli} {et~al.}(2010){Ceccarelli}, {Bacmann}, {Boogert},
  {Caux}, {Dominik}, {Lefloch}, {Lis}, {Schilke}, {van der Tak}, {Caselli},
  {Cernicharo}, {Codella}, {Comito}, {Fuente}, {Baudry}, {Bell}, {Benedettini},
  {Bergin}, {Blake}, {Bottinelli}, {Cabrit}, {Castets}, {Coutens}, {Crimier},
  {Demyk}, {Encrenaz}, {Falgarone}, {Gerin}, {Goldsmith}, {Helmich},
  {Hennebelle}, {Henning}, {Herbst}, {Hily-Blant}, {Jacq}, {Kahane}, {Kama},
  {Klotz}, {Langer}, {Lord}, {Lorenzani}, {Maret}, {Melnick}, {Neufeld},
  {Nisini}, {Pacheco}, {Pagani}, {Parise}, {Pearson}, {Phillips}, {Salez},
  {Saraceno}, {Schuster}, {Tielens}, {van der Wiel}, {Vastel}, {Viti},
  {Wakelam}, {Walters}, {Wyrowski}, {Yorke}, {Liseau}, {Olberg}, {Szczerba},
  {Benz}, \& {Melchior}}]{ceccarelli2010}
{Ceccarelli}, C., {Bacmann}, A., {Boogert}, A., {et~al.} 2010, \aap, 521, L22

\bibitem[{{Chandler} {et~al.}(1993){Chandler}, {Moore}, {Mountain}, \&
  {Yamashita}}]{chandler1993}
{Chandler}, C.~J., {Moore}, T.~J.~T., {Mountain}, C.~M., \& {Yamashita}, T.
  1993, \mnras, 261, 694

\bibitem[{{Charnley}(1997)}]{charnley1997}
{Charnley}, S.~B. 1997, \apj, 481, 396

\bibitem[{{Chavarr{\'{\i}}a} {et~al.}(2010){Chavarr{\'{\i}}a}, {Herpin},
  {Jacq}, {Braine}, {Bontemps}, {Baudry}, {Marseille}, {van der Tak},
  {Pietropaoli}, {Wyrowski}, {Shipman}, {Frieswijk}, {van Dishoeck},
  {Cernicharo}, {Bachiller}, {Benedettini}, {Benz}, {Bergin}, {Bjerkeli},
  {Blake}, {Bruderer}, {Caselli}, {Codella}, {Daniel}, {di Giorgio}, {Dominik},
  {Doty}, {Encrenaz}, {Fich}, {Fuente}, {Giannini}, {Goicoechea}, {de Graauw},
  {Hartogh}, {Helmich}, {Herczeg}, {Hogerheijde}, {Johnstone}, {J{\o}rgensen},
  {Kristensen}, {Larsson}, {Lis}, {Liseau}, {McCoey}, {Melnick}, {Nisini},
  {Olberg}, {Parise}, {Pearson}, {Plume}, {Risacher}, {Santiago-Garc{\'{\i}}a},
  {Saraceno}, {Stutzki}, {Szczerba}, {Tafalla}, {Tielens}, {van Kempen},
  {Visser}, {Wampfler}, {Willem}, \& {Y{\i}ld{\i}z}}]{chavarria2010}
{Chavarr{\'{\i}}a}, L., {Herpin}, F., {Jacq}, T., {et~al.} 2010, \aap, 521,
  L37+

\bibitem[{{Cortes}(2011)}]{cortes2011}
{Cortes}, P.~C. 2011, \apj, 743, 194

\bibitem[{{Csengeri} {et~al.}(2011){Csengeri}, {Bontemps}, {Schneider},
  {Motte}, {Gueth}, \& {Hora}}]{csengeri2011}
{Csengeri}, T., {Bontemps}, S., {Schneider}, N., {et~al.} 2011, \apjl, 740, L5

\bibitem[{{Daniel} {et~al.}(2011){Daniel}, {Dubernet}, \&
  {Grosjean}}]{daniel2011}
{Daniel}, F., {Dubernet}, M.-L., \& {Grosjean}, A. 2011, \aap, 536, A76

\bibitem[{{de Graauw} {et~al.}(2010){de Graauw}, {Helmich}, {Phillips},
  {Stutzki}, {Caux}, {Whyborn}, {Dieleman}, {Roelfsema}, {Aarts}, {Assendorp},
  {Bachiller}, {Baechtold}, {Barcia}, {Beintema}, {Belitsky}, {Benz}, {Bieber},
  {Boogert}, {Borys}, {Bumble}, {Ca{\"i}s}, {Caris}, {Cerulli-Irelli},
  {Chattopadhyay}, {Cherednichenko}, {Ciechanowicz}, {Coeur-Joly}, {Comito},
  {Cros}, {de Jonge}, {de Lange}, {Delforges}, {Delorme}, {den Boggende},
  {Desbat}, {Diez-Gonz{\'a}lez}, {di Giorgio}, {Dubbeldam}, {Edwards},
  {Eggens}, {Erickson}, {Evers}, {Fich}, {Finn}, {Franke}, {Gaier}, {Gal},
  {Gao}, {Gallego}, {Gauffre}, {Gill}, {Glenz}, {Golstein}, {Goulooze},
  {Gunsing}, {G{\"u}sten}, {Hartogh}, {Hatch}, {Higgins}, {Honingh}, {Huisman},
  {Jackson}, {Jacobs}, {Jacobs}, {Jarchow}, {Javadi}, {Jellema}, {Justen},
  {Karpov}, {Kasemann}, {Kawamura}, {Keizer}, {Kester}, {Klapwijk}, {Klein},
  {Kollberg}, {Kooi}, {Kooiman}, {Kopf}, {Krause}, {Krieg}, {Kramer},
  {Kruizenga}, {Kuhn}, {Laauwen}, {Lai}, {Larsson}, {Leduc}, {Leinz}, {Lin},
  {Liseau}, {Liu}, {Loose}, {L{\'o}pez-Fernandez}, {Lord}, {Luinge}, {Marston},
  {Mart{\'{\i}}n-Pintado}, {Maestrini}, {Maiwald}, {McCoey}, {Mehdi}, {Megej},
  {Melchior}, {Meinsma}, {Merkel}, {Michalska}, {Monstein}, {Moratschke},
  {Morris}, {Muller}, {Murphy}, {Naber}, {Natale}, {Nowosielski}, {Nuzzolo},
  {Olberg}, {Olbrich}, {Orfei}, {Orleanski}, {Ossenkopf}, {Peacock}, {Pearson},
  {Peron}, {Phillip-May}, {Piazzo}, {Planesas}, {Rataj}, {Ravera}, {Risacher},
  {Salez}, {Samoska}, {Saraceno}, {Schieder}, {Schlecht}, {Schl{\"o}der},
  {Schm{\"u}lling}, {Schultz}, {Schuster}, {Siebertz}, {Smit}, {Szczerba},
  {Shipman}, {Steinmetz}, {Stern}, {Stokroos}, {Teipen}, {Teyssier}, {Tils},
  {Trappe}, {van Baaren}, {van Leeuwen}, {van de Stadt}, {Visser}, {Wildeman},
  {Wafelbakker}, {Ward}, {Wesselius}, {Wild}, {Wulff}, {Wunsch}, {Tielens},
  {Zaal}, {Zirath}, {Zmuidzinas}, \& {Zwart}}]{deGraauw2010}
{de Graauw}, T., {Helmich}, F.~P., {Phillips}, T.~G., {et~al.} 2010, \aap, 518,
  L6+

\bibitem[{{Emprechtinger} {et~al.}(2013){Emprechtinger}, {Lis}, {Rolffs},
  {Schilke}, {Monje}, {Comito}, {Ceccarelli}, {Neufeld}, \& {van der
  Tak}}]{emprechtinger2013}
{Emprechtinger}, M., {Lis}, D.~C., {Rolffs}, R., {et~al.} 2013, \apj, 765, 61

\bibitem[{{Fa{\'u}ndez} {et~al.}(2004){Fa{\'u}ndez}, {Bronfman}, {Garay},
  {Chini}, {Nyman}, \& {May}}]{faundez2004}
{Fa{\'u}ndez}, S., {Bronfman}, L., {Garay}, G., {et~al.} 2004, \aap, 426, 97

\bibitem[{{Fraser} {et~al.}(2001){Fraser}, {Collings}, {McCoustra}, \&
  {Williams}}]{fraser2001}
{Fraser}, H.~J., {Collings}, M.~P., {McCoustra}, M.~R.~S., \& {Williams}, D.~A.
  2001, \mnras, 327, 1165

\bibitem[{{Fuller} {et~al.}(2005){Fuller}, {Williams}, \&
  {Sridharan}}]{fuller2005}
{Fuller}, G.~A., {Williams}, S.~J., \& {Sridharan}, T.~K. 2005, \aap, 442, 949

\bibitem[{{Garay} {et~al.}(2007){Garay}, {Mardones}, {Brooks}, {Videla}, \&
  {Contreras}}]{garay2007}
{Garay}, G., {Mardones}, D., {Brooks}, K.~J., {Videla}, L., \& {Contreras}, Y.
  2007, \apj, 666, 309

\bibitem[{{Gerner} {et~al.}(2014){Gerner}, {Beuther}, {Semenov}, {Linz},
  {Vasyunina}, {Bihr}, {Shirley}, \& {Henning}}]{gerner2014}
{Gerner}, T., {Beuther}, H., {Semenov}, D., {et~al.} 2014, \aap, 563, A97

\bibitem[{{Girart} {et~al.}(2013){Girart}, {Frau}, {Zhang}, {Koch}, {Qiu},
  {Tang}, {Lai}, \& {Ho}}]{girart2013}
{Girart}, J.~M., {Frau}, P., {Zhang}, Q., {et~al.} 2013, \apj, 772, 69

\bibitem[{{Goldsmith} \& {Langer}(1999)}]{goldsmith1999}
{Goldsmith}, P.~F. \& {Langer}, W.~D. 1999, \apj, 517, 209

\bibitem[{{Harvey} {et~al.}(1986){Harvey}, {Joy}, {Lester}, \&
  {Wilking}}]{harvey1986}
{Harvey}, P.~M., {Joy}, M., {Lester}, D.~F., \& {Wilking}, B.~A. 1986, \apj,
  300, 737

\bibitem[{{Heitsch} {et~al.}(2008){Heitsch}, {Hartmann}, {Slyz}, {Devriendt},
  \& {Burkert}}]{heitsch2008}
{Heitsch}, F., {Hartmann}, L.~W., {Slyz}, A.~D., {Devriendt}, J.~E.~G., \&
  {Burkert}, A. 2008, \apj, 674, 316

\bibitem[{{Helmich} {et~al.}(1996){Helmich}, {van Dishoeck}, {Black}, {de
  Graauw}, {Beintema}, {Heras}, {Lahuis}, {Morris}, \&
  {Valentijn}}]{helmich1996}
{Helmich}, F.~P., {van Dishoeck}, E.~F., {Black}, J.~H., {et~al.} 1996, \aap,
  315, L173

\bibitem[{{Hennemann} {et~al.}(2012){Hennemann}, {Motte}, {Schneider},
  {Didelon}, {Hill}, {Arzoumanian}, {Bontemps}, {Csengeri}, {Andr{\'e}},
  {Konyves}, {Louvet}, {Marston}, {Men'shchikov}, {Minier}, {Nguyen Luong},
  {Palmeirim}, {Peretto}, {Sauvage}, {Zavagno}, {Anderson}, {Bernard}, {Di
  Francesco}, {Elia}, {Li}, {Martin}, {Molinari}, {Pezzuto}, {Russeil}, {Rygl},
  {Schisano}, {Spinoglio}, {Sousbie}, {Ward-Thompson}, \&
  {White}}]{hennemann2012}
{Hennemann}, M., {Motte}, F., {Schneider}, N., {et~al.} 2012, \aap, 543, L3

\bibitem[{{Herpin} {et~al.}(2012){Herpin}, {Chavarr{\'{\i}}a}, {van der Tak},
  {Wyrowski}, {van Dishoeck}, {Jacq}, {Braine}, {Baudry}, {Bontemps}, \&
  {Kristensen}}]{herpin2012}
{Herpin}, F., {Chavarr{\'{\i}}a}, L., {van der Tak}, F., {et~al.} 2012, \aap,
  542, A76

\bibitem[{{Herpin} {et~al.}(2009){Herpin}, {Marseille}, {Wakelam}, {Bontemps},
  \& {Lis}}]{herpin2009}
{Herpin}, F., {Marseille}, M., {Wakelam}, V., {Bontemps}, S., \& {Lis}, D.~C.
  2009, \aap, 504, 853

\bibitem[{{Hogerheijde} \& {van der Tak}(2000)}]{hogerheijde2000}
{Hogerheijde}, M.~R. \& {van der Tak}, F.~F.~S. 2000, \aap, 362, 697

\bibitem[{{Hosokawa} {et~al.}(2010){Hosokawa}, {Yorke}, \&
  {Omukai}}]{hosokawa2010}
{Hosokawa}, T., {Yorke}, H.~W., \& {Omukai}, K. 2010, \apj, 721, 478

\bibitem[{{Hunter} {et~al.}(2014){Hunter}, {Brogan}, {Cyganowski}, \&
  {Young}}]{hunter2014}
{Hunter}, T.~R., {Brogan}, C.~L., {Cyganowski}, C.~J., \& {Young}, K.~H. 2014,
  \apj, 788, 187

\bibitem[{{Hunter} {et~al.}(2006){Hunter}, {Brogan}, {Megeath}, {Menten},
  {Beuther}, \& {Thorwirth}}]{hunter2006}
{Hunter}, T.~R., {Brogan}, C.~L., {Megeath}, S.~T., {et~al.} 2006, \apj, 649,
  888

\bibitem[{{Jacq} {et~al.}(1990){Jacq}, {Walmsley}, {Henkel}, {Baudry},
  {Mauersberger}, \& {Jewell}}]{jacq1990}
{Jacq}, T., {Walmsley}, C.~M., {Henkel}, C., {et~al.} 1990, \aap, 228, 447

\bibitem[{{Johnstone} {et~al.}(2010){Johnstone}, {Fich}, {McCoey}, {van
  Kempen}, {Fuente}, {Kristensen}, {Cernicharo}, {Caselli}, {Visser}, {Plume},
  {Herczeg}, {van Dishoeck}, {Wampfler}, {Bachiller}, {Baudry}, {Benedettini},
  {Bergin}, {Benz}, {Bjerkeli}, {Blake}, {Bontemps}, {Braine}, {Bruderer},
  {Codella}, {Daniel}, {di Giorgio}, {Dominik}, {Doty}, {Encrenaz}, {Giannini},
  {Goicoechea}, {de Graauw}, {Helmich}, {Herpin}, {Hogerheijde}, {Jacq},
  {J{\o}rgensen}, {Larsson}, {Lis}, {Liseau}, {Marseille}, {Melnick},
  {Neufeld}, {Nisini}, {Olberg}, {Parise}, {Pearson}, {Risacher},
  {Santiago-Garc{\'{\i}}a}, {Saraceno}, {Shipman}, {Tafalla}, {van der Tak},
  {Wyrowski}, {Y{\i}ld{\i}z}, {Caux}, {Honingh}, {Jellema}, {Schieder},
  {Teyssier}, \& {Whyborn}}]{johnstone2010}
{Johnstone}, D., {Fich}, M., {McCoey}, C., {et~al.} 2010, \aap, 521, L41+

\bibitem[{{Karska} {et~al.}(2014){Karska}, {Herpin}, {Bruderer}, {Goicoechea},
  {Herczeg}, {van Dishoeck}, {San Jos{\'e}-Garc{\'{\i}}a}, {Contursi},
  {Feuchtgruber}, {Fedele}, {Baudry}, {Braine}, {Chavarr{\'{\i}}a},
  {Cernicharo}, {van der Tak}, \& {Wyrowski}}]{karska2014}
{Karska}, A., {Herpin}, F., {Bruderer}, S., {et~al.} 2014, \aap, 562, A45

\bibitem[{{Ka{\'z}mierczak-Barthel} {et~al.}(2015){Ka{\'z}mierczak-Barthel},
  {Semenov}, {van der Tak}, {Chavarr{\'{\i}}a}, \& {van der
  Wiel}}]{kazmierczak2015}
{Ka{\'z}mierczak-Barthel}, M., {Semenov}, D.~A., {van der Tak}, F.~F.~S.,
  {Chavarr{\'{\i}}a}, L., \& {van der Wiel}, M.~H.~D. 2015, \aap, 574, A71

\bibitem[{{Klaassen} {et~al.}(2012){Klaassen}, {Testi}, \&
  {Beuther}}]{klassen2012}
{Klaassen}, P.~D., {Testi}, L., \& {Beuther}, H. 2012, \aap, 538, A140

\bibitem[{{Kristensen} {et~al.}(2010){Kristensen}, {Visser}, {van Dishoeck},
  {Y{\i}ld{\i}z}, {Doty}, {Herczeg}, {Liu}, {Parise}, {J{\o}rgensen}, {van
  Kempen}, {Brinch}, {Wampfler}, {Bruderer}, {Benz}, {Hogerheijde}, {Deul},
  {Bachiller}, {Baudry}, {Benedettini}, {Bergin}, {Bjerkeli}, {Blake},
  {Bontemps}, {Braine}, {Caselli}, {Cernicharo}, {Codella}, {Daniel}, {de
  Graauw}, {di Giorgio}, {Dominik}, {Encrenaz}, {Fich}, {Fuente}, {Giannini},
  {Goicoechea}, {Helmich}, {Herpin}, {Jacq}, {Johnstone}, {Kaufman}, {Larsson},
  {Lis}, {Liseau}, {Marseille}, {McCoey}, {Melnick}, {Neufeld}, {Nisini},
  {Olberg}, {Pearson}, {Plume}, {Risacher}, {Santiago-Garc{\'{\i}}a},
  {Saraceno}, {Shipman}, {Tafalla}, {Tielens}, {van der Tak}, {Wyrowski},
  {Beintema}, {de Jonge}, {Dieleman}, {Ossenkopf}, {Roelfsema}, {Stutzki}, \&
  {Whyborn}}]{kristensen2010}
{Kristensen}, L.~E., {Visser}, R., {van Dishoeck}, E.~F., {et~al.} 2010, \aap,
  521, L30+

\bibitem[{{Krumholz} \& {Bonnell}(2009)}]{krumholz2009}
{Krumholz}, M.~R. \& {Bonnell}, I.~A. 2009, {Models for the formation of
  massive stars}, ed. {Chabrier, G.} (Cambridge University Press), 288

\bibitem[{{Krumholz} {et~al.}(2005){Krumholz}, {McKee}, \&
  {Klein}}]{krumholz2005}
{Krumholz}, M.~R., {McKee}, C.~F., \& {Klein}, R.~I. 2005, \apjl, 618, L33

\bibitem[{{Kuiper} {et~al.}(2010){Kuiper}, {Klahr}, {Beuther}, \&
  {Henning}}]{kuiper2010}
{Kuiper}, R., {Klahr}, H., {Beuther}, H., \& {Henning}, T. 2010, \apj, 722,
  1556

\bibitem[{{Kuiper} {et~al.}(2015){Kuiper}, {Yorke}, \& {Turner}}]{kuiper2015}
{Kuiper}, R., {Yorke}, H.~W., \& {Turner}, N.~J. 2015, \apj, 800, 86

\bibitem[{{Lester} {et~al.}(1985){Lester}, {Dinerstein}, {Werner}, {Harvey},
  {Evans}, \& {Brown}}]{lester1985}
{Lester}, D.~F., {Dinerstein}, H.~L., {Werner}, M.~W., {et~al.} 1985, \apj,
  296, 565

\bibitem[{{Leurini} {et~al.}(2007){Leurini}, {Beuther}, {Schilke}, {Wyrowski},
  {Zhang}, \& {Menten}}]{leurini2007}
{Leurini}, S., {Beuther}, H., {Schilke}, P., {et~al.} 2007, \aap, 475, 925

\bibitem[{{Marseille} {et~al.}(2010{\natexlab{a}}){Marseille}, {van der Tak},
  {Herpin}, \& {Jacq}}]{marseille2010a}
{Marseille}, M.~G., {van der Tak}, F.~F.~S., {Herpin}, F., \& {Jacq}, T.
  2010{\natexlab{a}}, \aap, 522, A40+

\bibitem[{{Marseille} {et~al.}(2010{\natexlab{b}}){Marseille}, {van der Tak},
  {Herpin}, {Wyrowski}, {Chavarr{\'{\i}}a}, {Pietropaoli}, {Baudry},
  {Bontemps}, {Cernicharo}, {Jacq}, {Frieswijk}, {Shipman}, {van Dishoeck},
  {Bachiller}, {Benedettini}, {Benz}, {Bergin}, {Bjerkeli}, {Blake}, {Braine},
  {Bruderer}, {Caselli}, {Caux}, {Codella}, {Daniel}, {Dieleman}, {di Giorgio},
  {Dominik}, {Doty}, {Encrenaz}, {Fich}, {Fuente}, {Gaier}, {Giannini},
  {Goicoechea}, {de Graauw}, {Helmich}, {Herczeg}, {Hogerheijde}, {Jackson},
  {Javadi}, {Jellema}, {Johnstone}, {J{\o}rgensen}, {Kester}, {Kristensen},
  {Larsson}, {Laauwen}, {Lis}, {Liseau}, {Luinge}, {McCoey}, {Megej},
  {Melnick}, {Neufeld}, {Nisini}, {Olberg}, {Parise}, {Pearson}, {Plume},
  {Risacher}, {Roelfsema}, {Santiago-Garc{\'{\i}}a}, {Saraceno}, {Siegel},
  {Stutzki}, {Tafalla}, {van Kempen}, {Visser}, {Wampfler}, \&
  {Y{\i}ld{\i}z}}]{marseille2010}
{Marseille}, M.~G., {van der Tak}, F.~F.~S., {Herpin}, F., {et~al.}
  2010{\natexlab{b}}, \aap, 521, L32+

\bibitem[{{McCutcheon} {et~al.}(2000){McCutcheon}, {Sandell}, {Matthews},
  {Kuiper}, {Sutton}, {Danchi}, \& {Sato}}]{mccutcheon2000}
{McCutcheon}, W.~H., {Sandell}, G., {Matthews}, H.~E., {et~al.} 2000, \mnras,
  316, 152

\bibitem[{{McKee} \& {Tan}(2003)}]{mckee2003}
{McKee}, C.~F. \& {Tan}, J.~C. 2003, \apj, 585, 850

\bibitem[{{Motte} {et~al.}(2007){Motte}, {Bontemps}, {Schilke}, {Schneider},
  {Menten}, \& {Brogui{\`e}re}}]{motte2007}
{Motte}, F., {Bontemps}, S., {Schilke}, P., {et~al.} 2007, \aap, 476, 1243

\bibitem[{{Motte} {et~al.}(2003){Motte}, {Schilke}, \& {Lis}}]{motte2003}
{Motte}, F., {Schilke}, P., \& {Lis}, D.~C. 2003, \apj, 582, 277

\bibitem[{{Mottram} {et~al.}(2011){Mottram}, {Hoare}, {Davies}, {Lumsden},
  {Oudmaijer}, {Urquhart}, {Moore}, {Cooper}, \& {Stead}}]{mottram2011}
{Mottram}, J.~C., {Hoare}, M.~G., {Davies}, B., {et~al.} 2011, \apjl, 730, L33+

\bibitem[{{Mottram} {et~al.}(2014){Mottram}, {Kristensen}, {van Dishoeck},
  {Bruderer}, {San Jos{\'e}-Garc{\'{\i}}a}, {Karska}, {Visser}, {Santangelo},
  {Benz}, {Bergin}, {Caselli}, {Herpin}, {Hogerheijde}, {Johnstone}, {van
  Kempen}, {Liseau}, {Nisini}, {Tafalla}, {van der Tak}, \&
  {Wyrowski}}]{mottram2014}
{Mottram}, J.~C., {Kristensen}, L.~E., {van Dishoeck}, E.~F., {et~al.} 2014,
  \aap, 572, A21

\bibitem[{{Mu{\~n}oz} {et~al.}(2007){Mu{\~n}oz}, {Mardones}, {Garay},
  {Rebolledo}, {Brooks}, \& {Bontemps}}]{munoz2007}
{Mu{\~n}oz}, D.~J., {Mardones}, D., {Garay}, G., {et~al.} 2007, \apj, 668, 906

\bibitem[{{Neufeld} {et~al.}(2014){Neufeld}, {Gusdorf}, {G{\"u}sten},
  {Herczeg}, {Kristensen}, {Melnick}, {Nisini}, {Ossenkopf}, {Tafalla}, \& {van
  Dishoeck}}]{neufeld2014}
{Neufeld}, D.~A., {Gusdorf}, A., {G{\"u}sten}, R., {et~al.} 2014, \apj, 781,
  102

\bibitem[{{Ott}(2010)}]{ott2010}
{Ott}, S. 2010, in Astronomical Society of the Pacific Conference Series, Vol.
  434, Astronomical Data Analysis Software and Systems XIX, ed. {Y.~Mizumoto,
  K.-I.~Morita, \& M.~Ohishi}, 139--+

\bibitem[{{Palau} {et~al.}(2014){Palau}, {Estalella}, {Girart}, {Fuente},
  {Fontani}, {Commer{\c c}on}, {Busquet}, {Bontemps}, {S{\'a}nchez-Monge},
  {Zapata}, {Zhang}, {Hennebelle}, \& {di Francesco}}]{palau2014}
{Palau}, A., {Estalella}, R., {Girart}, J.~M., {et~al.} 2014, \apj, 785, 42

\bibitem[{{Pearson} {et~al.}(1991){Pearson}, {De Lucia}, {Anderson}, {Herbst},
  \& {Helminger}}]{pearson1991}
{Pearson}, J.~C., {De Lucia}, F.~C., {Anderson}, T., {Herbst}, E., \&
  {Helminger}, P. 1991, \apjl, 379, L41

\bibitem[{{Peters} {et~al.}(2010){Peters}, {Klessen}, {Mac Low}, \&
  {Banerjee}}]{peters2010}
{Peters}, T., {Klessen}, R.~S., {Mac Low}, M.-M., \& {Banerjee}, R. 2010, \apj,
  725, 134

\bibitem[{{Pilbratt} {et~al.}(2010){Pilbratt}, {Riedinger}, {Passvogel},
  {Crone}, {Doyle}, {Gageur}, {Heras}, {Jewell}, {Metcalfe}, {Ott}, \&
  {Schmidt}}]{Pilbratt2010}
{Pilbratt}, G.~L., {Riedinger}, J.~R., {Passvogel}, T., {et~al.} 2010, \aap,
  518, L1+

\bibitem[{{Ragan} {et~al.}(2012){Ragan}, {Henning}, {Krause}, {Pitann},
  {Beuther}, {Linz}, {Tackenberg}, {Balog}, {Hennemann}, {Launhardt}, {Lippok},
  {Nielbock}, {Schmiedeke}, {Schuller}, {Steinacker}, {Stutz}, \&
  {Vasyunina}}]{ragan2012}
{Ragan}, S., {Henning}, T., {Krause}, O., {et~al.} 2012, \aap, 547, A49

\bibitem[{{Richardson} {et~al.}(1994){Richardson}, {Sandell}, {Cunningham}, \&
  {Davies}}]{richardson1994}
{Richardson}, K.~J., {Sandell}, G., {Cunningham}, C.~T., \& {Davies}, S.~R.
  1994, \aap, 286, 555

\bibitem[{{Roberts} {et~al.}(2010){Roberts}, {Rawlings}, \&
  {Stace}}]{roberts2010}
{Roberts}, J.~F., {Rawlings}, J.~M.~C., \& {Stace}, H.~A. 2010, \mnras, 408,
  2426

\bibitem[{{Rodr{\'{\i}}guez} {et~al.}(2007){Rodr{\'{\i}}guez}, {Zapata}, \&
  {Ho}}]{rodriguez2007}
{Rodr{\'{\i}}guez}, L.~F., {Zapata}, L.~A., \& {Ho}, P.~T.~P. 2007, \apjl, 654,
  L143

\bibitem[{{Roelfsema} {et~al.}(2012){Roelfsema}, {Helmich}, {Teyssier},
  {Ossenkopf}, {Morris}, {Olberg}, {Shipman}, {Risacher}, {Akyilmaz},
  {Assendorp}, {Avruch}, {Beintema}, {Biver}, {Boogert}, {Borys}, {Braine},
  {Caris}, {Caux}, {Cernicharo}, {Coeur-Joly}, {Comito}, {de Lange},
  {Delforge}, {Dieleman}, {Dubbeldam}, {de Graauw}, {Edwards}, {Fich},
  {Flederus}, {Gal}, {di Giorgio}, {Herpin}, {Higgins}, {Hoac}, {Huisman},
  {Jarchow}, {Jellema}, {de Jonge}, {Kester}, {Klein}, {Kooi}, {Kramer},
  {Laauwen}, {Larsson}, {Leinz}, {Lord}, {Lorenzani}, {Luinge}, {Marston},
  {Mart{\'{\i}}n-Pintado}, {McCoey}, {Melchior}, {Michalska}, {Moreno},
  {M{\"u}ller}, {Nowosielski}, {Okada}, {Orlea{\'n}ski}, {Phillips}, {Pearson},
  {Rabois}, {Ravera}, {Rector}, {Rengel}, {Sagawa}, {Salomons},
  {S{\'a}nchez-Su{\'a}rez}, {Schieder}, {Schl{\"o}der}, {Schm{\"u}lling},
  {Soldati}, {Stutzki}, {Thomas}, {Tielens}, {Vastel}, {Wildeman}, {Xie},
  {Xilouris}, {Wafelbakker}, {Whyborn}, {Zaal}, {Bell}, {Bjerkeli}, {De Beck},
  {Cavali{\'e}}, {Crockett}, {Hily-Blant}, {Kama}, {Kaminski}, {Lefl{\'o}ch},
  {Lombaert}, {de Luca}, {Makai}, {Marseille}, {Nagy}, {Pacheco}, {van der
  Wiel}, {Wang}, \& {Y{\i}ld{\i}z}}]{roelfsema2012}
{Roelfsema}, P.~R., {Helmich}, F.~P., {Teyssier}, D., {et~al.} 2012, \aap, 537,
  A17

\bibitem[{{San Jos{\'e}-Garc{\'{\i}}a} {et~al.}(2013){San
  Jos{\'e}-Garc{\'{\i}}a}, {Mottram}, {Kristensen}, {van Dishoeck},
  {Y{\i}ld{\i}z}, {van der Tak}, {Herpin}, {Visser}, {McCoey}, {Wyrowski},
  {Braine}, \& {Johnstone}}]{sanjose2013}
{San Jos{\'e}-Garc{\'{\i}}a}, I., {Mottram}, J.~C., {Kristensen}, L.~E.,
  {et~al.} 2013, \aap, 553, A125

\bibitem[{{San Jose-Garcia} {et~al.}(2015){San Jose-Garcia}, {Mottram}, {van
  Dishoeck}, {Kristensen}, {van der Tak}, {Braine}, {Herpin}, {Johnstone}, {van
  Kempen}, \& {Wyrowski}}]{sanjose2015a}
{San Jose-Garcia}, I., {Mottram}, J.~C., {van Dishoeck}, E.~F., {et~al.} 2015,
  ArXiv e-prints

\bibitem[{{Sandell}(2000)}]{sandell2000}
{Sandell}, G. 2000, \aap, 358, 242

\bibitem[{{Schlingman} {et~al.}(2011){Schlingman}, {Shirley}, {Schenk},
  {Rosolowsky}, {Bally}, {Battersby}, {Dunham}, {Ellsworth-Bowers}, {Evans},
  {Ginsburg}, \& {Stringfellow}}]{schlingman2011}
{Schlingman}, W.~M., {Shirley}, Y.~L., {Schenk}, D.~E., {et~al.} 2011, \apjs,
  195, 14

\bibitem[{{Schmalzl} {et~al.}(2014){Schmalzl}, {Visser}, {Walsh}, {Albertsson},
  {van Dishoeck}, {Kristensen}, \& {Mottram}}]{schmalzl2014}
{Schmalzl}, M., {Visser}, R., {Walsh}, C., {et~al.} 2014, \aap, 572, A81

\bibitem[{{Schmeja} \& {Klessen}(2004)}]{schmeja2004}
{Schmeja}, S. \& {Klessen}, R.~S. 2004, \aap, 419, 405

\bibitem[{{Shipman} {et~al.}(2014){Shipman}, {van der Tak}, {Wyrowski},
  {Herpin}, \& {Frieswijk}}]{shipman2014}
{Shipman}, R.~F., {van der Tak}, F.~F.~S., {Wyrowski}, F., {Herpin}, F., \&
  {Frieswijk}, W. 2014, \aap, 570, A51

\bibitem[{{Shirley} {et~al.}(2003){Shirley}, {Evans}, {Young}, {Knez}, \&
  {Jaffe}}]{shirley2003}
{Shirley}, Y.~L., {Evans}, II, N.~J., {Young}, K.~E., {Knez}, C., \& {Jaffe},
  D.~T. 2003, \apjs, 149, 375

\bibitem[{{Shu}(1977)}]{shu1977}
{Shu}, F.~H. 1977, \apj, 214, 488

\bibitem[{{Snell} {et~al.}(2000){Snell}, {Howe}, {Ashby}, {Bergin}, {Chin},
  {Erickson}, {Goldsmith}, {Harwit}, {Kleiner}, {Koch}, {Neufeld}, {Patten},
  {Plume}, {Schieder}, {Stauffer}, {Tolls}, {Wang}, {Winnewisser}, {Zhang}, \&
  {Melnick}}]{snell2000}
{Snell}, R.~L., {Howe}, J.~E., {Ashby}, M.~L.~N., {et~al.} 2000, \apjl, 539,
  L101

\bibitem[{{Suutarinen} {et~al.}(2014){Suutarinen}, {Kristensen}, {Mottram},
  {Fraser}, \& {van Dishoeck}}]{suutarinen2014}
{Suutarinen}, A.~N., {Kristensen}, L.~E., {Mottram}, J.~C., {Fraser}, H.~J., \&
  {van Dishoeck}, E.~F. 2014, \mnras, 440, 1844

\bibitem[{{Tan} {et~al.}(2014){Tan}, {Beltr{\'a}n}, {Caselli}, {Fontani},
  {Fuente}, {Krumholz}, {McKee}, \& {Stolte}}]{tan2014}
{Tan}, J.~C., {Beltr{\'a}n}, M.~T., {Caselli}, P., {et~al.} 2014, Protostars
  and Planets VI, 149

\bibitem[{{Tan} \& {McKee}(2002)}]{mckee2002}
{Tan}, J.~C. \& {McKee}, C.~F. 2002, in Astronomical Society of the Pacific
  Conference Series, Vol. 267, Hot Star Workshop III: The Earliest Phases of
  Massive Star Birth, ed. P.~{Crowther}, 267--+

\bibitem[{{Thomas} \& {Fuller}(2008)}]{thomas2008}
{Thomas}, H.~S. \& {Fuller}, G.~A. 2008, \aap, 479, 751

\bibitem[{{Vall{\'e}e} \& {Fiege}(2006)}]{vallee2006}
{Vall{\'e}e}, J.~P. \& {Fiege}, J.~D. 2006, \apj, 636, 332

\bibitem[{{van der Tak} {et~al.}(2007){van der Tak}, {Black}, {Sch{\"o}ier},
  {Jansen}, \& {van Dishoeck}}]{vandertak2007}
{van der Tak}, F.~F.~S., {Black}, J.~H., {Sch{\"o}ier}, F.~L., {Jansen}, D.~J.,
  \& {van Dishoeck}, E.~F. 2007, \aap, 468, 627

\bibitem[{{van der Tak} {et~al.}(2013){van der Tak}, {Chavarr{\'{\i}}a},
  {Herpin}, {Wyrowski}, {Walmsley}, {van Dishoeck}, {Benz}, {Bergin},
  {Caselli}, {Hogerheijde}, {Johnstone}, {Kristensen}, {Liseau}, {Nisini}, \&
  {Tafalla}}]{vandertak2013}
{van der Tak}, F.~F.~S., {Chavarr{\'{\i}}a}, L., {Herpin}, F., {et~al.} 2013,
  \aap, 554, A83

\bibitem[{{van der Tak} {et~al.}(2010){van der Tak}, {Marseille}, {Herpin},
  {Wyrowski}, {Baudry}, {Bontemps}, {Braine}, {Doty}, {Frieswijk}, {Melnick},
  {Shipman}, {van Dishoeck}, {Benz}, {Caselli}, {Hogerheijde}, {Johnstone},
  {Liseau}, {Bachiller}, {Benedettini}, {Bergin}, {Bjerkeli}, {Blake},
  {Bruderer}, {Cernicharo}, {Codella}, {Daniel}, {di Giorgio}, {Dominik},
  {Encrenaz}, {Fich}, {Fuente}, {Giannini}, {Goicoechea}, {de Graauw},
  {Helmich}, {Herczeg}, {J{\o}rgensen}, {Kristensen}, {Larsson}, {Lis},
  {McCoey}, {Neufeld}, {Nisini}, {Olberg}, {Parise}, {Pearson}, {Plume},
  {Risacher}, {Santiago}, {Saraceno}, {Tafalla}, {van Kempen}, {Visser},
  {Wampfler}, {Y{\i}ld{\i}z}, {Ravera}, {Roelfsema}, {Siebertz}, \&
  {Teyssier}}]{vandertak2010}
{van der Tak}, F.~F.~S., {Marseille}, M.~G., {Herpin}, F., {et~al.} 2010, \aap,
  518, L107+

\bibitem[{{van der Tak} {et~al.}(2000){van der Tak}, {van Dishoeck}, {Evans},
  \& {Blake}}]{vandertak2000a}
{van der Tak}, F.~F.~S., {van Dishoeck}, E.~F., {Evans}, II, N.~J., \& {Blake},
  G.~A. 2000, \apj, 537, 283

\bibitem[{{van der Tak} {et~al.}(2006){van der Tak}, {Walmsley}, {Herpin}, \&
  {Ceccarelli}}]{vandertak2006}
{van der Tak}, F.~F.~S., {Walmsley}, C.~M., {Herpin}, F., \& {Ceccarelli}, C.
  2006, \aap, 447, 1011

\bibitem[{{van der Wiel} {et~al.}(2013){van der Wiel}, {Pagani}, {van der Tak},
  {Ka{\'z}mierczak}, \& {Ceccarelli}}]{wiel2013}
{van der Wiel}, M.~H.~D., {Pagani}, L., {van der Tak}, F.~F.~S.,
  {Ka{\'z}mierczak}, M., \& {Ceccarelli}, C. 2013, \aap, 553, A11

\bibitem[{{van Dishoeck} \& {Helmich}(1996)}]{vandishoeck1996}
{van Dishoeck}, E.~F. \& {Helmich}, F.~P. 1996, \aap, 315, L177

\bibitem[{{van Dishoeck} {et~al.}(2013){van Dishoeck}, {Herbst}, \&
  {Neufeld}}]{vandishoeck2013}
{van Dishoeck}, E.~F., {Herbst}, E., \& {Neufeld}, D.~A. 2013, Chemical
  Reviews, 113, 9043

\bibitem[{{van Dishoeck} {et~al.}(2011){van Dishoeck}, {Kristensen}, {Benz},
  {Bergin}, {Caselli}, {Cernicharo}, {Herpin}, {Hogerheijde}, {Johnstone},
  {Liseau}, {Nisini}, {Shipman}, {Tafalla}, {van der Tak}, {Wyrowski},
  {Aikawa}, {Bachiller}, {Baudry}, {Benedettini}, {Bjerkeli}, {Blake},
  {Bontemps}, {Braine}, {Brinch}, {Bruderer}, {Chavarr{\'{\i}}a}, {Codella},
  {Daniel}, {de Graauw}, {Deul}, {di Giorgio}, {Dominik}, {Doty}, {Dubernet},
  {Encrenaz}, {Feuchtgruber}, {Fich}, {Frieswijk}, {Fuente}, {Giannini},
  {Goicoechea}, {Helmich}, {Herczeg}, {Jacq}, {J{\o}rgensen}, {Karska},
  {Kaufman}, {Keto}, {Larsson}, {Lefloch}, {Lis}, {Marseille}, {McCoey},
  {Melnick}, {Neufeld}, {Olberg}, {Pagani}, {Pani{\'c}}, {Parise}, {Pearson},
  {Plume}, {Risacher}, {Salter}, {Santiago-Garc{\'{\i}}a}, {Saraceno},
  {St{\"a}uber}, {van Kempen}, {Visser}, {Viti}, {Walmsley}, {Wampfler}, \&
  {Y{\i}ld{\i}z}}]{vandishoeck2011}
{van Dishoeck}, E.~F., {Kristensen}, L.~E., {Benz}, A.~O., {et~al.} 2011,
  \pasp, 123, 138

\bibitem[{{van Kempen} {et~al.}(2014){van Kempen}, {McCoey}, {Tisi},
  {Johnstone}, \& {Fich}}]{kempen2014}
{van Kempen}, T.~A., {McCoey}, C., {Tisi}, S., {Johnstone}, D., \& {Fich}, M.
  2014, \aap, 569, A53

\bibitem[{{V{\'a}zquez-Semadeni} {et~al.}(2007){V{\'a}zquez-Semadeni},
  {G{\'o}mez}, {Jappsen}, {Ballesteros-Paredes}, {Gonz{\'a}lez}, \&
  {Klessen}}]{vazquez2007}
{V{\'a}zquez-Semadeni}, E., {G{\'o}mez}, G.~C., {Jappsen}, A.~K., {et~al.}
  2007, \apj, 657, 870

\bibitem[{{Villanueva} {et~al.}(2012){Villanueva}, {DiSanti}, {Mumma}, \&
  {Xu}}]{villanueva2012}
{Villanueva}, G.~L., {DiSanti}, M.~A., {Mumma}, M.~J., \& {Xu}, L.-H. 2012,
  \apj, 747, 37

\bibitem[{{Visser} {et~al.}(2013){Visser}, {J{\o}rgensen}, {Kristensen}, {van
  Dishoeck}, \& {Bergin}}]{visser2013}
{Visser}, R., {J{\o}rgensen}, J.~K., {Kristensen}, L.~E., {van Dishoeck},
  E.~F., \& {Bergin}, E.~A. 2013, \apj, 769, 19

\bibitem[{{Walsh} {et~al.}(1998){Walsh}, {Burton}, {Hyland}, \&
  {Robinson}}]{walsh1998}
{Walsh}, A.~J., {Burton}, M.~G., {Hyland}, A.~R., \& {Robinson}, G. 1998,
  \mnras, 301, 640

\bibitem[{{Wilson} \& {Rood}(1994)}]{wilson1994}
{Wilson}, T.~L. \& {Rood}, R. 1994, \araa, 32, 191

\bibitem[{{Woody} {et~al.}(1989){Woody}, {Scott}, {Scoville}, {Mundy},
  {Sargent}, {Padin}, {Tinney}, \& {Wilson}}]{woody1989}
{Woody}, D.~P., {Scott}, S.~L., {Scoville}, N.~Z., {et~al.} 1989, \apjl, 337,
  L41

\bibitem[{{Wu} {et~al.}(2010){Wu}, {Evans}, {Shirley}, \& {Knez}}]{wu2010}
{Wu}, J., {Evans}, II, N.~J., {Shirley}, Y.~L., \& {Knez}, C. 2010, \apjs, 188,
  313

\bibitem[{{Wyrowski} {et~al.}(2010){Wyrowski}, {van der Tak}, {Herpin},
  {Baudry}, {Bontemps}, {Chavarria}, {Frieswijk}, {Jacq}, {Marseille},
  {Shipman}, {van Dishoeck}, {Benz}, {Caselli}, {Hogerheijde}, {Johnstone},
  {Liseau}, {Bachiller}, {Benedettini}, {Bergin}, {Bjerkeli}, {Blake},
  {Braine}, {Bruderer}, {Cernicharo}, {Codella}, {Daniel}, {di Giorgio},
  {Dominik}, {Doty}, {Encrenaz}, {Fich}, {Fuente}, {Giannini}, {Goicoechea},
  {de Graauw}, {Helmich}, {Herczeg}, {J{\o}rgensen}, {Kristensen}, {Larsson},
  {Lis}, {McCoey}, {Melnick}, {Nisini}, {Olberg}, {Parise}, {Pearson}, {Plume},
  {Risacher}, {Santiago}, {Saraceno}, {Tafalla}, {van Kempen}, {Visser},
  {Wampfler}, {Y{\i}ld{\i}z}, {Black}, {Falgarone}, {Gerin}, {Roelfsema},
  {Dieleman}, {Beintema}, {de Jonge}, {Whyborn}, {Stutzki}, \&
  {Ossenkopf}}]{wyrowski2010}
{Wyrowski}, F., {van der Tak}, F., {Herpin}, F., {et~al.} 2010, \aap, 521, L34

\bibitem[{{Yorke} \& {Sonnhalter}(2002)}]{yorke2002}
{Yorke}, H.~W. \& {Sonnhalter}, C. 2002, \apj, 569, 846

\bibitem[{{Zapata} {et~al.}(2012){Zapata}, {Loinard}, {Su}, {Rodr{\'{\i}}guez},
  {Menten}, {Patel}, \& {Galv{\'a}n-Madrid}}]{zapata2012}
{Zapata}, L.~A., {Loinard}, L., {Su}, Y.-N., {et~al.} 2012, \apj, 744, 86

\bibitem[{{Zinnecker} \& {Yorke}(2007)}]{zinnecker2007}
{Zinnecker}, H. \& {Yorke}, H.~W. 2007, \araa, 45, 481

\end{thebibliography}
\bibliographystyle{aa}

\Online

\begin{appendix}

\section{AOR list}

\label{AOR}
\begin{table*}
\caption{AORs list of Herschel observations for each source.}             
\label{AOR_list}      
\centering                          
\begin{tabular}{lcccc}        
\hline\hline                 
Transitions & NGC6334I(N) & DR21(OH) & IRAS16272 & IRAS05358 \\   
 \hline                        
   \hoA$^a$    & 1342205282 & 1342210764& 1342205524& 1342205275 \\
   \hoB    &   1342205279 & 1342192361 & 1342191556&1342194488 \\     
   \hoF     &   1342204518 &  1342195025 &1342203167 & 1342204509 \\
   \hoG    &  1342206384 & 1342196427 & 1342214418 & 1342206123\\
   \hoI    &  1342206383 & 1342197974/1342194794 & 1342214417/1342192584 &1342206126/1342206124 \\
   \hoJ    &  1342206384 & 1342196427 & 1342214418 & 1342206123\\
   \hoM    &   1342214455 & 1342192569 &1342192584 & 1342203954\\
\hline
   \hoC$^a$    &  1342205282 & 1342210764 & 1342205524 & 1342205275 \\
   \hoD    & 1342205847 & 1342194574 & 1342205845 & 1342194684\\
     \hoP    &   & 1342223425 & & \\
   \hoE     &  1342204519 & 1342195026 & 1342203168 & 1342204510\\
   \hoH    &   1342206384 & 1342196427 & 1342214418 &1342206123\\
   \hoK    & 1342206383 & 1342197974/1342194794 &1342214419/1342192584 &1342206126/1342206124\\
   \hoL    & 1342214455 &  1342192569& 1342192584 &1342203954\\
   \hoN    & 1342214455 & 1342192569 & 1342192584 & 1342203954\\
\hline                                  
\end{tabular}
\end{table*}



\section{By-product lines}
\label{sec:byproducts}

\begin{figure*}
\centering
\includegraphics[width=9cm]{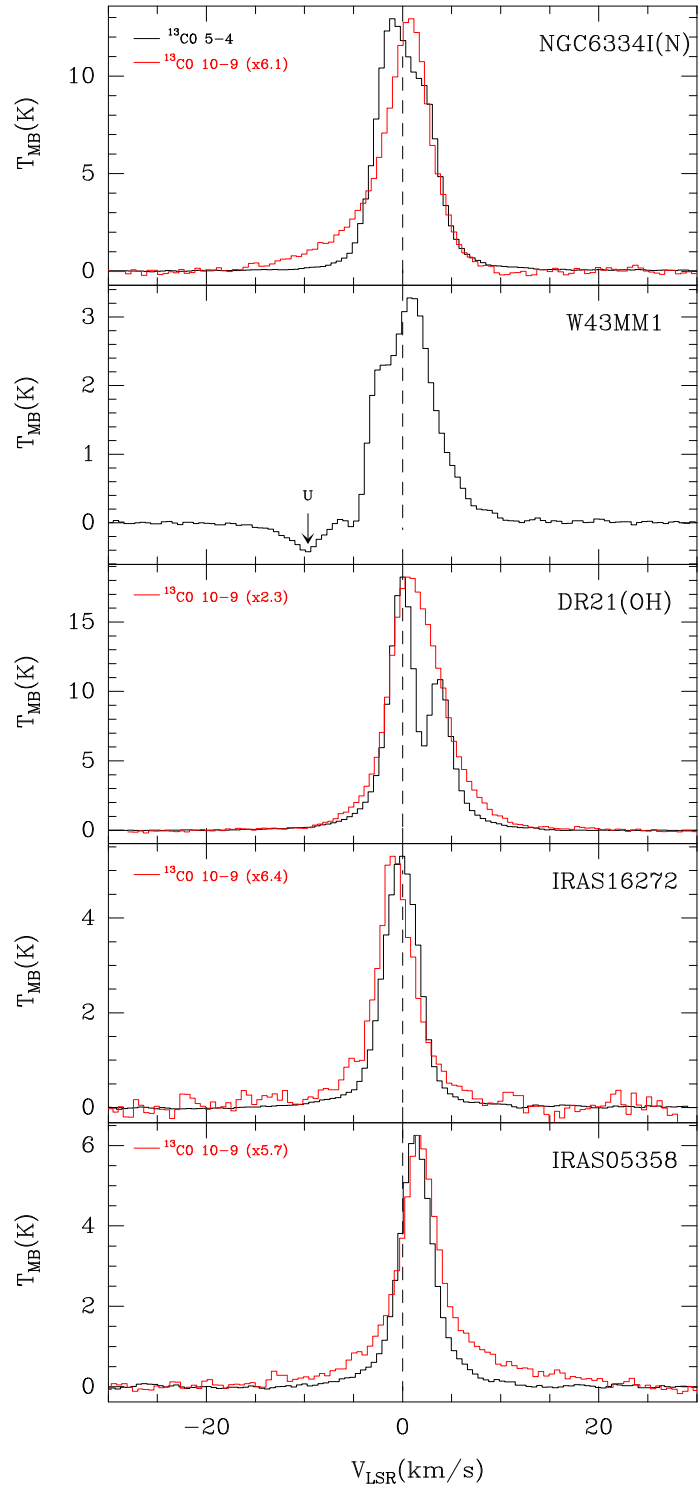}
\caption{HIFI spectra of $^{13}$CO J=5-4 and 10-9 lines (respectively in black and red). The spectra have been smoothed to 0.2 \kms. Vertical dotted lines indicate the \vlsr.}
\label{Figbyproducts13CO}%
\end{figure*}

\begin{figure*}
\centering
\includegraphics[width=9cm]{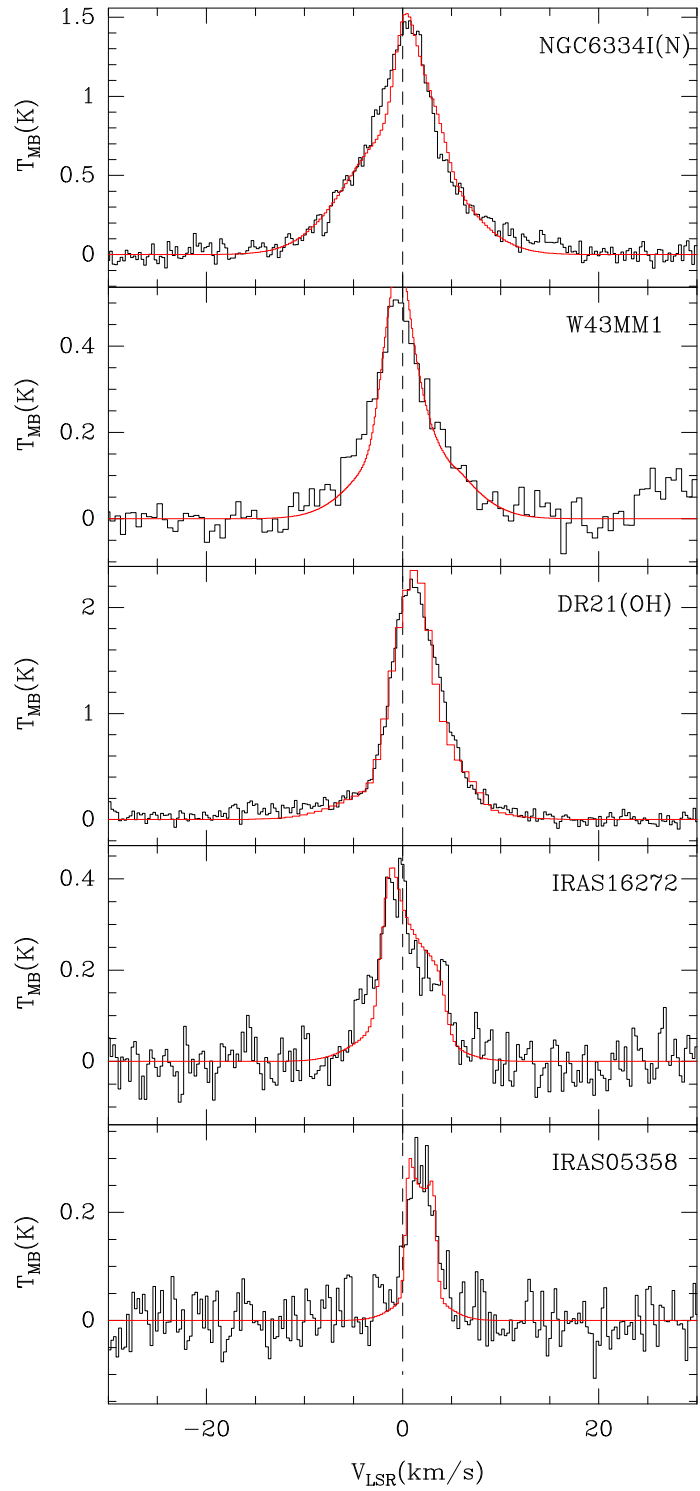}
\caption{HIFI spectra of the CS J=11-10 line. The best-fit model is shown as red over the spectra. Vertical dotted lines indicate the \vlsr. The spectra have been smoothed to 0.2 \kms.}
\label{FigbyproductsCS}%
\end{figure*}

\begin{figure*}
\centering
\includegraphics[width=9cm]{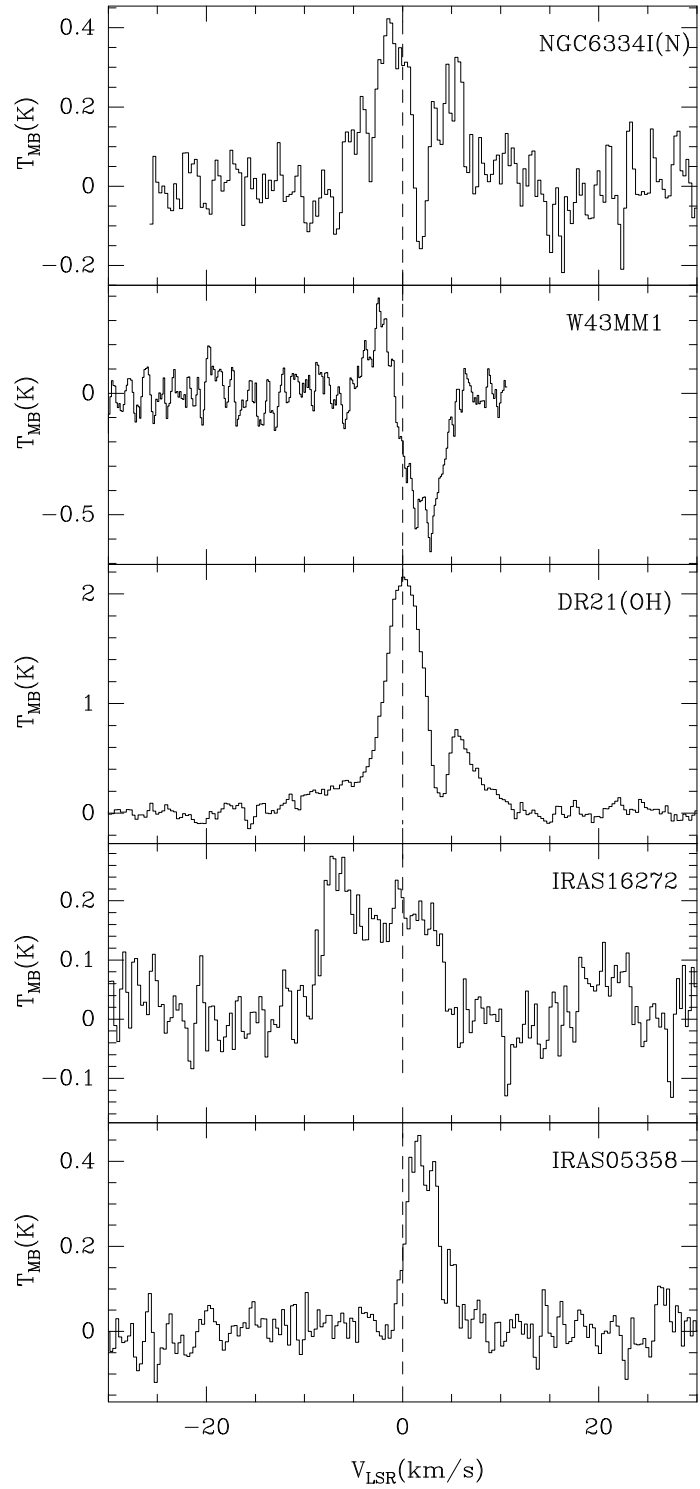}
\caption{HIFI spectra of H$_2$S $3_{0,3}-2_{1,2}$ line (blended with the H$_2$S $3_{0,3}-2_{1,2}$ line at 993.09701 GHz). The spectra have been smoothed to 0.2 \kms. Vertical dotted lines indicate the \vlsr.}
\label{FigbyproductsH2S}%
\end{figure*}

\begin{figure*}
\centering
\includegraphics[width=7.cm]{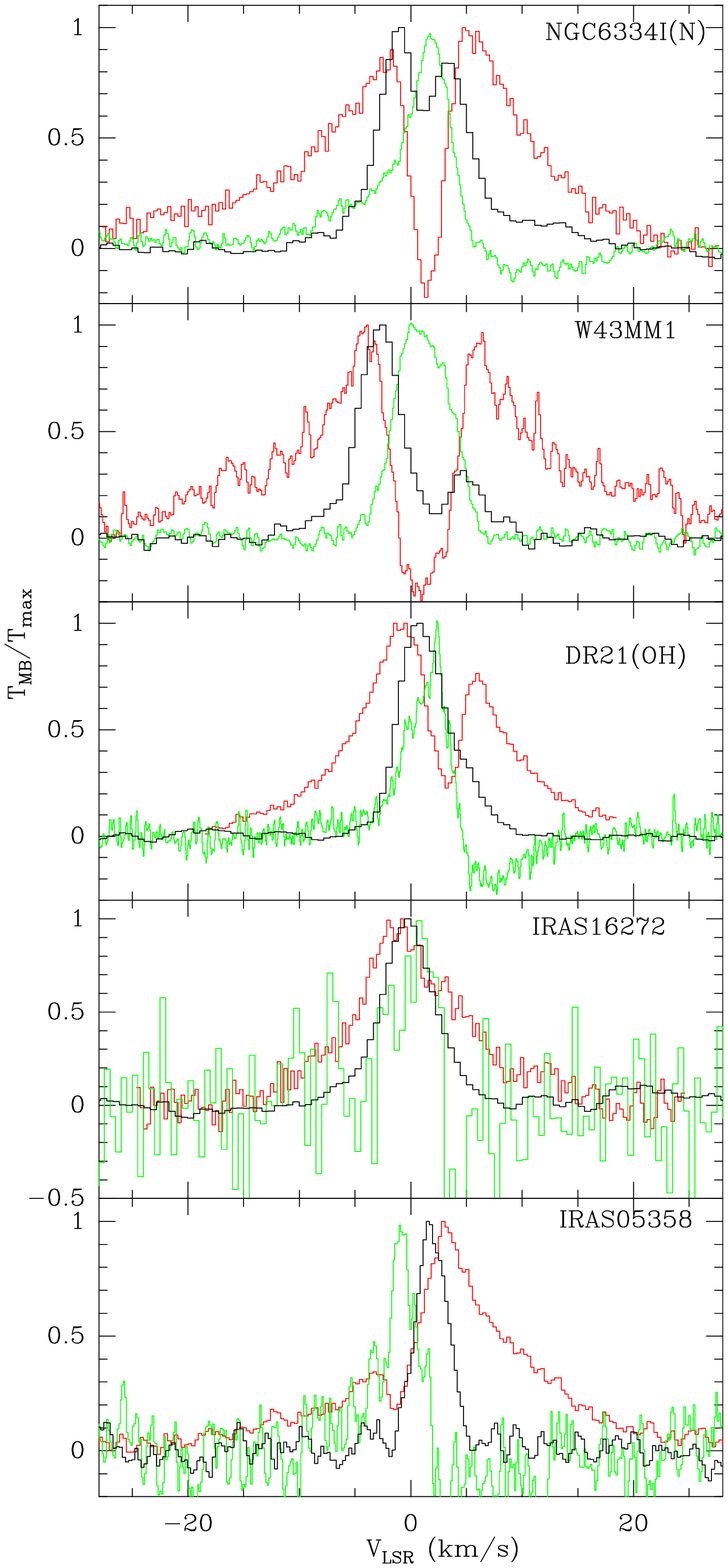}
\caption{HIFI nomalized spectra of CH$_3$OH $5_{1,5}-4_{0,4}~A_+-A_+$  line (in black) overplotted on the \hoE~ (red) and \hoI~(green and inverted) lines.}
\label{FigbyproductsCH3OH}%
\end{figure*}


\begin{table*}
  \caption{Observed line emission parameters for the detected "by-product" lines toward IRAS05358. $\varv_{LSR}$ is the Gaussian component peak velocity. $\Delta\varv$ are the velocity full width at half-maximum (FWHM) of the narrow, medium and broad components. }
\begin{center}
\label{byproduct_05358}      
\begin{tabular}{lccccccccc} \hline \hline
{\bf Species}  & Freq & $E_u$ & $\int T d\nu$ & $V_{nar}$ & $\Delta V_{nar}$ & $V_{med}$ & $\Delta V_{med}$ & $V_{br}$ & $\Delta V_{br}$  \\ 
 Line & [GHz] & [K] & [K.km$/$s] & [km$/$s] & [km$/$s] & [km$/$s] & [km$/$s] &  [km$/$s] &[km$/$s]   \\ 
\hline                        
 {\bf CH$_3$OH}    &  &   &  &  &  &  &  &  &  \\
$5_{1,5}-4_{0,4}~A_+-A_+$   & 538.57058 & 49.1  & 1.3$\pm$0.2 &  -15.7$\pm$0.5& 3.4$\pm$0.2 &  &  &  &  \\
 $6_{3,4}-5_{2,3}~A_+-A_+$   & 542.00098 & 98.5  & 0.89$\pm$0.2 & -16.0$\pm$0.5 & 4.0$\pm$0.5 & &  & &  \\
$6_{3,3}-5_{2,4}~A_--A_-$   & 542.08194 & 98.5  & 0.91$\pm$0.2 & -16.0$\pm$0.5 & 4.5$\pm$0.5 & &  & &  \\
 $8_{1,7}-7_{0,7}~E$   & 553.14630 & 104.6  & 0.6$\pm$0.1 &  &  & -14.2$\pm$0.5 & 6.2$\pm$0.9 & &  \\
$12_{1,11}-11_{0,11}~E$   & 751.56340 & 202.1  & 0.58$\pm$0.1 &  &  & -14.8$\pm$0.5 & 5.2$\pm$0.9 & &  \\
 $7_{4,4}-6_{3,3}~A_--A_-$   & 974.87674 & 145.3  & 1.1$\pm$0.2 & -16.3$\pm$0.3 & 4.8$\pm$0.6 & &  & &  \\
 $10_{2,8}-9_{1,9}~A_+-A_+$   & 986.09797 & 165.4  & 0.6$\pm$0.2 & -15.3$\pm$0.3 & 3.0$\pm$0.5 & &  & &  \\
 $5_{-5,0}-4_{-4,1}~E$   & 994.21725 & 158.8  & 0.7$\pm$0.2 &  &  & -16.0$\pm$0.4 & 5.8$\pm$0.8 & &  \\
\hline
{\bf $^{13}$CO}    &  &   &  &  &  &  &  &  &  \\
 $5-4$ & 550.92629 & 79.3 & 27$\pm$4 & -16.2$\pm$0.5& 3.3$\pm$0.5& -16.2$\pm$0.5 & 8.5$\pm$0.6& &  \\
$10-9$ & 1101.3497 & 290.8 & 7$\pm$1 & -15.9$\pm$0.3& 3.8$\pm$0.3&  & & -14.9$\pm$0.3& 14.6$\pm$0.6 \\
\hline
 {\bf  C$^{18}$O}    &  &   &  &  &  &  &  &  &  \\
 $9-8$  & 987.5604 & 237.0 & 1.0$\pm$0.8 & -16.3$\pm$0.3& 4.6$\pm$0.5& & & &   \\
$10-9$ & 1097.16288  & 289.7& 0.4$\pm$0.1 & -16.0$\pm$0.3& 3.5$\pm$0.3&  & & &  \\
\hline
 {\bf  CS}   &  &   &  &  &  &  &  &  &  \\
 $11-10$   & 538.6888 & 155.1  & 1.2$\pm$0.3 & -15.7$\pm$0.2 & 3.7$\pm$0.3 & & & &   \\     
\hline
{\bf  H$_2$O$^+$}    &  &   &  &  &  &  &  &  &  \\
$1_{11}-0_{00}~J=3/2-1/2$ & 1115.204 &  & -0.5$\pm$0.2 &  &  & & & -15.1$\pm$1.2$^a$ & 11$\pm$3  \\
\hline
{\bf  H$_2$S}    &  &   &  &  &  &  &  &  &  \\
$3_{0,3}-2_{1,2}$$^b$ & 993.10825 & 102.8 & 1.8$\pm$0.3 & -15.5$\pm$0.3 & 3.8$\pm$0.3 & & & &  \\
 \hline 
\end{tabular}
\end{center}

\tablefoot{$^a$ in absorption, $^b$ blended with H$_2$S $5_{2,3}-5_{1,4}$.}

\end{table*}

\begin{table*}
  \caption{Observed line emission parameters for the detected "by-product" lines toward IRAS16272. $\varv_{LSR}$ is the Gaussian component peak velocity. $\Delta\varv$ are the velocity full width at half-maximum (FWHM) of the narrow, medium and broad components. }
\begin{center}
\label{byproduct_16272}      
\begin{tabular}{lccccccccc} \hline \hline
{\bf Species}  & Freq & $E_u$ & $\int T d\nu$ & $V_{nar}$ & $\Delta V_{nar}$ & $V_{med}$ & $\Delta V_{med}$ & $V_{br}$ & $\Delta V_{br}$  \\ 
Line & [GHz] & [K] & [K.km$/$s] & [km$/$s] & [km$/$s] & [km$/$s] & [km$/$s] &  [km$/$s] &[km$/$s]   \\ 
\hline                        
 {\bf CH$_3$OH}    &  &   &  &  &  &  &  &  &  \\
$5_{1,5}-4_{0,4}~A_+-A_+$   & 538.57058 & 49.1  & 3.4$\pm$0.6 &  &  & -46.4$\pm$0.5 & 5.7$\pm$0.5 &  &  \\
 $6_{3,4}-5_{2,3}~A_+-A_+$   & 542.00098 & 98.5  & 1.8$\pm$0.4 & &  & -46.7$\pm$0.5 & 5.2$\pm$0.7 & &  \\
$6_{3,3}-5_{2,4}~A_--A_-$   & 542.08194 & 98.5  & 2.0$\pm$0.6 & & &  -46.6$\pm$0.6 & 5$\pm$01  & &  \\
$8_{1,7}-7_{0,7}~E$   & 553.14630 & 104.6  & 1.3$\pm$0.2 &  &  & -47.1$\pm$0.6 & 5.1$\pm$0.6 & &  \\
$18_{2,17}-18_{1,18}~A_--A_+$   & 553.57085 & 434.2  & 0.4$\pm$0.1 & -47.6$\pm$0.6 & 3.8$\pm$0.9 &  &  & &  \\
$12_{1,11}-11_{2,9}~E$   & 554.05552 & 202.1  & 2.6$\pm$0.6 & -42.9$\pm$0.6 & 2.9$\pm$0.6 & -43.7$\pm$0.6& 8.4$\pm$0.6 & &  \\
$12_{1,11}-11_{0,11}~E$   & 751.56340 & 202.1  & 0.8$\pm$0.1 & -41.8$\pm$0.5 & 3.8$\pm$0.5 &  &  & &  \\
$17_{-3,15}-17_{-2,16}~E$$^{b}$   & 753.86643 & 417.7  & 0.4$\pm$0.1 & -47.7$\pm$0.4 & 3.5$\pm$0.9 &  &  & &  \\
$7_{4,4}-6_{3,3}~A_--A_-$   & 974.87674 & 145.3  & 0.8$\pm$0.2 & -47.8$\pm$0.3 & 4.6$\pm$0.7 & &  & &  \\
$10_{2,8}-9_{1,9}~A_+-A_+$   & 986.09797 & 165.4  & 1.3$\pm$0.4 &  &  & -46.6$\pm$0.3& 6.7$\pm$0.8 & &  \\
$5_{-5,0}-4_{-4,1}~E$   & 994.21725 & 158.8  & 0.5$\pm$0.1 &  -47.2$\pm$0.4&  3.5$\pm$0.5&  &  & &  \\
$9_{7,3}-10_{6,5}~E$$^b$   & 995.92310 & 354.2  & 0.4$\pm$0.1 & -46.7$\pm$0.4 & 2.6$\pm$0.7 &  &  & &  \\
\hline
 {\bf $^{13}$CO}    &  &   &  &  &  &  &  &  &  \\
$5-4$ & 550.92629 & 79.3 & 26$\pm$4 & -46.3$\pm$0.5& 4.0$\pm$0.5&  & & -47.4$\pm$0.5&  12.$\pm$1\\
$10-9$ & 1101.3497 & 290.8 & 3.5$\pm$0.8 & -47.0$\pm$0.3& 3.4$\pm$0.3&  & & -46.8$\pm$0.3& 10.5$\pm$0.9 \\
\hline
 {\bf  C$^{18}$O}    &  &   &  &  &  &  &  &  &  \\
$9-8$   & 987.5604 & 237.0 & 2.$\pm$2 & & & -46.4$\pm$0.3 & 9.9$\pm$0.9  \\
\hline
 {\bf  CS}    &  &   &  &  &  &  &  &  &  \\
 $11-10$   & 538.6888 & 155.1 & 2.3$\pm$0.4& && -46$\pm$0.1& 7.0$\pm$1&  \\     
\hline
 {\bf  H$_2$O$^+$}    &  &   &  &  &  &  &  &  &  \\
     $1_{11}-0_{00}~J=3/2-1/2$ & 1115.204 &  & -8$\pm$1. &  &  & & & -34.5.1$\pm$0.5$^a$ & 23.7$\pm$0.7  \\
\hline
 {\bf  H$_2$S}    &  &   &  &  &  &  &  &  &  \\
    H$_2$S $3_{0,3}-2_{1,2}$$^c$ & 993.10825 & 102.8 & 0.9$\pm$0.4 & -48.6$\pm$0.4 & 3.7$\pm$0.7 & & & &  \\
 \hline 
\end{tabular}
\end{center}

\tablefoot{$^a$ in absorption, $^b$ band edge, $^c$ blended with H$_2$S $5_{2,3}-5_{1,4}$.}

\end{table*}

\begin{table*}
  \caption{Observed line emission parameters for the detected "by-product" lines toward NGC6334I(N). $\varv_{LSR}$ is the Gaussian component peak velocity. $\Delta\varv$ are the velocity full width at half-maximum (FWHM) of the narrow, medium and broad components. }
\begin{center}
\label{byproduct_6334}      
\begin{tabular}{lccccccccc} \hline \hline
{\bf Species}  & Freq & $E_u$ & $\int T d\nu$ & $V_{nar}$ & $\Delta V_{nar}$ & $V_{med}$ & $\Delta V_{med}$ & $V_{br}$ & $\Delta V_{br}$  \\ 
Line & [GHz] & [K] & [K.km$/$s] & [km$/$s] & [km$/$s] & [km$/$s] & [km$/$s] &  [km$/$s] &[km$/$s]   \\ 
\hline                        
 {\bf CH$_3$OH}    &  &   &  &  &  &  &  &  &  \\
$5_{1,5}-4_{0,4}~A_+-A_+$   & 538.57058 & 49.1  & 7.3$\pm$1.2 &  &  & -3.5$\pm$0.6 & 10$\pm$1 &  &  \\
$15_{0,15}-14_{1,13}~E$$^a$   &  540.92223 & 290.7  & 0.6$\pm$0.2 &  &  & -3.2$\pm$0.6 & 6.9$\pm$0.7 & &  \\
 $6_{3,4}-5_{2,3}~A_+-A_+$   & 542.00098 & 98.5  & 5.4$\pm$0.4 & -3.7$\pm$0.5& 4.9$\pm$0.5 &  &  & &  \\
$6_{3,3}-5_{2,4}~A_--A_-$   & 542.08194 & 98.5  & 5.2$\pm$1.1 & -3.8$\pm$0.5& 4.8$\pm$0.6 &   &   & &  \\
 $8_{1,7}-7_{0,7}~E$   & 553.14630 & 104.6  & 4.5$\pm$0.2 &  &  & -3.8$\pm$0.6 & 5.1$\pm$0.6 & &  \\
$12_{1,11}-11_{0,11}~E$   & 751.56340 & 202.1  & 14.4$\pm$0.4 &  &  & -6.8$\pm$0.5 & 7.6$\pm$0.5 & -7.6$\pm$0.5& 35$\pm$1 \\
 $7_{4,4}-6_{3,3}~A_--A_-$   & 974.87674 & 145.3  & 1.6$\pm$0.5 & -4.2$\pm$0.3 & 4.9$\pm$0.5 & &  & &  \\
 $10_{2,8}-9_{1,9}~A_+-A_+$   & 986.09797 & 165.4  & 1.1$\pm$0.3 & -3.9$\pm$0.3 & 4.0$\pm$0.4 & &  & &  \\
\hline
 {\bf CH$_3$OCHO}    &  &   &  &  &  &  &  &  &  \\
     $47_{6,42}-46_{6,41}$$^b$ & 551.18200  & 682.2 & 0.7$\pm$0.4 & -7.3$\pm$0.5 & 3.6$\pm$0.5 &  &   & &  \\
\hline
 {\bf $^{13}$CO}    &  &   &  &  &  &  &  &  &  \\
    $5-4$ & 550.92629 & 79.3 & 85$\pm$12 & & & -4.5$\pm$0.5 & 6.3$\pm$0.5 & &  \\
 $10-9$ & 1101.3497 & 290.8 & 14$\pm$2 & -3.7$\pm$0.3& 4.4$\pm$0.3&  & & -6.4$\pm$0.3& 11.6$\pm$0.3 \\
\hline
 {\bf C$^{18}$O}    &  &   &  &  &  &  &  &  &  \\
    $9-8$  & 987.5604 & 237.0& 1.7$\pm$0.5& -3.6$\pm$0.1& 2.6$\pm$0.3& & & &  \\
   $10-9$ & 1097.16288  & 289.7& 2.7$\pm$1.1 & & & -3.7$\pm$0.9 & 7$\pm$2& &  \\
\hline
  {\bf H$_2$S}    &  &   &  &  &  &  &  &  &  \\
   $3_{0,3}-2_{1,2}$$^c$ & 993.10825 & 102.8 & 1.4$\pm$0.4 & -5.7$\pm$0.4 & 3.3$\pm$0.4 & & & &  \\
 \hline 
   {\bf HDO}    &  &   &  &  &  &  &  &  &  \\
  $3_{1,2}-3_{0,3}$$^a$ & 753.41115 & 167.6 & 0.5$\pm$0.4 & -3.7$\pm$0.5& 4.1$\pm$0.7 &  & & &  \\
 \hline 
  {\bf CS}    &  &   &  &  &  &  &  &  &  \\
  $11-10$   & 538.6888 & 155.1 & 12.4$\pm$1.1& -4.0$\pm$0.1& 4.5$\pm$0.3& & & -4.7$\pm$0.2& 13.3$\pm$0.6 \\     
  \hline 
\end{tabular}
\end{center}

\tablefoot{$^a$ band edge, $^b$ blended with CH$_3$OCHO $47_{5,42}-46_{5,41}$, $^c$ blended with H$_2$S $5_{2,3}-5_{1,4}$.}

\end{table*}

\begin{table*}
  \caption{Observed line emission parameters for the detected "by-product" lines toward W43-MM1. $\varv_{LSR}$ is the Gaussian component peak velocity. $\Delta\varv$ are the velocity full width at half-maximum (FWHM) of the narrow, medium and broad components. }
\begin{center}
\label{byproduct_W43}      
\begin{tabular}{lccccccccc} \hline \hline
{\bf Species}  & Freq & $E_u$ & $\int T d\nu$ & $V_{nar}$ & $\Delta V_{nar}$ & $V_{med}$ & $\Delta V_{med}$ & $V_{br}$ & $\Delta V_{br}$  \\ 
Line & [GHz] & [K] & [K.km$/$s] & [km$/$s] & [km$/$s] & [km$/$s] & [km$/$s] &  [km$/$s] &[km$/$s]   \\ 
\hline                        
  {\bf CH$_3$OH}    &  &   &  &  &  &  &  &  &  \\
$5_{1,5}-4_{0,4}~A_+-A_+$   & 538.57058 & 49.1  & 3.0$\pm$1.0 &  &  & 98.0$\pm$1 & 5$\pm$1 &  &  \\
$6_{3,4}-5_{2,3}~A_+-A_+$   & 542.00098 & 98.5  & 3.4$\pm$0.9 & &  & 99.3$\pm$0.6 & 6$\pm$1 & &  \\
$6_{3,3}-5_{2,4}~A_--A_-$   & 542.08194 & 98.5  & 3.0$\pm$0.9 & &  &  99.2$\pm$0.6 & 6$\pm$1  & &  \\
 $8_{1,7}-7_{0,7}~E$   & 553.14630 & 104.6  & 2.9$\pm$0.2 &  &  & 98.8$\pm$0.5 & 7.1$\pm$0.5 & &  \\
$vt=1~5_{1,5}-5_{2,4}~A_+-A_-$   & 553.20160 & 359.9 & 0.6$\pm$0.1 &  &  & 100$\pm$1 & 8$\pm$3 & &  \\
$vt=1~4_{1,4}-4_{2,3}~A_+-A_-$   & 553.43748 & 348.4  & 0.6$\pm$0.1 &  &  & 99.5$\pm$0.5 & 7$\pm$1 & &  \\
 $18_{2,17}-18_{1,18}~A_--A_+$   & 553.57085 & 434.2  & 0.4$\pm$0.1 & &  & 102.1$\pm$0.6  & 6$\pm$1 & &  \\
$12_{1,11}-11_{2,9}~E$   & 554.05552 & 202.1  & 1.5$\pm$0.4 &  &  & 99.5 $\pm$0.6& 8$\pm$1 & &  \\
$12_{1,11}-11_{0,11}~E$   & 751.56340 & 202.1  & 1.6$\pm$0.4 &  &  & 104.4$\pm$0.5 & 7.7$\pm$0.6 & &  \\
$7_{4,4}-6_{3,3}~A_--A_-$   & 974.87674 & 145.3  & 2.8$\pm$0.6 &  & & 97.1$\pm$0.3& 6.9$\pm$0.4 & &  \\
$10_{2,8}-9_{1,9}~A_+-A_+$   & 986.09797 & 165.4  & 3.1$\pm$0.6 &  &  & 98.4$\pm$0.3& 5.5$\pm$0.3 & 96.7$\pm$0.4& 13.1$\pm$0.3 \\
$5_{-5,0}-4_{-4,1}~E$   & 994.21725 & 158.8  & 0.30$\pm$0.07 &  98.1$\pm$0.3&  2.6$\pm$0.3&  &  & &  \\
\hline
 {\bf  CH$_3$OCHO}    &  &   &  &  &  &  &  &  &  \\
    $47_{6,42}-46_{6,41}$$^c$ & 551.18200  & 682.2 & 0.5$\pm$0.1 &  &  & 98.1$\pm$0.5 & 5$\pm$1  & &  \\
\hline
 {\bf  $^{13}$CO}    &  &   &  &  &  &  &  &  &  \\
    $5-4$ & 550.92629 & 79.3 & 22$\pm$5 & & & 99.3$\pm$0.6 & 6.5$\pm$0.6 & &  \\
$10-9$$^b$  & 1101.3497 & 290.8 &  & & &  & & &  \\
\hline
 {\bf  C$^{18}$O}    &  &   &  &  &  &  &  &  &  \\
   $9-8$  & 987.5604 & 237.0&  1.7$\pm$0.9& & & 99.4$\pm$0.2& 5.6$\pm$0.3 & &  \\
$10-9$ & 1097.16288  & 289.7& 0.7$\pm$0.1 & 99.9$\pm$0.3 & 4.5$\pm$0.4&  & & &  \\
\hline
 {\bf  CS}    &  &   &  &  &  &  &  &  &  \\
  $11-10$   & 538.6888 & 155.1& 3.5$\pm$0.8 & 98.3$\pm$0.3& 2.9$\pm$0.9& & & 98.8$\pm$0.4& 10$\pm$1 \\     
\hline
 {\bf  H$_2$S}    &  &   &  &  &  &  &  &  &  \\
     $3_{0,3}-2_{1,2}$$^d$ & 993.10825 & 102.8 & -0.9$\pm$0.1 & &  &98.1$\pm$0.5  &7$\pm$1 & & \\
\hline
 {\bf  H$_2$O$^+$}    &  &   &  &  &  &  &  &  &  \\
     $1_{11}-0_{00}~J=3/2-1/2$$^a$ & 1115.204 &  & -8.2$\pm$1.2 & 96.7$\pm$0.3 & 3.2$\pm$0.3 & 100.6$\pm$0.3& 9.3$\pm$0.3& &   \\
\hline
 {\bf  H$_3$O$^+$}    &  &   &  &  &  &  &  &  &  \\
     $0(0)-1(0),0^--0^+$$^a$ & 984.70866 & 54.6 & -2.1$\pm$0.4 &  &  &98.0$\pm$0.3 & 7.8$\pm$0.5& &  \\
\hline 
 {\bf  HDO}    &  &   &  &  &  &  &  &  &  \\
   $3_{0,3}-2_{1,2}$ & 995.41150 & 131.4 & 0.4$\pm$0.7 & 98.8$\pm$0.2 & 2.7$\pm$0.4&  & &  &  \\
  \hline 
\hline 
\end{tabular}
\end{center}

\tablefoot{$^a$ in absorption, $^b$ blended with H$_2$O$^+$, $^c$ blended with CH$_3$OCHO $47_{5,42}-46_{5,41}$, $^d$ blended with H$_2$S $5_{2,3}-5_{1,4}$.}

\end{table*}

\begin{table*}
  \caption{Observed line emission parameters for the detected "by-product" lines toward DR21(OH). $\varv_{LSR}$ is the Gaussian component peak velocity. $\Delta\varv$ are the velocity full width at half-maximum (FWHM) of the narrow, medium and broad components. }
\begin{center}
\label{byproduct_DR21OH}      
\begin{tabular}{lccccccccc} \hline \hline
{\bf Species}  & Freq & $E_u$ & $\int T d\nu$ & $V_{nar}$ & $\Delta V_{nar}$ & $V_{med}$ & $\Delta V_{med}$ & $V_{br}$ & $\Delta V_{br}$  \\ 
 Line & [GHz] & [K] & [K.km$/$s] & [km$/$s] & [km$/$s] & [km$/$s] & [km$/$s] &  [km$/$s] &[km$/$s]   \\ 
\hline                        
   {\bf CH$_3$OH}    &  &   &  &  &  &  &  &  &  \\
$5_{1,5}-4_{0,4}~A_+-A_+$   & 538.57058 & 49.1  & 8.5$\pm$1.2 &  &  & -3.3$\pm$0.5 & 5.5$\pm$0.7 &  &  \\
$6_{3,4}-5_{2,3}~A_+-A_+$   & 542.00098 & 98.5  & 6.1$\pm$1.2 & &  & -2.8$\pm$0.5 & 5.9$\pm$0.9 & &  \\
$6_{3,3}-5_{2,4}~A_--A_-$   & 542.08194 & 98.5  & 6.0$\pm$1. & &  &  -2.9$\pm$0.5 & 6.1$\pm$0.5  & &  \\
$8_{1,7}-7_{0,7}~E$   & 553.14630 & 104.6  & 3.8$\pm$0.2 &  &  & -3.0$\pm$0.3 & 5.3$\pm$0.3 & &  \\
 $vt=1~5_{1,5}-5_{2,4}~A_+-A_-$   & 553.20160 & 359.9 & 0.4$\pm$0.1 & & & -4$\pm$1 & 5$\pm$2 & &   \\
 $18_{2,17}-18_{1,18}~A_--A_+$   & 553.57085 & 434.2  & 0.7$\pm$0.2 & -3.6$\pm$0.5 & 2.5$\pm$0.7 &   & &1$\pm$1 &  11$\pm$2 \\
$12_{1,11}-11_{2,9}~E$   & 554.05552 & 202.1  & 0.8$\pm$0.1 &  &  & -2.5 $\pm$0.3 & 5.3$\pm$0.5 & &  \\
$12_{1,11}-11_{0,11}~E$   & 751.56340 & 202.1  & 2.6$\pm$0.5 &  &  & 2.3$\pm$0.3 & 6.1$\pm$0.4 & &  \\
$11_{-2,10}-10_{-1,10}~E$   & 959.34565 & 179.2  & 2.4$\pm$1.2 &  &  &  -2.8$\pm$0.9 & 6$\pm$2 & &  \\
$9_{-4,6}-8_{-3,6}~E$   & 959.90049 & 192.3 & 3.2$\pm$0.7 &  &  &  -2.7$\pm$0.4 & 6.3$\pm$0.4 & &  \\
$20_{-2,19}-19_{-2,18}~E$   & 970.83461 & 514.2  & 0.8$\pm$0.2 &  &  &  -2.9$\pm$0.4 & 5.5$\pm$0.8 & &  \\
 $15_{3,13}-14_{2,12}~A_+-A_+$   & 974.67316 & 328.3  & 2.4$\pm$1.2 &  &  & -3$\pm$1& 7$\pm$3 & &  \\
$7_{4,4}-6_{3,3}~A_--A_-$   & 974.87674 & 145.3  & 3.5$\pm$0.6 & -3.8$\pm$0.3 & 4.6$\pm$0.5& &  & &  \\
$10_{2,8}-9_{1,9}~A_+-A_+$   & 986.09797 & 165.4  & 3.4$\pm$0.6 &  &  & -3.0$\pm$0.6& 5$\pm$1 & &  \\
$5_{-5,0}-4_{-4,1}~E$   & 994.21725 & 158.8  & 6.5$\pm$0.07 & -3.4$\pm$0.3&  4.0$\pm$0.4&  &  & -2.3$\pm$0.3& 13.8$\pm$0.5 \\
\hline
  {\bf CH$_3$OCH$_3$}    &  &   &  &  &  &  &  &  &  \\
    $19_{7,13}-19_{6,12}$$^e$ & 740.13792  & 241.8 & 2.2$\pm$0.1 &  &  &   -4.2$\pm$0.5& 10$\pm$1 &&\\
\hline
  {\bf CH$_3$OCHO}    &  &   &  &  &  &  &  &  &  \\
     $47_{6,42}-46_{6,41}$$^f$ & 551.18200  & 682.2 & 2.1$\pm$0.5 &  &  & -3.9$\pm$0.4 & 8.1$\pm$0.9  & &  \\
\hline
   {\bf $^{13}$CO}    &  &   &  &  &  &  &  &  &  \\
   $5-4$ & 550.92629 & 79.3 & 89$\pm$18 & -2.5$\pm$0.5 & 3.1$\pm$0.5 & -2.5$\pm$0.5 & 9.2$\pm$0.6 & &  \\
   $10-9$ & 1101.3497 & 290.8 & 57$\pm$12 & -3.8$\pm$0.3& 3.9$\pm$0.3& -2.6$\pm$0.3 & 8.1$\pm$0.3& &  \\
\hline
   {\bf C$^{18}$O}    &  &   &  &  &  &  &  &  &  \\
    $9-8$  & 987.5604 & 237.0& 12.1$\pm$1.5& & & -3.1$\pm$0.1& 5.6$\pm$0.1 & &  \\
  $10-9$ & 1097.16288  & 289.7& 8$\pm$2 & & & -3.2$\pm$0.7 & 6$\pm$1&  &   \\
\hline
   {\bf $^{13}$CS}    &  &   &  &  &  &  &  &  &  \\
   $24-23$ & 1107.51176  & 665.1& 1.5$\pm$0.4 & & & -5.2$\pm$0.3 & 9.9$\pm$0.7 & &  \\
   {\bf CS}    &  &   &  &  &  &  &  &  &  \\
  $11-10$   & 538.6888 & 155.1&13.3$\pm$1.1& & & -3.2$\pm$0.2& 6.1$\pm$0.2& &  \\     
\hline
   {\bf H$_2$O$^+$}    &  &   &  &  &  &  &  &  &  \\
     $1_{11}-0_{00}~J=3/2-1/2$$^a$ & 1115.204 &  & -37$\pm$7 &  &  & 7.6$\pm$0.3& 6.9$\pm$0.3& 2.5$\pm$0.3& 33.7$\pm$0.3  \\
\hline
   {\bf CH$^+$}    &  &   &  &  &  &  &  &  &  \\
     $2-1$$^a$ & 1669.28129 & 120.2 & -3.2$\pm$0.7 & -2.3$\pm$0.9 & 2.6$\pm$0.2 & & & &   \\
\hline
   {\bf OH$^+$}    &  &   &  &  &  &  &  &  &  \\
     $1-0$$^a$ & 971.800 &  & -64$\pm$7 & &  & & & 1.0$\pm$0.5&  20.6$\pm$0.3  \\
\hline
   {\bf H$_2$S}    &  &   &  &  &  &  &  &  &  \\
     $3_{0,3}-2_{1,2}$$^g$ & 993.10825 & 102.8 & 9$\pm$2 & -4.3$\pm$0.5& 4.0$\pm$0.4 &  & & & \\
  \hline 
   {\bf H$_2$CO}    &  &   &  &  &  &  &  &  &  \\
     $13_{1,12}-12_{1,11}$ & 970.19918 & 339.0 & 2.2$\pm$0.6 & & &  -2.3$\pm$0.6& 10$\pm$1& &  \\
\hline 
   {\bf HDO}    &  &   &  &  &  &  &  &  &  \\
   $3_{0,3}-2_{1,2}$ & 995.41150 & 131.4 & 4.0$\pm$0.7 & & &  & & -1.3$\pm$0.3 & 11.4$\pm$0.4 \\
    $3_{1,2}-3_{0,3}$$^c$ & 753.41115 & 167.6 & 1.4$\pm$0.2 & &  &  -1.5$\pm$0.3& 7.2$\pm$0.5& &  \\
\hline
   {\bf OS$^{18}$O}    &  &   &  &  &  &  &  &  &  \\
     $20_{8,13}-19_{7,12}$$^d$ & 1108.60656 & 352.3 & 1.4$\pm$0.3 & &  & -3.4$\pm$0.4 & 7.6$\pm$0.9 & &  \\
\hline
   {\bf SO$_2$}    &  &   &  &  &  &  &  &  &  \\
     $10_{6,4}-9_{5,5}$ & 753.06035 & 138.8 & 0.9$\pm$0.4 & &  & -0.9$\pm$0.3 & 5.3$\pm$0.5 & &  \\
    $13_{9,5}-12_{8,4}$$^c$ & 1113.50582 & 281.9 & 0.5$\pm$0.4 & &  & -0.3$\pm$0.3 & 5$\pm$1 & &  \\
     $12_{9,3}-11_{8,4}$ & 1094.32950 & 269.9 & 3.5$\pm$1. & 2.7$\pm$0.4& 4.0$\pm$0.6 & -7.6$\pm$0.3 & 10.0$\pm$0.6 & &  \\
    $11_{8,4}-10_{7,3}$ & 	974.60251 & 217.4 & 1.4$\pm$1.0 & & & 0$\pm$1 & 5$\pm$2 &  &   \\
    $12_{8,4}-11_{7,5}$ & 	993.76628 & 228.5 & 1.8$\pm$1.0 &  &  & -1.8$\pm$0.4 & 8.9$\pm$0.9 & &  \\
   $16_{7,9}-15_{6,10}$ & 	969.22389 & 245.1 & 0.9$\pm$0.2 &  &  & -0.4$\pm$0.4 & 5.8$\pm$0.6 & &  \\
   $20_{14,6}-21_{13,9}$ & 	956.82941 & 670.3 & 0.5$\pm$0.2 &  -2.5$\pm$0.4&2.8$\pm$0.7  &  & & &  \\
     $31_{1,31}-30_{0,30}$$^c$ & 554.21278 & 431.5 & 1.1$\pm$0.6 & &  & 1.3$\pm$0.5 & 9$\pm$2 & &  \\
\hline
   {\bf $^{34}$SO}    &  &   &  &  &  &  &  &  &  \\
     $26_{26}-25_{25}$ & 1095.06010 & 723.3 & 0.6$\pm$0.4 & &  & -3.3$\pm$0.6 & 6$\pm$1 & &  \\
  \hline 
 \hline 
\end{tabular}
\end{center}

\tablefoot{$^a$ in absorption, $^c$ band edge, $^d$ blended with OS$^{18}$O $20_{8,12}-19_{7,13}$, $^e$ blended with CH$_3$OCH$_3$ $19_{7,12}-19_{6,13}$, $^f$ blended with CH$_3$OCHO $47_{5,42}-46_{5,41}$, $^g$ blended with H$_2$S $5_{2,3}-5_{1,4}$.}

\end{table*}


\section{Gaussian fit components for IRAS05358}
\label{sec:gaussian}

\begin{figure*}
\centering
\includegraphics[width=9cm]{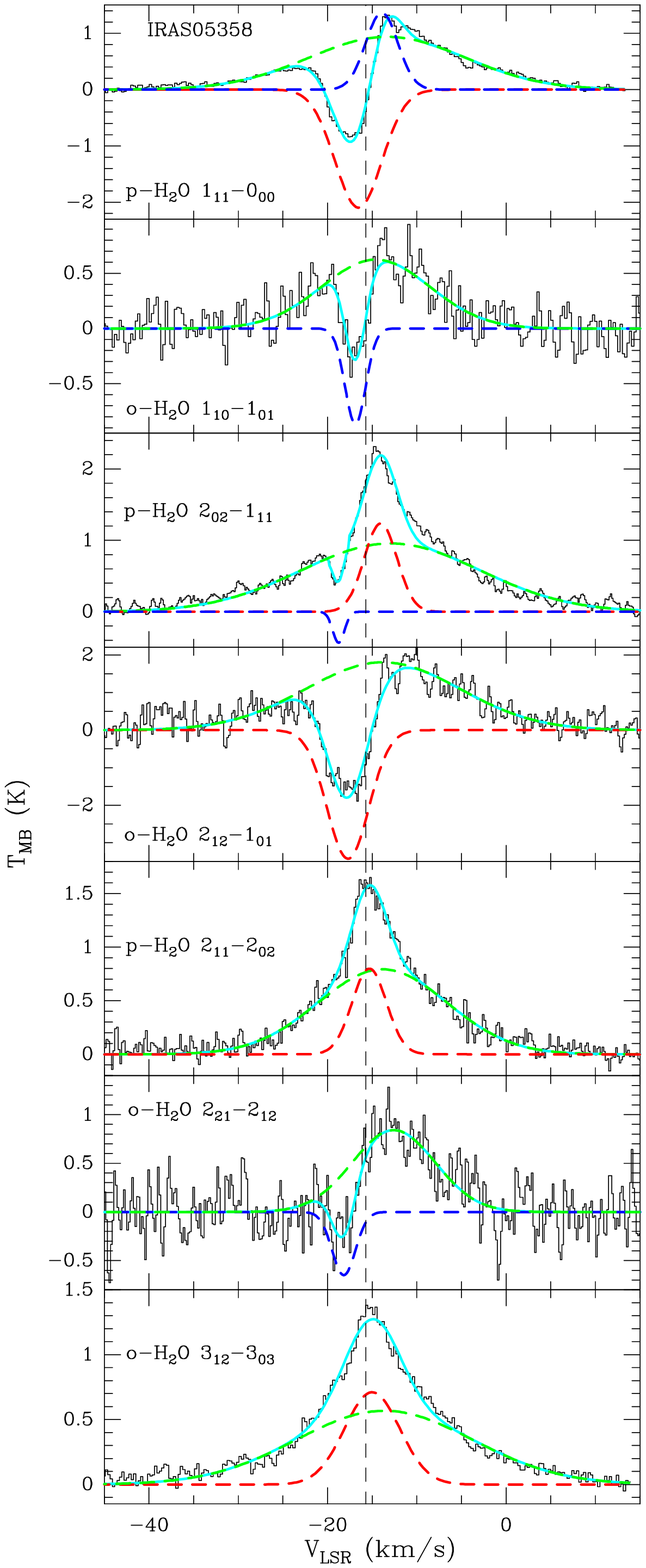}
\caption{Gaussian fit components (red, blue, and green) for \water~lines in IRAS05358. The light blue curve stands for the total fit.}
\label{FigGaussian}%
\end{figure*}

\section{CS distribution}
\label{sec:CS_dis}

We have also applied the model to the CS 11-10 line. In addition we tried to reproduce other CS lines observed by other authors:  J=7-6 for W43-MM1 \citep[][]{herpin2009} and NGC6334I(N) \citep[]{mccutcheon2000}, J=5-4 for IRAS05358 and W43-MM1 \citep[][]{herpin2009}, DR21(OH) \citep[][]{richardson1994}, J=3-2 for IRAS05358 \citep[][]{herpin2009}, and J=2-1 for IRAS16272 (Mopra observation, Herpin et al., unpublished). Hence, the sources IRAS05358 and W43-MM1 are the best constrained due to the number of CS lines available. Results for CS 11-10 line are shown on Fig.\ref{FigbyproductsCS}. For all sources but DR21(OH), a jump of abundance is necessary to reproduce the lower-J CS lines (7-6, 5-4, 3-2, and 2-1). The CS inner abundance  is $4\times10^{-8}$ for IRAS05358, IRAS16272, and DR21(OH) while the abundance is lower for NGC6334I(N) (an order of magnitude: $4\times10^{-9}$) and W43-MM1 ($9\times10^{-9}$).  All parameters are given in Table \ref{output_modCS}. We have first modeled the CS 11-10 line solely with constant turbulent and infall velocities, but, as for water, adopting a turbulence increasing with the radius (but constant infall) improves the line fitting, specially when including the other CS lines. The resulting $V_{turb}$ profiles are overplotted on Fig. \ref{Vturb_plot} with turbulence inferred from water lines, but should be considered with caution as the CS model is less constrained than water's one due to a significantly lower number of lines involved in the process. The variation of these "two" turbulent velocities is not strictly identical, the obvious difference being a lower turbulence for CS lines. This could be consistent with CS not tracing exactly the same gas than water lines, being located more in the inner part of the enveloppe where the turbulence is lower: as proposed in Sect. \ref{meth}, water is produced by gas-phase synthesis of H$_2$O from shocked material in the outflow cavity and CS likely not.

Compared to previous publications, our CS abundances are larger by almost two orders of magnitude for IRAS05358 \citep[]{herpin2009} and DR21(OH) \citep[][]{richardson1994}, but consistent with \citet{cortes2011} for W43-MM1. These CS abundances are in agreement with predictions from chemical model \citep[][]{roberts2010}: the lower CS abundance in the less evolved mid-IR quiet sources, NGC6334I(N) and W43-MM1, might indicate than CS is depleted on the grain mantles in these younger objects. The abundance of CS is dependent on the evaporation of sulfur from the ice mantles by radiation from the central star.

\begin{table*}
\caption{CS parameters derived from model. The turbulent and infall velocities are the nominal constant values for modeling the CS 11-10 line solely. }             
\label{output_modCS}      
\centering                          
\begin{tabular}{lccccc}        
\hline\hline                 
Object &  $\chi_{in}$  &   $\chi_{out}$ &$V_{turb}$ & $V_{inf}$ & $\dot{M}_{acc}$ \\   
                      & $10^{-8}$ & $10^{-8}$ & [\kms] &   [\kms] &  [M$_{\odot}$.yr$^{-1}$]\\
                      \hline
NGC6334I(N)  & 0.4  & 0.1 & 1.5 & -2.4 & $2.5\times10^{-4}$\\
W43-MM1  & 0.9 & 0.03 & 2.4 & -1.9 & $2\times10^{-2}$\\
DR21(OH)  & 4.0 & 3.0 & 1.6 & -2.0   & $6.6\times10^{-5}$\\ 
IRAS16272  & 3.5 & 0.4 & 1.6 & -0.8  & $3.7\times10^{-5}$\\
IRAS05358  & 4.0 & 0.7 &0.8 & 0.0  & \\
\hline                                  
\end{tabular}
\end{table*}

\section{Methanol lines}

\begin{figure*}
\centering
\includegraphics[width=9.cm]{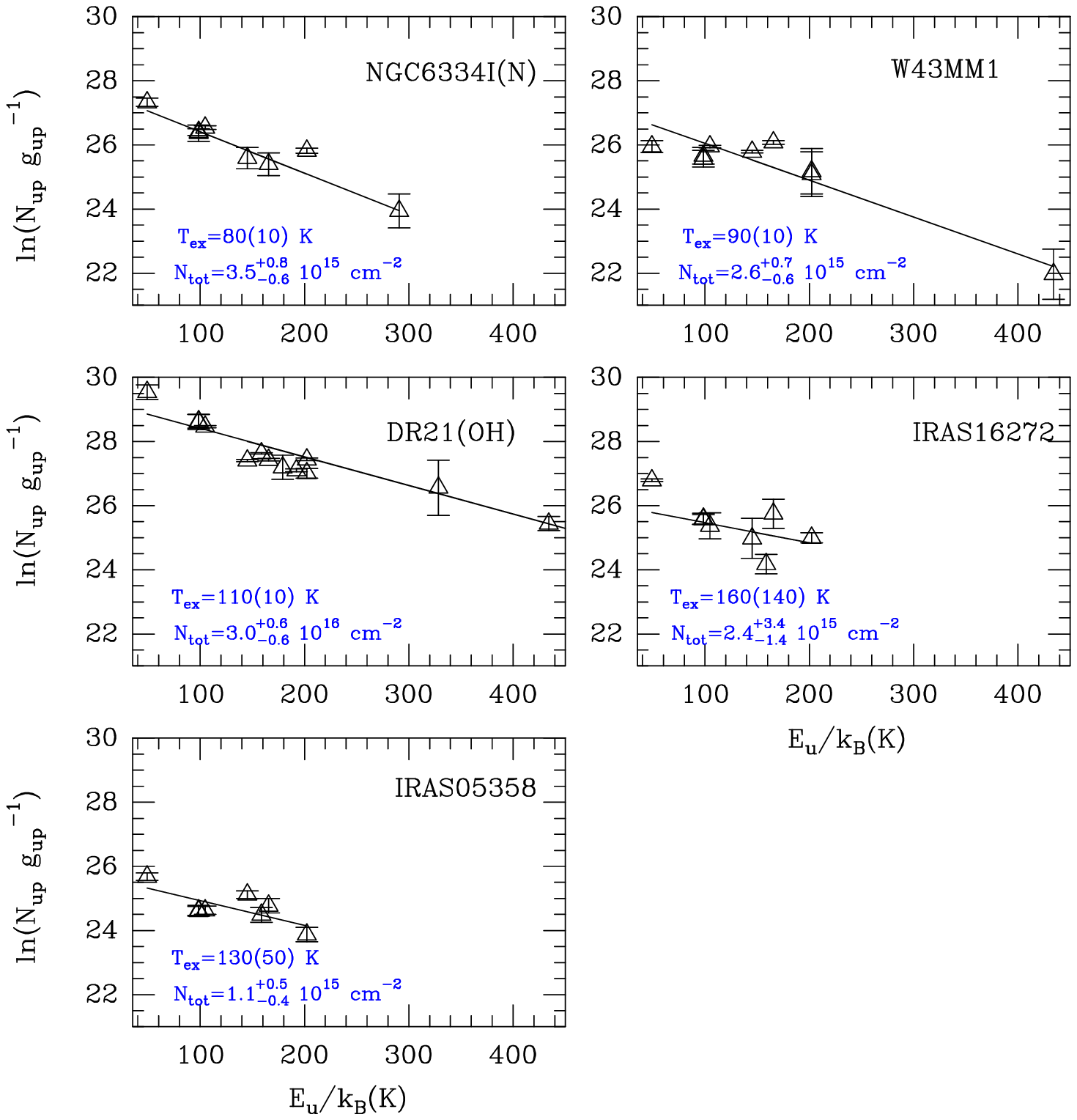}
\caption{Rotational diagrams for methanol lines ($v_t=0$). Sources are ordered following the evolutionary sequence.}
\label{Diag_rot}
\end{figure*}

 From the CH$_3$OH ($v_t=0$) integrated lines fluxes, corrected from the beam dilution (reference beam is $21.1\arcsec$), and assuming LTE, we have made the rotational diagrams for all sources (see Fig. \ref{Diag_rot}) up to E$_{up}/$k$_B=440$ K for DR21(OH) and W43-MM1, and up to E$_{up}/$k$_B=200$ K for IRAS16272 and IRAS05358. We have estimated the rotational temperature and the total methanol column density, using the partition function from \citet{villanueva2012}. All rotational diagrams can be described by a single temperature, but $\textrm{ln}(N_{up}/g_{up})$ values are scattered for W43-MM1 and IRAS16272 either because of opacity effects or of non-LTE conditions. Even for our very limited sample of methanol transitions, we have attempted to iteratively correct individual $N_{up}/g_{up}$ values by muliplying by the optical depth correction factor after the method of \citet{goldsmith1999}. We find that the corrected optical depth is always $<<1$, and the correction factors are less than $1\%$. Taking into account the uncertainties, we note that the rotational temperature possibly increase with the evolutionary sequence assumed in this paper, hence underlining the increasing temperature in the protostellar envelope. Beam-averaged methanol column densities are a few $10^{15}$ cm$^{-2}$ for all sources, except for DR21(OH) where the column density is one order of magnitude larger. The fact that we detect torsionally excited methanol lines (v=1) in DR21(OH) and W43-MM1 and that the number of detected CH$_3$OH lines increases with the source luminosity reveal the role of the IR pumping on the methanol emission \citep[][]{leurini2007}.
 
\end{appendix}

\end{document}